\documentclass[12pt, a4paper]{article}
\usepackage[utf8]{inputenc} % if your are submitting a pdflatex (i.e. if you have
\usepackage{chet}
\usepackage{graphicx}
\usepackage{authblk}
\usepackage{slashed}
\usepackage{indentfirst}
\usepackage{amsmath}
\usepackage{amsfonts}
\usepackage{amssymb}
\usepackage{courier}
\usepackage{hyperref}

\let\ampersand\&
\renewcommand*\&{and}

\usepackage[T1]{fontenc} % if needed
\usepackage{xcolor}

\definecolor{mygreen}{rgb}{0.0,1.0,0.0}
\definecolor{britishracinggreen}{rgb}{0.0, 0.26, 0.15}
\definecolor{cornsilk}{rgb}{1.0,1.0,1.0}
\definecolor{cream}{rgb}{1.0, 0.99, 0.82}
\definecolor{daffodil}{rgb}{1.0, 0.9,0}
\definecolor{cordovan}{rgb}{0.54, 0.25, 0.27}
\definecolor{carnelian}{rgb}{0.7, 0.11, 0.11}

\numberwithin{equation}{section}

\title{Supersymmetric String Configurations in
Ad$\mathbb{S}_5\times\mathbb{S}^5$}
\author{Lucas N. S. Martins$^a$\footnote[1]{\texttt{lucas\_nmartins@hotmail.com}}}
\affil{$^a$\textit{International Institute of Physics, UFRN - Univ. Federal do Rio Grande do Norte}\\
\textit{Anel Viário da UFRN - Lagoa Nova, Natal, RN, Brazil}}

\abstract{In this work, we investigate off-shell supersymmetric string configurations in Euclidean Ad$\mathbb{S}_5\times \mathbb{S}^5$ background. By extending the Green-Schwarz supersymmetric equations using the pure spinor formalism, we demonstrate that while the solutions to the Green-Schwarz equations are necessarily minimal surfaces, the solutions to the extended equations form infinite-dimensional families, even with fixed boundary conditions, thus preventing direct derivations of exact results via supersymmetric localization. For concreteness, we have selected a supercharge that localizes the off-shell supersymmetric configurations to an Ad$\mathbb{S}_4 \times \mathbb{S}^2$ slice and observed that the corresponding action functional reduces to an integral of a two-form.}

\begin{document} 
\maketitle
\section{Introduction}

What gives superstring theory its predictive power is its lack of free parameters. Especially for a theory that lacks direct experimental contact due to current technological limitations, too many parameters could become vacuous. One could introduce free parameters at will to align with the desires of the scientific community, for example, to accommodate predictions derived from Maldacena's conjecture that the four-dimensional maximally supersymmetric gauge theory on $\mathbb{S}^4$ is dual to type IIB superstrings in the Ad$\mathbb{S}_5 \times \mathbb{S}^5$ background \cite{Maldacena:1997re}.  
\vskip 0.3cm
Unfortunately, our current understanding of superstring theory in Ramond-Ramond backgrounds, such as Ad$\mathbb{S}_5 \times \mathbb{S}^5$, is precarious, especially due to the complex manner in which spacetime supersymmetries of flat background are represented in the Ramond-Neveu-Schwarz formalism \cite{Friedan:1985ge}. Additionally, each alternative formalism suffers from its own issues. For instance, the Green-Schwarz formalism cannot be covariantly quantized \cite{Green:1983wt}\cite{Berkovits:2002zv}, while the pure spinor formalism \cite{Berkovits:2000fe}\cite{Berkovits:2005bt}, despite recent developments \cite{Berkovits:2022qhc}\cite{Berkovits:2022dbm}\cite{Chandia:2023eel}, remains somewhat obscure. Nevertheless, if one wishes to investigate off-shell superstrings in the Ad$\mathbb{S}_5 \times \mathbb{S}^5$ background, the pure spinor formalism appears to be the best option available to date, and we will proceed within this framework. 
\vskip 0.3cm
There are several reasons why this formalism might be the best option, and here we mention just a few. First, the spacetime supersymmetries are manifest in the formalism, allowing for the straightforward construction of Ramond-Ramond integrated vertex operators with zero ghost number, which are readily usable to deform the flat background sigma model action \cite{Berkovits:2000fe}. Second, the formalism performs well in flat backgrounds, including the computation of scattering amplitudes involving fermions and Ramond-Ramond exciations \cite{Berkovits:2004px}\cite{Mafra:2011nv}\cite{Gomez:2013sla}. Third, the back-reaction of deformations from massless integrated vertex operators agrees with supergravity in an elegant superspace formulation \cite{Berkovits:2001ue}.
\vskip 0.3cm
In the Ad$\mathbb{S}_5 \times \mathbb{S}^5$ background, the pure spinor formalism is still in a rudimentary stage \cite{Berkovits:2008ga}\cite{Berkovits:2004xu}\cite{Berkovits:2012ps}\cite{Azevedo:2014rva}\cite{Berkovits:2019ulm}\cite{Berkovits:2019rwq}. For example, an explicit expression for a single half-BPS vertex operator was constructed in \cite{Fleury:2021ieo}, and only recently has this result been generalized to multiple half-BPS vertex operators \cite{toappear}. In parallel to these recent developments, this paper will explore the fundamental aspects concerning the holographic duals of supersymmetric Wilson loops \cite{Maldacena:1998im}\cite{Drukker:1999zq}.
\vskip 0.3cm
In the gauge theory side of the duality, the situation is much improved. At weak coupling, one can rely on Feynman rules and twistorial methods \cite{Siegel:1992xp}\cite{Witten:2003nn}\cite{Cachazo:2004kj}\cite{Mason:2005zm}\cite{Boels:2007qn}\cite{Costello:2022wso}\cite{Costello:2023vyy}\cite{Dixon:2024mzh}. At strong coupling, integrability \cite{Beisert:2003tq}\cite{Beisert:2003yb}\cite{Gromov:2013pga}\cite{Basso:2015zoa}\cite{Fleury:2016ykk}\cite{Caetano:2023zwe} and supersymmetric localization \cite{Pestun:2007rz}\cite{Pestun:2009nn}\cite{Giombi:2009ds}\cite{Gomis:2011pf}\cite{Correa:2012at}\cite{Giombi:2017cqn}\cite{Giombi:2018qox}\cite{Binder:2019jwn} are available. Strong coupling, in conjunction with Maldacena's conjecture, allows the gauge theory to make predictions about superstring theory. For example, combining all available results, the Virasoro-Shapiro amplitude for superstrings in Ad$\mathbb{S}_5\times\mathbb{S}^5$ was predicted in \cite{Alday:2023mvu}, fixing more than one hundred parameters, thus providing an important opportunity to test Maldacena's conjecture. The work of \cite{toappear} aims to test this prediction.
\vskip 0.3cm
In this paper, we will focus on another classic prediction made in \cite{Drukker:2000rr}, which involves taking the planar limit of the exact formula for the expectation value of a circular half-BPS Wilson loop, where holography relates this to a worldsheet disk partition function. The exact formula was proven using supersymmetric localization in \cite{Pestun:2007rz}, and was later extended in \cite{Giombi:2012ep} to include many Wilson loops and local operators.  
\vskip 0.3cm
It is straightforward to compare the leading approximation at large 't Hooft coupling with the semiclassical approximation of the superstring partition function, where matches have been found \cite{Drukker:2000ep}\cite{Frolov:2002av}\cite{Frolov:2003qc}. An important direction to pursue is to enhance the status of these matches by moving beyond the semiclassical approximation. A significant step in this direction, which includes both half and one-quarter BPS Wilson loops, was made in \cite{Drukker:2006ga}. It was identified that, in addition to the true minimal area of the embedded worldsheet in Ad$\mathbb{S}_5\times\mathbb{S}^5$, which governs the leading saddle approximation in the large 't Hooft coupling expansion of the planar formula, there is also another critical point for the action corresponding to the next leading saddle approximation. Specifically, the planar limit of the exact result is given by
\begin{equation*}
\langle W_{\frac{1}{4}\text{- BPS}}\rangle_{N\rightarrow\infty} = \sqrt{\frac{2}{\pi}}\frac{e^{\sqrt{\lambda'}}}{\lambda'^{3/4}} \sum_{k=0}^{\infty} \left(\frac{-1}{2 \sqrt{\lambda'}}\right)^{k} \frac{\Gamma\left(\frac{3}{2} + k\right)}{\Gamma\left(\frac{3}{2} - k\right)} - i \sqrt{\frac{2}{\pi}}\frac{e^{-\sqrt{\lambda'}}}{\lambda'^{3/4}} \sum_{k=0}^{\infty}  \left(\frac{+1}{2 \sqrt{\lambda'}}\right)^{k} \frac{\Gamma\left(\frac{3}{2} + k\right)}{\Gamma\left(\frac{3}{2} - k\right)}\,,
\end{equation*}
where $\lambda'=\lambda\cos^2\theta_0 $, $\lambda$ is the t' Hooft coupling, and $\theta_0 \in [0,\pi)$ is a parameter interpolating between the half-BPS Wilson loop at $\theta_0=0$ and the Zarembo Wilson loop at $\theta_0=\frac{\pi}{2}$. Besides the leading contribution marked by the pesence of the factor $e^{\sqrt{\lambda'}}$, together with its perturbative corrections, there is a subleading contribution coming from the term which contains $e^{-\sqrt{\lambda'}}$. In \cite{Drukker:2006ga}, it was demonstrated that just as the exponent $\sqrt{\lambda'}$ correspond to a minimal surface whose boundary conditions match the configuration of the supersymmetric Wilson loop, the exponent of the subleading term, $-\sqrt{\lambda'}$, corresponds to another critical point of the sigma model action. This was further presented as evidence for the existence of supersymmetric localization in the worldsheet theory, giving hope that one might obtain exact formulas, at least at the planar level, through supersymmetric localization in the worldsheet theory.
\vskip 0.3cm
The first step in establishing exact formulas using supersymmetric localization is to choose a supercharge and determine its off-shell supersymmetric configurations, the loci where the path integral is expected to localize. This is achieved by acting with $Q-q_{\varepsilon}$ on all fermionic variables and requiring the final result to vanish, where $Q$ is the BRST generator and $q_{\varepsilon}$ is the supercharge, both closing off-shel algebras \cite{Pestun:2016zxk}. It is crucial that both the supersymmetric algebra and BRST nilpotency close off-shell since the localization argument involves integrating over the $Q-q_{\varepsilon}$ fermionic vector fields within the domain of path integration. Motivated by this, we analyze off-shell supersymmetric string configurations in Ad$\mathbb{S}_5\times\mathbb{S}^5$, focusing particularly on those holographically dual to supersymmetric Wilson loops.
\vskip 0.3cm
The on-shell analysis of supersymmetric string configurations in Ad$\mathbb{S}_5 \times \mathbb{S}^5$ have been carried out in \cite{Drukker:2007qr}\cite{Giombi:2012ep}. The supersymmetric equations governing these configurations are obtained in Green-Schwarz formalism and can be writen as follows
\begin{equation}
\label{on intro}
j^{a}(\gamma_{a}\varepsilon)_{\alpha}=0\,,\qquad \overline j^{a}(\gamma_{a}\widehat\varepsilon)_{\widehat\alpha} = 0\,,
\end{equation}
where the Greek indices $\alpha, \beta$ and $\widehat\alpha, \widehat\beta$ range over 16 values and correspond to chiral $\text{Spin}(10)$ spinorial components, while the Latin indices $a$ range over 10 values and correspond to $SO(10)$ vector components. The $(j^{a},\overline{j}^{a})$ are the holomorphic and anti-holomorphic currents for the sigma model, and $(\varepsilon^{\alpha}, \widehat\varepsilon^{\widehat\alpha})$ are spacetime Killing spinors. In this paper, we generalize these supersymmetric equations to include off-shell supersymmetric string configurations.
\vskip 0.3cm
Our first result is the derivation, within the pure spinor formalism, of the equations governing off-shell supersymmetric string configurations:
\begin{equation}
\label{off - intro}
( j _{a}-\Omega_{a})(\gamma^{a}\lambda)_{\alpha}=0\,,\qquad  (\overline{ j }_{a}-\widehat \Omega_{a})(\gamma^{a}\widehat\lambda)_{\alpha}=0 \,,  
\end{equation}
\begin{equation*}
\lambda^{\alpha}=\varepsilon^{\alpha}\,,\qquad \Omega_{a} (\gamma^{a}\widehat\lambda)_{\alpha}=0\,,\qquad \widehat\Omega_{a}(\gamma^{a}\lambda)_{\alpha}=0\,,\qquad \widehat\lambda^{\widehat\alpha} = \widehat\varepsilon^{\widehat\alpha}\,,
\end{equation*}
where $\lambda^{\alpha}$ and $\widehat\lambda^{\widehat\alpha}$ are the worldsheet variables obeying the pure spinors constraints 
\begin{equation}
(\lambda\gamma^{a}\lambda)=0\,,\qquad (\widehat\lambda\gamma^{a}\widehat\lambda) = 0 \,,  
\end{equation}
and $\Omega_{a}$ and $\widehat\Omega_{a}$ are ghost-dependent variables, obtained from redefinitions of the canonical conjugates of the pure spinor variables. These equations differ from those obtained in the Green-Schwarz formalism. Our second result demonstrates that the Green-Schwarz supersymmetric equations imply the equations of motion (in Euclidean signature):
\begin{equation}
j^{a}(\gamma_{a}\varepsilon)_{\alpha}=0\,,\qquad \overline j^{a}(\gamma_{a}\widehat\varepsilon)_{\widehat\alpha} = 0\quad\implies    \quad \nabla \overline j^{a} +\overline\nabla j^{a} = 0\,.
\end{equation}
We then briefly comment that, when evaluated at the on-shell supersymmetric configurations, the sigma model action can be written as an integral of a two-form $S = \frac{r^2}{2\pi\alpha'}\int j^{a}\overline j^{b}B_{ab}$, where $B_{ab} = \frac{\overline\varepsilon\gamma_{ab}\varepsilon}{2\varepsilon\overline\varepsilon}$ and $\overline\varepsilon_{\alpha}$ is the Euclidean complex conjugate of the Killing spinor $\varepsilon^{\alpha}$. 
\vskip 0.3cm
One puzzle we encounter with equations \ref{off - intro}, which also appears in the flat background, is that in the pure spinor formalism, the supersymmetric equations lack solutions for certain supercharges. Specifically, when the corresponding Killing spinor is impure everywhere in space-time, the constraints on the pure spinor variables $\lambda^{\alpha}$ and $\widehat\lambda^{\widehat\alpha}$ lead to no solutions for the equations $\varepsilon^{\alpha} = \lambda^{\alpha}$ and $\widehat\varepsilon^{\widehat\alpha} = \widehat\lambda^{\widehat\alpha}$. As we will see in subsection \ref{Derivation of the supersymmetric equations} and \ref{bps implies eom}, if we simply disregard the pure spinor constraints in deriving the supersymmetric equations, these equations default to the on-shell ones obtained in the Green-Schwarz formalism shown in \ref{on intro}, thus implying equations of motion. A possible workaround could be to integrate this discussion within the so-called B-RNS-GSS formalism, a framework that bridges the pure spinor and Ramond-Neveu-Schwarz formalisms \cite{Berkovits:2022qhc}. In this framework, the pure spinor variables are replaced with unconstrained bosonic spinors, potentially circumventing these constraints. It has recently been generalized to an Ad$\mathbb{S}_5\times\mathbb{S}^5$ background \cite{Chandia:2023eel}. We leave this exploration for future work.
\vskip 0.3cm
Finally, our last result involves solving the off-shell supersymmetric equations \ref{off - intro} for a choice of supercharge that preserves a particular circular and half-BPS boundary condition:
\begin{equation}
\varepsilon_{c} = i \gamma_{129}\varepsilon_{s}\,,
\end{equation}
where the constant spinors $\varepsilon_{s}$ and $\varepsilon_{c}$ parametrize the 16 + 16 super-Poincaré and superconformal supercharges, respectively. See appendix \ref{AdS Killing spinors} and reference \cite{Dymarsky:2009si} for notation on Killing spinors. For our particular choice of supercharge (particular choice of $\varepsilon_{s}$), the supersymmetric equations of \ref{off - intro} localizes the embedded surfaces to an Ad$\mathbb{S}_4 \times \mathbb{S}^2$ slice of Ad$\mathbb{S}_5 \times \mathbb{S}^5$, where the corresponding Killing spinors are pure spinors. Within Ad$\mathbb{S}_4 \times \mathbb{S}^2$, the above equations for off-shell supersymmetric configurations are completely solved for $\Omega_a$ and $\widehat\Omega_a$ in a particular gauge choice. The solution is
\begin{equation}
\Omega^{a} = \frac{(\overline\lambda\gamma^{a}\gamma^{b}\lambda)}{2\lambda\overline\lambda}j_{b}\,,\qquad   \widehat\Omega^{a} = \frac{(\lambda\gamma^{a}\gamma^{b}\overline\lambda)}{2(\overline\lambda\lambda)}\overline j_{b}\,,
\end{equation}
where $\overline\lambda_{\alpha} = (\gamma_{358}\widehat\lambda)_{\alpha}$ is the complex conjugate to $\lambda^{\alpha}$ within the Ad$\mathbb{S}_4\times \mathbb{S}^2$ slice. However, beyond this, the corresponding supersymmetric locus will remain infinite-dimensional, thus unfortunately preventing the direct derivation of exact results.
\vskip 0.3cm
By substituting our solution into the sigma model action, we find that the action also simplifies to an integral of a two-form, analogous to the on-shell supersymmetric case. The expression is:
\begin{equation}
S = r^2 \int \frac{d^2\sigma}{2\pi\alpha'}\left[j^{a}\overline j^{b}\delta_{ab} - \Omega^{a}\widehat\Omega^{b}\delta_{ab}\right]=r^2 \int \frac{d^2\sigma}{2\pi\alpha'}j^{a}\overline j^{b}B_{ab}\,,\qquad B_{ab} = \frac{\overline\lambda\gamma_{ab}\lambda}{2\overline\lambda\lambda}\,,    
\end{equation}
where $B_{ab}$ is the same two-form that would be obtained in the Green-Schwarz formalism with this choice of Killing spinor, and explicitly given by
\begin{equation*}
B_{\mu\nu} = i\frac{\varepsilon_{\mu\nu\rho}y_{(\rho+1)} - x_{\mu}y_{(\nu+1)} + x_{\nu}y_{(\mu+1)}}{|x|^2 + |y|^2 + 1}\,, \qquad  B_{(4+i)(4+j)} =i \frac{-\varepsilon_{ijk}y_{k} - x_{(i-1)}y_{j}+x_{(j-1)}y_{i}}{|x|^2 + |y|^2 + 1}\,,
\end{equation*}
\begin{equation*}
B_{\mu (4+i)} = \frac{i}{2}\delta_{(\mu+1)i}\left[\frac{|x|^2 + |y|^2 - 1 - (x_{\mu}^2 + y_{i}^2)}{|x|^2 + |y|^2 + 1}\right] +i\left[\frac{-\varepsilon_{\mu\nu\rho}x_{\rho} - x_{\mu}x_{(i-1)} - y_{i}y_{(\mu+1)}}{|x|^2 + |y|^2 + 1}\right]\,,  
\end{equation*}
where
\begin{equation}
\mu,\nu,\rho = 1,2,4\qquad i,j,k = 2,3,5 \,.    
\end{equation}
We demonstrate that this form is not closed, indicating that the action generally depends on variations of the embedded surface in Ad$\mathbb{S}_4\times\mathbb{S}^2$.
\vskip 0.3cm
This paper is structured as follows: In section \ref{review}, we review the pure spinor formalism applied to Ad$\mathbb{S}_5\times\mathbb{S}^5$ background, taking the opportunity to introduce fermionic auxiliary variables that render the BRST transformation nilpotent without needing to impose equations of motion. In section \ref{Supersymmetric string configurations}, we begin our analysis of supersymmetric string configurations. We briefly discuss the nature of off-shell supersymmetric configurations and their relevance to supersymmetric localization, and then we demonstrate the derivation of the equations. This task is delicate, as the Ad$\mathbb{S}_5\times\mathbb{S}^5$ pure Killing spinors are highly singular objects, causing most of the usual Fierz identities to degenerate. We compare these equations with those of the Green-Schwarz formalism and demonstrate that the action, when evaluated on both on-shell and off-shell supersymmetric configurations, becomes invariant under reparametrization. Specifically, the Virasoro constraints, both with and without ghost contributions, are implied by the off-shell and on-shell supersymmetric equations, respectively.
\vskip 0.3cm
In section \ref{the AdS slice}, we explain that for a particular choice of supercharge, the supersymmetric configurations localize to an Ad$\mathbb{S}_4\times\mathbb{S}^2$ slice. This aims to solve the supersymmetric equations that equate the Killing spinors with the worldsheet variables $\lambda^{\alpha}$ and $\widehat\lambda^{\widehat\alpha}$ obeying the pure spinor constraints. We verify that, when $\Omega_{a}$ and $\widehat\Omega_{a}$ are suppressed, the well-known Ad$\mathbb{S}_2$ minimal solution satisfies the supersymmetric equations. Then, in section \ref{off-shell}, we solve the supersymmetric equations with $\Omega_{a}$ and $\widehat\Omega_{a}$, uncovering an infinite-dimensional family of embedded surfaces in Ad$\mathbb{S}_4\times\mathbb{S}^2$. We demonstrate that the sigma model action is an integral of a two-form and that this two-form is not closed. Finally, in section \ref{conclusion}, we conclude.

\vskip 0.3cm
We provide further technical details in the appendices. In appendix \ref{coordinates}, we establish our coordinate system for the Ad$\mathbb{S}_5\times\mathbb{S}^5$ spacetime. In appendix \ref{Conventions on spinors}, we define our conventions for spinors. In appendix \ref{Killin spinors appendix}, we derive the Killing spinor equations in Ramond-Ramond backgrounds from superspace. In appendix \ref{AdS Killing spinors}, we solve these equations for the Ad$\mathbb{S}_5\times\mathbb{S}^5$ background. In Appendix \ref{pure killing spinors - appendix}, we derive the equations governing pure Killing spinors. In appendix \ref{fixing sigma model action}, we fix the sigma model action using BRST symmetry and PSU(2,2|4) symmetry. Finally, in appendix \ref{Gamma matrices}, we explicitly present our gamma matrices. 

\section{Reviewing the pure spinor formalism}
\label{review}
The pure spinor formalism has advantages over the RNS formalism in that it can describe Ramond-Ramond backgrounds such as Ad$\mathbb{S}_5\times \mathbb{S}^{5}$, and it has advantages over the GS formalism in that the sigma model action is quantizable. In this section, we will quickly review the pure spinor formalism for the superstring in Ad$\mathbb{S}_5\times \mathbb{S}^{5}$ background to establish notation.

\subsection{The worldsheet variables}
The worldsheet variables for the superstring in Ad$\mathbb{S}_5\times \mathbb{S}^{5}$ using the pure spinor formalism consist of Green-Schwarz-like supercoset variables \cite{Metsaev:1998it}:
\begin{equation}
\label{supercoset}
g \in \frac{PSU(2,2|4)}{USp(2,2)\times USp(4)}\,,
\end{equation}
and Type IIB pure spinor variables together with their gauge-transforming canonical conjugates \cite{Berkovits:2000fe}:
\begin{equation*}
\lambda^{\alpha}\gamma^{a}_{\alpha\beta}\lambda^{\beta}=0\,,\qquad  \widehat\lambda^{\widehat\alpha}\gamma^{a}_{\widehat\alpha\widehat\beta}\widehat\lambda^{\widehat\beta}=0\,, 
\end{equation*}
\begin{equation}
\label{w gauge transf}
\delta w_{\alpha}=v^{a}(\gamma_{a}\lambda)_{\alpha}\,,\qquad \delta \widehat w_{\widehat\alpha}=\widehat v^{a}(\gamma_{a}\widehat\lambda)_{\widehat\alpha}\,,  
\end{equation}
where the gauge invariances are generated by the pure spinor constraints. The Greek indices $\alpha, \beta$ and $\widehat\alpha, \widehat\beta$ range over 16 values and correspond to chiral $Spin(10)$ spinorial components, while the Latin indices $a$ range over 10 values and correspond to $SO(10)$ vector components. The supercoset variables become concrete once we choose a particular parametrization of the supercoset in \ref{supercoset}. We will fix a parametrization that is convenient for our work in the next section. For now, we will keep this choice abstract to maintain generality.

\subsection{The sigma model action}
The sigma model action for the superstring in Ad$\mathbb{S}_5\times \mathbb{S}^{5}$, using the pure spinor formalism, is completely fixed by $PSU(2,2|4)$ invariance and BRST symmetry (see appendix \ref{fixing sigma model action}). It is quantizable \cite{Berkovits:2004xu} and is given by:
\begin{equation*}
S=r^2\int \frac{d^2\sigma}{2 \pi\alpha'}\left[\frac{1}{2} j^{a}\overline j^{b}\delta_{ab}+\overline\nabla\lambda^{\alpha}w_{\alpha}+\nabla\widehat\lambda^{\widehat\alpha}\widehat w_{\widehat\alpha}-\frac{1}{4}R_{abcd}N^{ab}\widehat N^{cd}\right. -   
\end{equation*}
\begin{equation}
\label{sigma model action}
\left.-( j^{\alpha} \overline  j^{\widehat\alpha}+\overline  j^{\alpha} j^{\widehat\alpha})\eta_{\widehat\alpha\alpha}+\frac{1}{2}( j^{\alpha} \overline  j\,^{\widehat\alpha}-\overline  j\,^{\alpha}  j^{\widehat\alpha})\eta_{\widehat\alpha\alpha} -w^{*}_{\alpha}\widehat w^{*}_{\widehat\alpha}\eta^{\widehat\alpha \alpha}\right]\,.    
\end{equation}
Just as before, the Latin indices $a$ range over 10 values and correspond to $SO(10)$ vector components, while the Greek indices $\alpha$ and $\widehat\alpha$ range over 16 values and correspond to chiral $Spin(10)$ spinorial components. Of course, $SO(10)$ is not a symmetry of Ad$\mathbb{S}_5\times \mathbb{S}^{5}$ , as evidenced by the presence of the symbols $\eta_{\alpha\widehat\alpha}=-\eta_{\widehat\alpha\alpha}$ and $R_{abcd}$ in the sigma model action. These symbols denote, respectivelly, the inverse of the self-dual Ramond-Ramond bispinor, which breaks $SO(10)$ down to $SO(4,1) \times SO(5)$, and the Ad$\mathbb{S}_5\times \mathbb{S}^{5}$ curvature tensor. 
\subsection{The worldsheet currents}
The holomorphic and anti-holomorphic Ad$\mathbb{S}_5\times \mathbb{S}^{5}$  left-invariant currents are denoted by $j^{a}$ and $\overline{j}^{a}$, respectively. These currents can be expressed in terms of the supercoset elements $g$, defined in \ref{supercoset}, as follows:
\begin{equation}
 j^{a}=(g^{-1}\partial g)^{a}\,,\qquad \overline j^{a}=(g^{-1}\overline\partial g)^{a}    \,,
\end{equation}
\begin{equation*}
 j^{\alpha}=(g^{-1}\partial g)^{\alpha}\,,\qquad \overline j^{\alpha}=(g^{-1}\overline\partial g)^{\alpha}  \,,    
\end{equation*}
\begin{equation*}
 j^{\widehat\alpha}=(g^{-1}\partial g)^{\widehat\alpha}\,,\qquad \overline j^{\widehat\alpha}=(g^{-1}\overline\partial g)^{\widehat\alpha}      \,,
\end{equation*}
and the holomorphic and anti-holomorphic derivatives are given by
\begin{equation} 
\partial=\partial_{\tau} - i\partial_{\sigma}\,,\qquad \overline\partial = \partial_{\tau} + i\partial_{\sigma}\,, 
\end{equation} 
where $\partial_{\tau}$ is an uncompact vector field on the worldsheet, and $\partial_{\sigma}$ is a compact vector field. The spin connections in $\nabla$ and $\overline\nabla$ can be expressed in terms of the supercoset elements $g$, defined in \ref{supercoset}, as follows
\begin{equation}
\nabla\widehat \lambda^{\alpha}=\partial \widehat \lambda^{\widehat \alpha} +\frac{1}{4} (g^{-1}\partial g)^{ab}(\gamma_{ab}\widehat \lambda)^{\widehat \alpha} \,,\qquad \overline\nabla\lambda^{\alpha}=\overline\partial \lambda^{\alpha}+\frac{1}{4} (g^{-1}\overline\partial g)^{ab}(\gamma_{ab}\lambda)^{\alpha}\,.
\end{equation}
The $N_{ab}$ and $\widehat N_{ab}$ are contributions to the Lorentz current, arising from combinations of pure spinors and their canonical conjugates:
\begin{equation}
N_{ab}=\frac{1}{2}(w\gamma_{ab}\lambda)\,,\qquad \widehat N_{ab}=\frac{1}{2}(\widehat w\gamma_{ab}\widehat\lambda)\,.   
\end{equation}
\subsection{The auxilliary fermions and the BRST transformations}
The last term of the sigma model action \ref{sigma model action} contains the \emph{auxilliary fermionic variables} $w_{\alpha}^{*}$ and $\widehat w_{\widehat\alpha}^{*}$, introduced to ensure BRST nilpotency without imposing equations of motion. They modify the BRST transformations of the canonical conjugates of the pure spinor variables, while their own BRST transformations correspond to the equations of motion derived from varying the pure spinor canonical conjugates. The off-shell BRST transformations are:
\begin{equation} 
\label{BRST modifications}
Qg = g (\lambda^{\alpha}t_{\alpha}+\widehat\lambda^{\widehat\alpha}t_{\widehat\alpha})\,,
\end{equation}
\begin{equation*}
Qw_{\alpha}=2 j^{\widehat\alpha}\eta_{\widehat\alpha\alpha}+w_{\alpha}^{*}\,,\qquad Q\widehat w_{\widehat\alpha}=2\overline  j^{\alpha}\eta_{\alpha\widehat\alpha}+\widehat w_{\widehat \alpha}^{*}\,.  
\end{equation*}
\begin{equation*}
Q w_{ \alpha}^{*}= \eta_{\alpha\widehat\alpha}\nabla\widehat\lambda^{\widehat\alpha}-\frac{1}{8}R_{abcd} N^{dc}(\eta\gamma^{ba}\widehat\lambda)_{\alpha}\,,
\qquad 
Q\widehat w_{\widehat \alpha}^{*}= \eta_{\widehat\alpha\alpha}\overline\nabla\lambda^{\alpha}-\frac{1}{8}R_{abcd}\widehat N^{dc}(\eta\gamma^{ba}\lambda)_{\widehat\alpha} \,,
\end{equation*}
where the off-shell nilpotency was demonstrated in \cite{Berkovits:2008ga}.
\vskip 0.3cm
Off-shell nilpotence of the BRST transformations are important in that localization arguments are only valid within fermionic generators that closes an algebra off-shell. For example, in \cite{Pestun:2007rz}, where supersymmetric localization were used to compute the vacuum expectation values of certain supersymmetric Wilson loops, the auxilliary variables of \cite{Berkovits:1993hx}\cite{Baulieu:2007ew} were essential.

\section{Supersymmetric string configurations in Ad$\mathbb{S}_5\times\mathbb{S}^{5}$}
\label{Supersymmetric string configurations}

In this paper, we assume that the correct characterization of supersymmetric string configurations in the Ad$\mathbb{S}_5\times \mathbb{S}^{5}$ background with respect to the supercharge $q_{\varepsilon}$, using the pure spinor formalism, is achieved by acting with $-q_{\varepsilon}+Q$, where $Q$ is the BRST charge, on all fermionic worldsheet variables, and imposing the final result to vanish. Although this proposal has been known for almost ten years, it has not been explicitly stated in any paper, nor has there been any careful implementation of this proposal.
\vskip 0.3cm
In the pure spinor formalism for the superstring in Ad$\mathbb{S}_5\times \mathbb{S}^{5}$, there are two types of fermions: the fermionic coordinates and the auxiliary variables required to ensure BRST nilpotency off-shell, as discussed in Section \ref{review}. By applying $-q_{\varepsilon}+Q$ to the fermionic coordinates, we demonstrate in subsection \ref{pure=killing} that the corresponding supersymmetric condition equates the pure spinor variables with the Ad$\mathbb{S}_5\times \mathbb{S}^{5}$ Killing spinors associated with supersymmetry. This poses a puzzle, as it seems to imply that the selection of supercharges is not arbitrary. For instance, selecting supercharges whose corresponding Killing spinors are impure everywhere in Ad$\mathbb{S}_5\times \mathbb{S}^{5}$ would result in supersymmetric equations without solutions. In the appendix \ref{pure killing spinors - appendix}, we review the analysis of \cite{Dymarsky:2009si} on the conditions a supercharge must meet to admit a Killing spinor that is pure somewhere, everywhere, or nowhere in Ad$\mathbb{S}_5\times \mathbb{S}^{5}$. This puzzle is also present in a flat background, where a Killing spinor is constant and consequently either pure everywhere or nowhere. In this paper, we select a supercharge whose Killing spinor is pure only within an $Ad\mathbb{S}_4\times \mathbb{S}^2$ slice of Ad$\mathbb{S}_5\times \mathbb{S}^{5}$. Consequently, the off-shell supersymmetric conditions derived from applying $-q_{\varepsilon}+Q$ on the fermionic coordinates, where $q_{\varepsilon}$ is our chosen supercharge as specified in \ref{Killing spinors and their pure region}, localize the superstrings to a $Ad\mathbb{S}_4\times \mathbb{S}^2$ region.
\vskip 0.3cm
By applying $-q_{\varepsilon}+Q$ to the auxiliary fermionic variables, as discussed in subsection \ref{Derivation of the supersymmetric equations}, the supersymmetric conditions enforce the equations of motion on the pure spinor variables. Together with the identification of the pure spinors with the Killing spinors, this approach leads to a system of partial differential equations. If not for the highly singular properties of the pure Killing spinors, this system would mirror that derived using the Green-Schwarz formalism, albeit with the supercharges restricted such that their Killing spinors are pure spinors. These singular properties have previously been overlooked \footnote{Private communications with Nathan Berkovits and Jaume Gomis.}.

\subsection{Off-shell configurations and BRST localization}
\label{off def}
In this section, we explain what we mean by \emph{off-shell configurations} and why this concept is crucial in computing the partition function of superstrings with prescribed boundary conditions. By off-shell configuration, we refer to configurations that yield zero after the BRST generators act on the fermionic worldsheet variables. This definition is essential because physical states are constrained to be BRST invariant. Consequently, the path integral will necessarily localize to a subdomain of integration, where the bosonic part of this subdomain corresponds to the \emph{off-shell configurations}. For instance, when the BRST generator acts on the fermionic coordinates, we obtain:
\begin{equation}
Q\theta^{\alpha} = \lambda^{\alpha}+O(\theta^2,\widehat\theta^2,\theta\widehat\theta)\,,\qquad Q\widehat\theta^{\widehat\alpha} = \widehat\lambda^{\alpha}+O(\theta^2,\widehat\theta^2,\theta\widehat\theta)\,.   
\end{equation}
This implies that the path integral will localize to configurations that obey:
\begin{equation}
\lambda^{\alpha} = 0\,,\qquad \widehat\lambda^{\widehat\alpha} = 0\,,
\end{equation}
and indeed, since these variables carry a ghost number $+1$, only homogeneous expressions in $\lambda^{\alpha}$ and $\widehat\lambda^{\widehat\alpha}$ have non-zero vacuum expectation values under fixed boundary conditions and topology. It follows from \ref{BRST modifications} that the conditions $Qw_{\alpha}^{*}=0$ and $Q\widehat w_{\widehat\alpha}^{*}=0$ are trivially satisfied, and from equation \ref{sigma model action} that the components $w_{\alpha}$ and $\widehat w_{\widehat\alpha}$ completely decouple from the sigma model action. The on-shell configurations are those which, besides being off-shell, are also critical points of the sigma model action. To generalize to \emph{off-shell supersymmetric configurations}, we simply replace $Q$ with $Q-q_{\varepsilon}$, where $q_{\varepsilon}$ is the supersymmetric generator.

\subsection{Off-shell supersymmetric configurations in flat background}

In flat background, the pure spinor formalism is slightly different from that in Ad$\mathbb{S}_5\times\mathbb{S}^5$ due to the absence of Ramond-Ramond flux, resulting in the bi-spinor $\eta^{\alpha\widehat\alpha}$ not being an invertible matrix. This indicates that the flat background, from the perspective of the pure spinor formalism, represents a singular limit. This qualitative difference is marked by the inclusion of Siegel's fermionic variables, $d_{\alpha}$ and $\widehat d_{\widehat\alpha}$, in the sigma model action for the flat background:
\begin{equation*}
S=\int \frac{d^2\sigma}{2 \pi\alpha'}\left[\frac{1}{2} \pi^{a}\overline \pi^{b}\delta_{ab}+\overline\pi^{\alpha}d_{\alpha}+\pi^{\widehat\alpha}\widehat d_{\widehat\alpha} + B_{II} + \overline\partial\lambda^{\alpha}w_{\alpha}+\partial\widehat\lambda^{\widehat\alpha}\widehat w_{\widehat\alpha}\right]\,,
\end{equation*}
where 
\begin{equation}
\pi^{a} = \partial x^{a} - \frac{1}{2}\partial\theta\gamma^{a}\theta - \frac{1}{2}\partial\widehat\theta\gamma^{a}\widehat\theta \,,\qquad  \overline\pi^{a}= \overline\partial x^{a} - \frac{1}{2}\overline\partial\widehat\theta\gamma^{a}\widehat\theta   - \frac{1}{2}\overline \partial\theta\gamma^{a}\theta\,,
\end{equation}
and $B_{II}$ is the B-superfield for type IIA or type IIB flat superspace. The BRST transformations in the flat background, generated by $Q$, along with the supersymmetric transformations, generated by $q_{\varepsilon}$, are given by
\begin{equation*}
Q\theta^{\alpha} = \lambda^{\alpha}\,,\qquad Qx^{a} = (\lambda\gamma^{a}\theta) + (\widehat\lambda\gamma^{a}\widehat\theta)\,\qquad Q\widehat\theta^{\widehat\alpha} = \widehat\lambda^{\widehat\alpha}\,,    
\end{equation*}
\begin{equation}
Q d_{\alpha} = \pi_{a}(\lambda\gamma^{a})_{\alpha}\,,\qquad 
Q w_{\alpha} = d_{\alpha}\,,\qquad Q\widehat w_{\widehat\alpha} =  \widehat d_{\widehat\alpha}\,,\qquad Q \widehat d_{\widehat\alpha} = \overline\pi_{a}(\widehat\lambda\gamma^{a})_{\widehat\alpha}   \,,
\end{equation}
and
\begin{equation}
q_{\varepsilon}\theta^{\alpha}=\varepsilon^{\alpha}\,,\qquad q_{\varepsilon} x^{a} = - (\varepsilon\gamma^{a}\theta) - (\widehat\varepsilon\gamma^{a}\widehat\theta)     \,,\qquad q_{\varepsilon}\widehat\theta^{\widehat\alpha} = \widehat\varepsilon^{\widehat\alpha}\,.
\end{equation}
Consequently, the conditions characterizing off-shell supersymmetric configurations in a flat background are given by:
\begin{equation}
\lambda^{\alpha} = \varepsilon^{\alpha}\,,\qquad\partial x^{a}(\lambda\gamma_{a})_{\alpha} =0\,,\qquad \overline\partial x^{a}(\widehat\lambda\gamma_{a})_{\widehat\alpha}=0\,,\qquad  \widehat\lambda^{\widehat\alpha} = \widehat\varepsilon^{\widehat\alpha} \,,  
\end{equation}
where $\varepsilon^{\alpha}$ and $\widehat\varepsilon^{\widehat\alpha}$ are constant spinors parametrizing the supercharges. If the supercharge is not nilpotent, these equations have no solution, since $\varepsilon^{\alpha}$ and $\widehat\varepsilon^{\widehat\alpha}$ will not satisfy the pure spinor constraints. A future direction worth exploring is the potential replacement of the pure spinor constraints on the worldsheet variables $\lambda^{\alpha}$ and $\widehat\lambda^{\widehat\alpha}$ with conditions derived from the nilpotency of $Q-q_{\varepsilon}$. This would involve generalizing the pure spinor formalism. \footnote{We thank Kevin Costello for pointing out that one can consider equivariant cohomology.}

\subsection{Pure Killing Spinors in Ad$\mathbb{S}_5\times\mathbb{S}^5$}
\label{pure=killing}
Here, we parametrize the Ad$\mathbb{S}_5\times\mathbb{S}^5$ superspace, the $\frac{PSU(2,2|4)}{USp(2,2)\times USp(4)}$ supercoset, in a convenient manner that places all fermions on the left and all bosons on the right (the $USp(2,2)\times USp(4)$ gauge transformations act on the right). Schematically, we have:
\begin{equation}
G(\theta, x) = F(\theta)g(x)    
\end{equation}
where $\theta$ comprises all 32 fermionic coordinates, and $x$ comprises all 10 bosonic coordinates. By acting with $-q_{\varepsilon}+Q$ on the fermionic coordinates, neglecting the fermions, and retaining terms only up to the first order in $\lambda$, $\widehat\lambda$, $\varepsilon_{s}$, and $\varepsilon_{c}$, we obtain:
\begin{equation*}
(1-q_{\varepsilon} +Q)G(\theta, x) = (1-\varepsilon_{s} - \varepsilon_{c}) F(\theta) g(x) (1+\lambda+\widehat\lambda) =
\end{equation*}
\begin{equation}
=  F(\theta) (1-\varepsilon_{s} - \varepsilon_{c}) g(x) (1+\lambda+\widehat\lambda) =   F(\theta) g(x) (1+\lambda - \varepsilon +\widehat\lambda- \widehat\varepsilon) 
\end{equation}
where $\lambda$, $\widehat\lambda$, $\varepsilon_{s}$ and $\varepsilon_{c}$ take values at the supercharges, and $\varepsilon^{\alpha}$ and $\widehat\varepsilon^{\widehat\alpha}$ are the Killing spinors, expressed in terms of $\varepsilon_{s}$ and $\varepsilon_{c}$ in appendix \ref{AdS Killing spinors}. The resulting supersymmetric condition is:
\begin{equation}
\label{l=e}
\lambda^{\alpha}=\varepsilon^{\alpha}\,,\qquad \widehat\lambda^{\alpha}=\widehat\varepsilon^{\alpha}\,.   
\end{equation}
In backgrounds with Ramond-Ramond flux, as derived in appendix \ref{Killin spinors appendix}, the Killing spinors obey different equations than in pure NS-NS backgrounds, namely:
\begin{equation}
\nabla_{a}\varepsilon^{\alpha}=\widehat\varepsilon^{\widehat\alpha}    T_{\widehat\alpha a}\,^{\alpha}\,,\qquad \nabla_{a}\varepsilon^{\widehat\alpha}=\varepsilon^{\alpha}T_{\alpha a}\,^{\widehat\alpha}\,,
\end{equation}
where $T_{\widehat\alpha a},^{\alpha}$ and $T_{\alpha a},^{\widehat\alpha}$ are the torsion components given by:
\begin{equation}
\label{RR-background-torsion}
T_{\widehat\alpha a}\,^{\alpha} =\frac{1}{2}(\gamma_{a}\eta)_{\widehat\alpha}\,^{\alpha}\,,\qquad T_{\alpha a}\,^{\widehat\alpha} = \frac{1}{2}(\gamma_{a}\eta)_{\alpha}\,^{\widehat\alpha}\,,    
\end{equation}
where $\eta^{\alpha\widehat\alpha}=-\eta^{\widehat\alpha\alpha}$ is the Ramond-Ramond bi-spinor. According to \ref{l=e} and \ref{RR-background-torsion}, in supersymmetric configurations, the pure spinor variables must obey the following relations:
\begin{equation}
\label{pure killing}
\overline\nabla\lambda^{\alpha}=-\frac{1}{2} \widehat  j ^{a}(\eta\gamma_{a})^{\alpha}\,_{\widehat\alpha}\widehat\lambda^{\widehat\alpha} \,,  \qquad \nabla\widehat\lambda^{\widehat\alpha}=-\frac{1}{2}   j ^{a}(\eta\gamma_{a})^{\widehat\alpha}\,_{\alpha}\lambda^{\alpha} \,.
\end{equation}
The equations in \ref{l=e} also impose constraints on the choice of supercharges $q_{\varepsilon}$, and possible the embedding coordinates, to ensure that the Killing spinor is a pure spinor. Pure Killing spinors, distinct from generic Killing spinors, have been investigated in previous studies, such as in \cite{Dymarsky:2009si}.

\subsection{Derivation of the supersymmetric equations in Ad$\mathbb{S}_5\times\mathbb{S}^5$}
\label{Derivation of the supersymmetric equations}

The remaining fermionic worldsheet variables are now the auxiliary ones, $w_{\alpha}^{*}$ and $\widehat w_{\widehat\alpha}^{*}$, as introduced in \ref{sigma model action} to ensure BRST nilpotency off-shell. In acting with $-q_{\varepsilon} + Q$, we find that the supersymmetric transformations do not contribute while the BRST transformation \ref{BRST modifications} implies that the pure spinors must be on-shell:
\begin{equation}
\eta_{\widehat\alpha\alpha}\overline\nabla\lambda^{\alpha}=\frac{1}{8}R_{abcd}\widehat N^{cd}(\eta\gamma^{ab}\lambda)_{\widehat\alpha} \,,  \qquad \eta_{\alpha\widehat\alpha}\nabla\widehat\lambda^{\widehat\alpha}=\frac{1}{8}R_{abcd} N^{cd}(\eta\gamma^{ab}\widehat\lambda)_{\alpha}\,.
\end{equation}
Combining the above equation with the fact that the pure spinor variables are pure Killing spinors, as stated in \ref{pure killing}, we can eliminate the covariant derivatives of the pure spinor variables to obtain the following equations:
\begin{equation}
\label{susy eq}
 j ^{a}(\gamma_{a}\lambda)_{\alpha}=-\frac{1}{4}R_{abcd} N^{cd}(\eta\gamma^{ab}\widehat\lambda)_{\alpha}\,,\qquad \overline  j ^{a}(\gamma_{a}\widehat\lambda)_{\widehat\alpha}=-\frac{1}{4}R_{abcd} \widehat N^{cd}(\eta\gamma^{ab}\lambda)_{\widehat\alpha} \,.
\end{equation}
To understand the right-hand side of these equations, consider the following identities:
\begin{equation}
R_{abcd}N^{cd}(\eta\gamma^{ab}\widehat\lambda)_{\alpha}=R_{abcd}N^{cd}(\gamma^{ab}\eta\widehat\lambda)_{\alpha}
\,,\qquad
\mu_{\alpha}=(\eta\widehat\lambda)_{\alpha}\implies (\mu\gamma^{a}\mu)=0\,,
\end{equation}
where we have defined the anti-chiral pure spinor $\mu_{\alpha}$. For any generic pair of pure spinors, $\lambda$ and $\widehat\lambda$, assuming that $(\lambda\mu) \neq 0$, we can use a Fierz identity
\begin{equation}
\delta_{\alpha}^{\beta}\delta_{\gamma}^{\delta}=-\frac{1}{4}\delta_{\alpha}^{\delta}\delta_{\gamma}^{\beta}+\frac{1}{8}\gamma_{ab}\,_{\alpha}\,^{\delta}\gamma^{ba}\,_{\gamma}\,^{\beta}+\frac{1}{2}\gamma^{a}_{\alpha\gamma}\gamma_{a}^{\delta\beta}    
\end{equation}
to decompose:
\begin{equation}
w_{\alpha}=w_{\alpha}\frac{(\lambda\mu)}{(\lambda\mu)}=(w\lambda)\left[-\frac{\mu_{\alpha}}{4(\lambda\mu)}\right]+(\lambda\gamma_{ab}w)\left[\frac{(\gamma^{ba}\mu)_{\alpha}}{8(\lambda\mu)}\right]+(w\gamma^{a}\mu)\left[\frac{(\gamma_{a}\lambda)_{\alpha}}{2(\lambda\mu)}\right]\,.    
\end{equation}
Contracting the equations \ref{susy eq} with $(\gamma_{b}\lambda)_{\beta}(\gamma^{b}\mu)^{\alpha}$ results in
\begin{equation}
 j ^{a}(\gamma_{a}\lambda)_{\alpha}(\gamma^{b}\mu)^{\alpha}(\gamma_{b}\lambda)_{\beta}= j ^{a}(\lambda\gamma_{a}\gamma_{b}\mu)(\gamma^{b}\lambda)_{\beta}=(\lambda\mu) j ^{a}(\gamma_{a}\lambda)_{\beta}\overset{(\lambda\mu\neq 0)}{\implies} 
\end{equation}
\begin{equation}
\overset{(\lambda\mu\neq 0)}{\implies}  j ^{a}(\gamma_{a}\lambda)_{\beta}=0\implies R_{abcd}N^{dc}(\gamma^{ba}\widehat\lambda)^{\widehat\alpha}=0 \,.   
\end{equation}
We observe a decoupling between the canonical conjugates of the pure spinors, contained in $N_{ab}=\frac{1}{2}(w\gamma_{ab}\lambda)$, and the embedding coordinates:
\begin{equation}
\label{old susy eq}
j^{a}(\gamma_{a}\lambda)_{\alpha} = 0\,,\qquad     \overline j^{a}(\gamma_{a}\widehat \lambda)_{\widehat \alpha} = 0\,.
\end{equation}
However, our pair of pure spinors, $\lambda$ and $\widehat\lambda$, are not generic; they are pure Killing spinors. The following singular property is characteristic of pure Killing spinors:
\begin{equation}
(\lambda\mu)=0\,,\qquad (\lambda\gamma^{ab}\mu)=0\,,
\end{equation}
which causes the manipulations described above to degenerate. These relations can be derived by applying a derivative to the pure spinor constraints and then using the Killing spinor equations to eliminate the derivatives:
\begin{equation}
0=\frac{1}{2}\nabla_{b}(\lambda\gamma_{a}\lambda) = -\frac{1}{2}(\lambda\gamma_{b}\eta\gamma_{a}\widehat\lambda)\quad\implies \quad(\lambda\mu)=0\,,\quad    (\lambda\gamma_{ab}\mu)=0\,,
\end{equation}
where $\mu_{\alpha}= (\eta\widehat\lambda)_{\alpha}$. For this degenerated case, we have another useful Fierz identity
\begin{equation}
\lambda^{\alpha}\mu_{\beta}=\frac{1}{2}(\gamma^{a}\lambda)_{\beta}(\gamma_{a}\mu)^{\alpha}\, 
\end{equation}
implying
\begin{equation*}
-\frac{1}{8}R_{abcd}(w\gamma^{dc}\lambda)(\gamma^{ba}\mu)_{\alpha}=-\frac{1}{8}(w\eta\gamma_{[a}\eta\gamma_{b]}\lambda)(\gamma^{ba}\mu)_{\alpha}=    
\end{equation*}
\begin{equation}
=\frac{1}{4}(w\eta\gamma_{a}\eta\gamma_{b}\lambda)(\gamma^{a}\gamma^{b}\mu)_{\alpha}=\frac{1}{2}(w\eta\gamma_{a}\eta\mu)(\gamma^{a}\lambda)_{\alpha}=\frac{1}{2}(w\eta\gamma_{a}\widehat\lambda)(\gamma^{a}\lambda)_{\alpha} \,.   
\end{equation}
such that
\begin{equation}
\left[ j ^{a}-\frac{1}{2}(w\eta\gamma^{a}\widehat\lambda)\right](\gamma_{a}\lambda)_{\alpha}=0\,,\qquad \left[\overline{ j }^{a}-\frac{1}{2}(\widehat w\eta\gamma^{a}\lambda)\right](\gamma_{a}\widehat \lambda)_{\widehat\alpha}=0\,,
\end{equation}
is the correct equation resulting from acting with $-q_{\varepsilon}+Q$ on worldsheet fermionic variables. We can re-interpret the variables $w_{\alpha}$ and $\widehat w_{\widehat\alpha}$ as solving the constraints:
\begin{equation}
\Omega_{a}(\gamma^{a}\widehat\lambda)_{\alpha}=0\,,\quad \widehat\Omega_{a}(\gamma^{a}\lambda)_{\alpha}=0\,\,\,\iff\, \Omega_{a}=\frac{1}{2}(w\eta\gamma_{a}\widehat\lambda)\,,\quad \widehat\Omega_{a}=\frac{1}{2}(\widehat w\eta\gamma_{a}\lambda)\,,
\end{equation}
such that the equations for supersymmetric configuration in pure spinor formalism become
\begin{equation*}
( j _{a}-\Omega_{a})(\gamma^{a}\lambda)_{\alpha}=0\,,\qquad  (\overline{ j }_{a}-\widehat \Omega_{a})(\gamma^{a}\widehat\lambda)_{\alpha}=0 \,,  
\end{equation*}
\begin{equation}
\label{new susy eq}
\Omega_{a} (\gamma^{a}\widehat\lambda)_{\alpha}=0\,,\qquad \widehat\Omega_{a}(\gamma^{a}\lambda)_{\alpha}=0\,. 
\end{equation}
In the next subsection, \ref{bps implies eom}, we will compare the above equations with those obtained within the Green-Schwarz formalism.

\subsection{Comparison with Green-Schwarz formalism}
\label{bps implies eom}

In Green-Schwarz formalism, supersymmetric string configurations are characterized by offsets in supersymmetric transformations due to kappa symmetry. In the $Ad\mathbb{S}_5\times \mathbb{S}^5$ superspace, $\frac{PSU(2,2|4)}{USp(2,2)\times USp(4)}$, the kappa symmetry transformations are defined as \cite{Metsaev:1998it}\cite{Berkovits:2002uc}:
\begin{equation*}
(1+q_{\kappa})G(\theta, x) = F(\theta) g(x) (1+\xi+\widehat\xi)
\end{equation*}
where $\xi$ and $\widehat\xi$ are valued at the supercharges and are constrained by:
\begin{equation}
j^{a} (\gamma_{a} \xi)_{\alpha} = 0\,,\qquad    \overline j^{a} (\gamma_{a} \widehat \xi)_{\widehat \alpha} = 0\,.
\end{equation}
One can solve the above constraints by introducing the unconstrained spinors $\kappa_{\alpha}$ and $\widehat\kappa_{\widehat\alpha}$, and expressing $\xi^{\alpha}$ and $\widehat\xi^{\widehat\alpha}$ as $\xi^{\alpha} =   j^{a} (\gamma_{a} \kappa)^{\alpha}$ and $\widehat\xi^{\widehat\alpha} =  \overline j^{a} (\gamma_{a} \widehat\kappa)^{\widehat\alpha}$. The supersymmetric equations in Green-Schwarz formalism are derived by ensuring that the Killing spinors satisfy the same constraints as those for kappa symmetry:
\begin{equation}
\label{GS equations}
j^{a} (\gamma_{a} \varepsilon)_{\alpha} = 0\,,\qquad    \overline j^{a} (\gamma_{a} \widehat \varepsilon)_{\widehat \alpha} = 0\,.
\end{equation}
Effectively, the kappa parameters play a role similar to that of the pure spinor variables.
\vskip 0.3cm
Now, pluggin the Green-Schwarz supersymmetric configurations into the sigma model action appering in equation \ref{sigma model action}, if we denote the Euclidean complex conjugate of $\varepsilon^{\alpha}$ to be $\overline\varepsilon_{\alpha}$, we find that the following manipulation follows:
\begin{equation}
j^{a}\overline j^{b}\delta_{ab} =    j^{a}\overline j^{b}\left[\frac{\overline\varepsilon\gamma^{a}\gamma^{b}\varepsilon + \overline\varepsilon\gamma^{b}\gamma^{a}\varepsilon}{2\varepsilon\overline\varepsilon}  \right] =  j^{a}\overline j^{b}\left[\frac{\overline\varepsilon\gamma^{a}\gamma^{b}\varepsilon - \overline\varepsilon\gamma^{b}\gamma^{a}\varepsilon}{2\varepsilon\overline\varepsilon} \right]\,,
\end{equation}
thus transforming the sigma model action into an integral of a two-form:
\begin{equation}
S=\frac{r^2}{2 \pi\alpha'}\int d^2z j^{a}\overline j^{b}B_{ab} \,,\qquad B_{ab}= \frac{\overline\varepsilon\gamma_{ab}\varepsilon}{2\varepsilon\overline\varepsilon} \,.
\end{equation}
A primary distinction between the supersymmetric equations obtained in Green-Schwarz and pure spinor formalisms is that in Green-Schwarz, there is no requirement on the Killing spinors to be pure. In fact, the Green-Schwarz supersymmetric equations for impure Killing spinors impose stricter requirements on the embedding surfaces than those for pure Killing spinors. This is because in the case of impure Killing spinors, the equations $j^{a} (\gamma_{a} \varepsilon)_{\alpha} = 0$ and $j^{a} (\overline\gamma_{a} \widehat\varepsilon)_{\widehat\alpha} = 0$ imply that there exist conformal factors $\rho(z,\overline z)$ and $\overline\rho(z,\overline z)$ such that
\begin{equation}
j^{a} = \frac{1}{2}\rho(z,\overline z) (\varepsilon\gamma^{a}\varepsilon)\,,  \qquad \overline j^{a} = \frac{1}{2}\overline \rho(z,\overline z) (\widehat\varepsilon\gamma^{a}\widehat\varepsilon)\,,    
\end{equation}
(see appendix A in \cite{Dymarsky:2009si}). Given that an impure Killing spinor can always be written as a sum of two pure Killing spinors:
\begin{equation}
\varepsilon^{\alpha} = \lambda_1^{\alpha}  +\lambda_2^{\alpha}\,,\qquad \widehat\varepsilon^{\widehat\alpha} = \widehat\lambda_1^{\widehat\alpha} + \widehat\lambda_{2}^{\widehat\alpha}\,, 
\end{equation}
it follows that $j^{a} = \rho (\lambda_1\gamma^{a}\lambda_2)$ and $\overline j^{a} = \overline \rho (\widehat\lambda_1\gamma^{a}\widehat\lambda_2)$, and consequently, the currents satisfy the following set of supersymmetric equations:
\begin{equation}
j^{a}(\gamma_{a}\lambda_1)_\alpha = 0\,,\qquad \overline j^{a}(\gamma_{a}\widehat\lambda_1)_\alpha =0\,,\qquad j^{a}(\gamma_{a}\lambda_2)_\alpha = 0\,,\qquad \overline j^{a}(\gamma_{a}\widehat\lambda_2)_\alpha =0\,, 
\end{equation}
which constitute an intersection of two supersymmetric configurations arising from two different choices of supercharges whose Killing spinors are pure.
\vskip 0.3cm
The second main difference between Green-Schwarz and pure spinor supersymmetric equations is that, even when the Killing spinors are pure, the two sets of equations are distinguished by the presence of the variables $\Omega_{a}$ and $\widehat\Omega_{a}$ in the pure spinor formalism, as in \ref{new susy eq}. Setting the ghost variables $\Omega_{a}$ and $\widehat\Omega_{a}$ to zero, we recover equations \ref{old susy eq}, which are essentially the Green-Schwarz equations \ref{GS equations} for pure Killing spinors.
\vskip 0.3cm
We will now demonstrate that equations \ref{GS equations} imply the equations of motion. In fact, we will establish that this holds true for any background without NS-NS flux, provided both the target space and the worldsheet have Euclidean signatures. In any curved background, by setting the spin connection with torsion components equal to the NS-NS flux as follows:
\begin{equation}
\label{eq:Holomorphic gauge}
T_{abc}=-iH_{abc}\,,
\end{equation}
and neglect ghosts and fermions, the equations of motion simplify to $\overline\nabla j^{a}= 0$. Here, the currents $j^{a}$ satisfy the Maurer-Cartan identities:
\begin{equation}
\label{MC}
\overline\nabla j^{a}-\nabla\overline j^{a}=\overline j^{c}j^{b}T_{bc}\,^{a}\,. 
\end{equation}
The supersymmetric equations with ghosts suppressed, as stated in \ref{old susy eq}, imply that the Ramond-Ramond correction in \ref{pure killing} is nullified, permitting the Killing spinor to return to the standard Killing spinor equation, $\overline\nabla\varepsilon^{\alpha}=0$. Consequently, applying a covariant derivative to \ref{old susy eq} yields $\overline\nabla j^{a}(\gamma_{a}\varepsilon)_{\alpha}=0$. For $\varepsilon^{\alpha}\neq 0$, $\overline \nabla j^{a}$ must be a null vector since:
\begin{equation}
\label{null}
\overline\nabla j^{b}\overline\nabla j^{a}(\gamma_{b}\gamma_{a}\varepsilon)^{\alpha}=\frac{1}{2}\overline\nabla j^{b}\overline\nabla j^{a}(\gamma_{(b}\gamma_{a)}\varepsilon)^{\alpha}=\overline\nabla j^{b}\overline\nabla j^{a}\delta_{ab}\varepsilon^{\alpha}\,.
\end{equation}
Using the connection specified in \ref{eq:Holomorphic gauge}, if $H_{abc}=0$, the Maurer-Cartan identity \ref{MC} implies $\overline\nabla j^{a}=\nabla \overline j^{a}$. Therefore, in Euclidean signature, $\overline \nabla j^{a}$ is real and null, and thus it vanishes.

\subsection{Reparametrization invariance}

We now turn our attention to reparametrization invariance. Before localizing to supersymmetric configurations, the sigma model action referenced in \ref{sigma model action} is only reparametrization invariant with the aid of the Virasoro constraints. In pure spinor formalism, it is implicitly assumed that the Virasoro constraints arise from the BRST localization, linked to the existence of a $b$-ghost satisfying $Qb = T$, where $Q$ is the BRST generator and $T$ is the stress-energy tensor. After localizing with $Q-q_{\varepsilon}$, where $q_{\varepsilon}$ is the supersymmetry generator, we will demonstrate that the action becomes invariant under worldsheet reparametrization because the stress-energy tensor vanishes when evaluated at supersymmetric configurations. 
\vskip 0.3cm
For on-shell supersymmetric configurations, where no ghost contributions are present in the Virasoro constraint, the stress-energy tensor, given by $T = j^{a}j^{b}\delta_{ab}$ and $\overline T = \overline j^{a}\overline j^{b}\delta_{ab}$, vanishes since
\begin{equation}
j^{a}(\gamma_{a}\varepsilon)_{\alpha}=0\implies j^{a}j^{b}\delta_{ab} = 0\,,\qquad \overline j^{a}(\gamma_{a}\widehat\varepsilon)_{\widehat\alpha}=0\implies \overline j^{a}\overline j^{b}\delta_{ab} = 0 \,.
\end{equation}
For the off-shell ones, the Virasoro constraint does receive contributions from the ghosts, as follows:
\begin{equation}
T = \frac{1}{2}j^{a}j^{b}\delta_{ab} + \nabla\lambda^{\alpha}w_{\alpha}\,,\qquad   \overline T = \frac{1}{2}\overline j^{a}\overline j^{b}\delta_{ab} + \overline \nabla\widehat\lambda^{\widehat\alpha}\widehat w_{\widehat\alpha} \,.
\end{equation}
Using the supersymmetric equations in \ref{new susy eq}, we can eliminate the covariant derivatives of the pure spinors, obtaning:
\begin{equation}
T =    \frac{1}{2}j^{a}j^{b}\delta_{ab} - j^{a}\Omega^{b}\delta_{ab}\,,\qquad  \overline T =    \frac{1}{2}\overline j^{a}\overline j^{b}\delta_{ab} - \overline j^{a}\widehat \Omega^{b}\delta_{ab}\,.
\end{equation}
With some manipulation, we can re-express the term $\frac{1}{2}\delta_{ab}j^{a}j^{b}$ as $j^{a}\Omega^{b}\delta_{ab}$, as follows:
\begin{equation}
\frac{1}{2}j^{a}\delta_{ab}j^{b} =  \frac{1}{2}j^{a}\left[\frac{\overline\lambda\gamma_{a}\gamma_{b}\lambda}{2(\overline\lambda\lambda)}+\frac{\overline\lambda\gamma_{b}\gamma_{a}\lambda}{2(\overline\lambda\lambda)}\right]j^{b} = \frac{1}{2}j^{a} \frac{\overline\lambda\gamma_{a}\gamma_{b}\lambda}{2(\overline\lambda\lambda)}\Omega^{b} +\frac{1}{2} \Omega^{a}\frac{\overline\lambda\gamma_{b}\gamma_{a}\lambda}{2(\overline\lambda\lambda)} j^{b} = 
\end{equation}
\begin{equation*}
= \frac{1}{2}j^{a} \left[\frac{\overline\lambda\gamma_{a}\gamma_{b}\lambda}{2(\overline\lambda\lambda)} + 0\right]\Omega^{b} + \frac{1}{2}\Omega^{a}\left[0 +\frac{\overline\lambda\gamma_{b}\gamma_{a}\lambda}{2(\overline\lambda\lambda)}\right] j^{b}  =  
\end{equation*}
\begin{equation*}
= \frac{1}{2}j^{a} \left[\frac{\overline\lambda\gamma_{a}\gamma_{b}\lambda}{2(\overline\lambda\lambda)}+\frac{\overline\lambda\gamma_{b}\gamma_{a}\lambda}{2(\overline\lambda\lambda)}\right]\Omega^{b} + \frac{1}{2}\Omega^{a}\left[\frac{\overline\lambda\gamma_{a}\gamma_{b}\lambda}{2(\overline\lambda\lambda)}+\frac{\overline\lambda\gamma_{b}\gamma_{a}\lambda}{2(\overline\lambda\lambda)}\right] j^{b}    = \frac{1}{2}j^{a}\delta_{ab}\Omega^{b} + \frac{1}{2}\Omega^{a}\delta_{ab}j^{b}\,,
\end{equation*}
and similar for the right-mover ones, $\overline j^{a}$ and $\widehat\Omega_{a}$. We obtain that the stress-energy tensor, with ghost corrections, vanishes at off-shell supersymmetric configurations:
\begin{equation}
T = 0 \,,\qquad \overline T = 0\,,
\end{equation}
thus the sigma model action, when evaluated at off-shell supersymmetric configurations, becomes reparametrization invariant.

\section{The Ad$\mathbb{S}_4\times\mathbb{S}^2$ slice}

\label{the AdS slice}

In this section we will fix a supercharge to further specify the supersymmetric string configurations we saw in the previous section. Our choice of supercharge is such that the resulting pure Killing spinor is pure only in a Ad$\mathbb{S}_4\times \mathbb{S}^2$ slice of Ad$\mathbb{S}_5\times \mathbb{S}^{5}$, so the supersymmetric string configurations will be localized at this slice. We check that the well-known classical solution, namely, the minimal Ad$\mathbb{S}_2$ configuration, obey the on-shell supersymmetric equations. Then, we consider the full off-shell supersymmetric equations, and show that essentially all surface embedding in the Ad$\mathbb{S}_4\times \mathbb{S}^2$ slice are off-shell supersymmetric, at the expense of determining the components of $\Omega_{a}$ and $\widehat\Omega_{a}$. Finaly, we evaluate the sigma model action at these surfaces and observe that it reduces to an integral of a two-form. The two-form is not closed in Ad$\mathbb{S}_4\times \mathbb{S}^2$. Our choice of supercharge will be one that preserves a circular half-BPS Wilson loop.

\subsection{Supersymmetric Wilson loops and boundary conditions}

Supersymmetric boundary conditions are holographically dual to supersymmetric Wilson loops according to \cite{Drukker:1999zq}. The supersymmetric Wilson loops are characterized by the configuration of the loop $x^{\mu}(\sigma)$, and a $SO(6)_{R}$ vector $\theta^{i}(\sigma)$ distributed along the loop. It is given by
\begin{equation}
W=\text{Tr}\text{P}\exp\int_0^{2\pi} id\sigma\left(\partial_{\sigma}x^{\mu}(\sigma)A_{\mu}(x(\sigma))+i\theta^{i}(\sigma)\phi_{i}(x(\sigma))\right)\,,  
\end{equation}
where $A_{\mu}$ is the gauge potential and $\phi_{i}$ the six-scalars of $\mathcal{N}=4$ d=4 super-Yang-Mills, both valued at the Lie algebra of the gauge group. Regarding the holographic map between the boundary conditions and supersymmetric Wilson loop $W$, while the configuration of the loop in $\mathbb{R}^4$ is essentially the values of $x^{\mu}(0,\sigma)$, the  $SO(6)_{R}$ vectors are related with the embedded surface as follows:
\begin{equation}
\theta^{i}=(y\partial_{\tau}-\partial_{\tau}y)\frac{y^{i}}{y}\,.
\end{equation}
Let us see how this dictionary is compatible with the Virasoro constraint. Close to the boundary, when $y\rightarrow 0$, the Virasoro constraints approach to
\begin{equation}
\frac{\partial x^{\mu}\partial x_{\mu}+\partial y^{i}\partial y_{i}}{y^{2}}=0\quad \xrightarrow{y\rightarrow 0} \quad\frac{-\partial_{\sigma} x^{\mu}\partial_{\sigma} x_{\mu}+\partial_{\tau} y^{i}\partial_{\tau} y_{i}}{y^{2}}=0\,,    
\end{equation}
due to the different nature in the boundary conditions for $x^{\mu}$ and $y^{i}$ variables. In using the holographic map between supersymmetric Wilson loops and boundary conditions, we have that the Virasoro constraint implies
\begin{equation}
\partial_{\sigma}x^{\mu}(\sigma)\partial_{\sigma}x_{\mu}(\sigma)=\theta^{i}(\sigma)\theta_{i}(\sigma)\,.    
\end{equation}
This is precisely the constraints for the supersymmetric Wilson loop to be locally supersymmetric \cite{Drukker:1999zq} \footnote{I am grateful to Nathan Berkovits for emphasizing this point.}. Using the coordinates of appendix \ref{coordinates}, we will eventually fix the circular half-BPS boundary conditions:
\begin{equation*}
x^{1}(0,\sigma)=\cos(\sigma)\,,\qquad x^{2}(0,\sigma)=\sin(\sigma)  \,,\qquad x^{3,4}(0,\sigma)=0\,,
\end{equation*}
\begin{equation}
\label{BC choice}
\partial_{\tau}y^{5}(0,\sigma) = 1\,,\qquad \partial_{\tau} y^{1,\dots,5}(\tau,\sigma)=0\,,
\end{equation}
together with the condition that as $\tau\rightarrow \infty$, the embedding coordinates converge to a single point contained in the bulk. If we relax this condition in $\tau\rightarrow \infty$, we obtain the holographic duals of correlations between supersymmetric Wilson loops and local operators, or between different supersymmetric Wilson loops.
\subsection{Pure Killing spinors and the Ad$\mathbb{S}_4\times\mathbb{S}^2$ slice}
\label{Killing spinors and their pure region}
Following the parametrization of the superchages estabilshed in the appendix \ref{AdS Killing spinors}, the $16$ superchages preserved by the boundary conditions in \ref{BC choice} are given by
\begin{equation}
\varepsilon_c = i\gamma_{129}\varepsilon_{s}\,.
\end{equation}
We can take the 16 spinorial components of $\varepsilon_{s}$ to parametrize these supercharges. Depending on the choice of $\varepsilon_{s}$, the corresponding Killing spinor can be pure everywhere in Ad$\mathbb{S}_5\times \mathbb{S}^{5}$, pure somewhere in Ad$\mathbb{S}_5\times \mathbb{S}^{5}$ or be impure. This have been studied before in the context of supersymmetric Wilson loops \cite{Dymarsky:2009si}, and in the appendix \ref{pure killing spinors - appendix} we review their studies. Here, we will focus in a particular choice of $\varepsilon_{s}$ where the corresponding Killing spinor is only pure in a Ad$\mathbb{S}_4\times \mathbb{S}^2$ slice of Ad$\mathbb{S}_5\times \mathbb{S}^{5}$, namely:
\begin{equation}
\label{pure region}
x_3=0\,,\qquad y_1=0\,,\qquad y_4 = 0 \,,\qquad y_6 = 0\,.  
\end{equation}
Our choice of supercharge, following the convertions of appendix \ref{Conventions on spinors} and \ref{Gamma matrices}, is characterized as taking $\varepsilon_{s}^{1}$ to be the only non-zero component of $\varepsilon_s^{\alpha}$, where $\alpha$ range over the $16$ chiral spinorial components. It is equivalently defined as
\begin{equation*}
(\gamma_1-i\gamma_6) \varepsilon_{s}=0   \,,
\qquad
(\gamma_2-i\gamma_7) \varepsilon_{s} =0  \,,
\qquad
(\gamma_3-i\gamma_8) \varepsilon_{s}  =0 \,,
\end{equation*}
\begin{equation}
(\gamma_4-i\gamma_9) \varepsilon_{s} =0  \,,
\qquad 
(\gamma_5-i\gamma_{10}) \varepsilon_{s}   =0\,.
\end{equation}
If we normalize the non-zero component of $\varepsilon_{s}$, the resulting Killing spinors are given by the non-zero components
\begin{equation*}
\varepsilon^{1} = y^{-\frac{1}{2}}\,,    
\end{equation*}
\begin{equation*}
 \varepsilon^{2} = y^{-\frac{1}{2}}\left(-x_4+iy_5\right)\qquad \varepsilon^{5} = y^{-\frac{1}{2}}\left(x_2-iy_3\right)\,
\end{equation*}
\begin{equation*}
\varepsilon^{6} =y^{-\frac{1}{2}}\left( -x_1+iy_2\right)\qquad \varepsilon^{8} = y^{-\frac{1}{2}}\left(x_3+iy_4\right) \qquad \varepsilon^{14} = y^{-\frac{1}{2}}\left(y_1-iy_6\right)\,,   
\end{equation*}
\begin{equation*}
-i \widehat\varepsilon^{8} = y^{-\frac{1}{2}}\,,    
\end{equation*}
\begin{equation*}
-i \widehat\varepsilon^{1} = y^{-\frac{1}{2}}\left(x_3-iy_4\right)\qquad -i \widehat\varepsilon^{3} = y^{-\frac{1}{2}}\left(-x_1-iy_2\right)\,,
\end{equation*}
\begin{equation}
\label{complex structure}
-i \widehat\varepsilon^{4} = y^{-\frac{1}{2}}\left(-x_2-iy_3\right)\qquad -i \widehat\varepsilon^{7} = y^{-\frac{1}{2}}\left(x_4+iy_5\right)\qquad -i \widehat\varepsilon^{11} = y^{-\frac{1}{2}}\left(-y_1+iy_6\right)\,.    
\end{equation}
Notice that components in $\varepsilon$ and $-i\widehat\varepsilon$ are complex conjugates to one another except for $y_5-iy_6$. Using the gamma matrices and charge conjugation matrices of appendices \ref{pure killing spinors - appendix} and \ref{Gamma matrices}, we can check that
\begin{equation*}
 (\varepsilon\gamma^{3}\varepsilon) = 2y^{-1}(y_1-iy_6)\,,\qquad (\varepsilon\gamma^{5}\varepsilon) = 2y^{-1}(x_3+iy_4)\,,     
\end{equation*}
\begin{equation*}
  (\varepsilon\gamma^{8}\varepsilon) = -2iy^{-1}(y_1-iy_6)\,,\qquad (\varepsilon\gamma^{10}\varepsilon) = -2iy^{-1}(x_3+iy_4) \,,
\end{equation*}
\begin{equation*}
 -(\widehat\varepsilon\gamma^{3}\widehat\varepsilon)=2y^{-1}(y_1-iy_6)\,,\qquad    -(\widehat\varepsilon\gamma^{5}\widehat\varepsilon)=2y^{-1}(x_3-iy_4)\,,    
\end{equation*}
\begin{equation}
-(\widehat\varepsilon\gamma^{8}\widehat\varepsilon)=2iy^{-1}(y_1-iy_6)\,,\qquad -(\widehat\varepsilon\gamma^{10}\widehat\varepsilon)=-2iy^{-1}(x_3+iy_4)\,,
\end{equation}
such that if we fix the signature to be Euclidean, the condition that the Killing spinor must coincide with the pure spinors living on the worldsheet restrict the embedded surface to be in a Ad$\mathbb{S}_4\times \mathbb{S}^2$ slice defined by equation \ref{pure region}. One can observe the degeneration observed in subsection \ref{Derivation of the supersymmetric equations} of pure Killing spinor by computing:
\begin{equation}
\mu_{\alpha}\varepsilon^{\alpha} = \widehat\varepsilon^{\widehat\alpha} \eta_{\widehat\alpha\alpha}\varepsilon^{\alpha}=2 x_3(y_1-iy_6),
\end{equation}
where at the Ad$\mathbb{S}_4\times \mathbb{S}^2$ slice defined by equation \ref{pure region}, $\mu_{\alpha}\varepsilon^{\alpha}$ vanishes.

\subsection{On-shell supersymmetric configuration}
\label{on}

\vskip 0.3cm
When the $\Omega_{a}$ and $\widehat\Omega_{a}$ are are supressed, in the Ad$\mathbb{S}_4\times\mathbb{S}^2$ slice, where the Killing spinors are pure, the supersymmetric equations are given by
\begin{equation}
\label{BPS}
\text{BPS}_1 = 0\,,\qquad \text{BPS}_2 = 0\,,\qquad \text{BPS}_3 = 0\,,\qquad \text{BPS}_4=0    
\end{equation}
\begin{equation*}
 \overline{\text{BPS}}_1 = 0\,,\qquad \overline{\text{BPS}}_2 = 0\,,\qquad \overline{\text{BPS}}_3 = 0\,,\qquad \overline{\text{BPS}}_4=0    
\end{equation*}
where
\begin{equation*}
\text{BPS}_{1} = -\partial_{\tau}x_1y_2^2 - \partial_{\tau}x_1y_3^2 - \partial_{\tau}x_1y_5^2 + \partial_{\tau}x_2x_4y_2^2 + \partial_{\tau}x_2x_4y_3^2 + \partial_{\tau}x_2x_4y_5^2 - 
\end{equation*}
\begin{equation*} 
i\partial_{\tau}x_2y_2^2y_5 - i\partial_{\tau}x_2y_3^2y_5 - i\partial_{\tau}x_2y_5^3 - \partial_{\tau}x_4x_2y_2^2 - \partial_{\tau}x_4x_2y_3^2 - \partial_{\tau}x_4x_2y_5^2 + i\partial_{\tau}x_4y_2^2y_3 + \end{equation*}
\begin{equation*} 
i\partial_{\tau}x_4y_3^3 + i\partial_{\tau}x_4y_3y_5^2 + 2i\partial_{\tau}y_2x_2y_2y_5 - 2i\partial_{\tau}y_2x_4y_2y_3 - i\partial_{\tau}y_2y_2^2 + i\partial_{\tau}y_2y_3^2 + i\partial_{\tau}y_2y_5^2 + \end{equation*}
\begin{equation*} 
2i\partial_{\tau}y_3x_2y_3y_5 + i\partial_{\tau}y_3x_4y_2^2 - i\partial_{\tau}y_3x_4y_3^2 + i\partial_{\tau}y_3x_4y_5^2 + \partial_{\tau}y_3y_2^2y_5 - 2i\partial_{\tau}y_3y_2y_3 + \partial_{\tau}y_3y_3^2y_5 + \end{equation*}
\begin{equation*} 
\partial_{\tau}y_3y_5^3 - i\partial_{\tau}y_5x_2y_2^2 - i\partial_{\tau}y_5x_2y_3^2 + i\partial_{\tau}y_5x_2y_5^2 - 2i\partial_{\tau}y_5x_4y_3y_5 - \partial_{\tau}y_5y_2^2y_3 - 2i\partial_{\tau}y_5y_2y_5 - \end{equation*}
\begin{equation*} 
\partial_{\tau}y_5y_3^3 - \partial_{\tau}y_5y_3y_5^2 + i\partial_{\sigma}x_1y_2^2 + i\partial_{\sigma}x_1y_3^2 + i\partial_{\sigma}x_1y_5^2 - i\partial_{\sigma}x_2x_4y_2^2 - i\partial_{\sigma}x_2x_4y_3^2 - \end{equation*}
\begin{equation*} 
i\partial_{\sigma}x_2x_4y_5^2 - \partial_{\sigma}x_2y_2^2y_5 - \partial_{\sigma}x_2y_3^2y_5 - \partial_{\sigma}x_2y_5^3 + i\partial_{\sigma}x_4x_2y_2^2 + i\partial_{\sigma}x_4x_2y_3^2 + i\partial_{\sigma}x_4x_2y_5^2 + \end{equation*}
\begin{equation*} 
\partial_{\sigma}x_4y_2^2y_3 + \partial_{\sigma}x_4y_3^3 + \partial_{\sigma}x_4y_3y_5^2 + 2\partial_{\sigma}y_2x_2y_2y_5 - 2\partial_{\sigma}y_2x_4y_2y_3 - \partial_{\sigma}y_2y_2^2 + \partial_{\sigma}y_2y_3^2 + \end{equation*}
\begin{equation*} 
\partial_{\sigma}y_2y_5^2 + 2\partial_{\sigma}y_3x_2y_3y_5 + \partial_{\sigma}y_3x_4y_2^2 - \partial_{\sigma}y_3x_4y_3^2 + \partial_{\sigma}y_3x_4y_5^2 - i\partial_{\sigma}y_3y_2^2y_5 - 2\partial_{\sigma}y_3y_2y_3 - \end{equation*}
\begin{equation*} 
i\partial_{\sigma}y_3y_3^2y_5 - i\partial_{\sigma}y_3y_5^3 - \partial_{\sigma}y_5x_2y_2^2 - \partial_{\sigma}y_5x_2y_3^2 + \partial_{\sigma}y_5x_2y_5^2 - 2\partial_{\sigma}y_5x_4y_3y_5 + i\partial_{\sigma}y_5y_2^2y_3 - \end{equation*}
\begin{equation*} 
2\partial_{\sigma}y_5y_2y_5 + i\partial_{\sigma}y_5y_3^3 + i\partial_{\sigma}y_5y_3y_5^2\,,
\end{equation*}
\begin{equation*}
\text{BPS}_{2} = -\partial_{\tau}x_1x_4y_2^2 - \partial_{\tau}x_1x_4y_3^2 - \partial_{\tau}x_1x_4y_5^2 + i\partial_{\tau}x_1y_2^2y_5 + 
\end{equation*}
\begin{equation*}
i\partial_{\tau}x_1y_3^2y_5 + i\partial_{\tau}x_1y_5^3 - \partial_{\tau}x_2y_2^2 - \partial_{\tau}x_2y_3^2 - \partial_{\tau}x_2y_5^2 + \partial_{\tau}x_4x_1y_2^2 + \partial_{\tau}x_4x_1y_3^2 + 
\end{equation*}
\begin{equation*}
\partial_{\tau}x_4x_1y_5^2 - i\partial_{\tau}x_4y_2^3 - i\partial_{\tau}x_4y_2y_3^2 - i\partial_{\tau}x_4y_2y_5^2 - 2i\partial_{\tau}y_2x_1y_2y_5 + i\partial_{\tau}y_2x_4y_2^2 - i\partial_{\tau}y_2x_4y_3^2 - 
\end{equation*}
\begin{equation*}
i\partial_{\tau}y_2x_4y_5^2 - \partial_{\tau}y_2y_2^2y_5 - 2i\partial_{\tau}y_2y_2y_3 - \partial_{\tau}y_2y_3^2y_5 - \partial_{\tau}y_2y_5^3 - 2i\partial_{\tau}y_3x_1y_3y_5 + 2i\partial_{\tau}y_3x_4y_2y_3 + 
\end{equation*}
\begin{equation*}
i\partial_{\tau}y_3y_2^2 - i\partial_{\tau}y_3y_3^2 + i\partial_{\tau}y_3y_5^2 + i\partial_{\tau}y_5x_1y_2^2 + i\partial_{\tau}y_5x_1y_3^2 - i\partial_{\tau}y_5x_1y_5^2 + 2i\partial_{\tau}y_5x_4y_2y_5 + 
\end{equation*}
\begin{equation*}
\partial_{\tau}y_5y_2^3 + \partial_{\tau}y_5y_2y_3^2 + \partial_{\tau}y_5y_2y_5^2 - 2i\partial_{\tau}y_5y_3y_5 + i\partial_{\sigma}x_1x_4y_2^2 + i\partial_{\sigma}x_1x_4y_3^2 + i\partial_{\sigma}x_1x_4y_5^2 + 
\end{equation*}
\begin{equation*}
\partial_{\sigma}x_1y_2^2y_5 + \partial_{\sigma}x_1y_3^2y_5 + \partial_{\sigma}x_1y_5^3 + i\partial_{\sigma}x_2y_2^2 + i\partial_{\sigma}x_2y_3^2 + i\partial_{\sigma}x_2y_5^2 - i\partial_{\sigma}x_4x_1y_2^2 - 
\end{equation*}
\begin{equation*}
i\partial_{\sigma}x_4x_1y_3^2 - i\partial_{\sigma}x_4x_1y_5^2 - \partial_{\sigma}x_4y_2^3 - \partial_{\sigma}x_4y_2y_3^2 - \partial_{\sigma}x_4y_2y_5^2 - 2\partial_{\sigma}y_2x_1y_2y_5 + \partial_{\sigma}y_2x_4y_2^2 - 
\end{equation*}
\begin{equation*}
\partial_{\sigma}y_2x_4y_3^2 - \partial_{\sigma}y_2x_4y_5^2 + i\partial_{\sigma}y_2y_2^2y_5 - 2\partial_{\sigma}y_2y_2y_3 + i\partial_{\sigma}y_2y_3^2y_5 + i\partial_{\sigma}y_2y_5^3 - 2\partial_{\sigma}y_3x_1y_3y_5 + 
\end{equation*}
\begin{equation*}
2\partial_{\sigma}y_3x_4y_2y_3 + \partial_{\sigma}y_3y_2^2 - \partial_{\sigma}y_3y_3^2 + \partial_{\sigma}y_3y_5^2 + \partial_{\sigma}y_5x_1y_2^2 + \partial_{\sigma}y_5x_1y_3^2 - \partial_{\sigma}y_5x_1y_5^2 + 
\end{equation*}
\begin{equation*}
2\partial_{\sigma}y_5x_4y_2y_5 - i\partial_{\sigma}y_5y_2^3 - i\partial_{\sigma}y_5y_2y_3^2 - i\partial_{\sigma}y_5y_2y_5^2 - 2\partial_{\sigma}y_5y_3y_5\,,
\end{equation*}
\begin{equation*}
\text{BPS}_{3} = \partial_{\tau}x_1x_2y_2^2 + \partial_{\tau}x_1x_2y_3^2 + \partial_{\tau}x_1x_2y_5^2 - 
\end{equation*}
\begin{equation*}
i\partial_{\tau}x_1y_2^2y_3 - i\partial_{\tau}x_1y_3^3 - i\partial_{\tau}x_1y_3y_5^2 - \partial_{\tau}x_2x_1y_2^2 - \partial_{\tau}x_2x_1y_3^2 - \partial_{\tau}x_2x_1y_5^2 + i\partial_{\tau}x_2y_2^3 + 
\end{equation*}
\begin{equation*}
i\partial_{\tau}x_2y_2y_3^2 + i\partial_{\tau}x_2y_2y_5^2 - \partial_{\tau}x_4y_2^2 - \partial_{\tau}x_4y_3^2 - \partial_{\tau}x_4y_5^2 + 2i\partial_{\tau}y_2x_1y_2y_3 - i\partial_{\tau}y_2x_2y_2^2 + 
\end{equation*}
\begin{equation*}
i\partial_{\tau}y_2x_2y_3^2 + i\partial_{\tau}y_2x_2y_5^2 + \partial_{\tau}y_2y_2^2y_3 - 2i\partial_{\tau}y_2y_2y_5 + \partial_{\tau}y_2y_3^3 + \partial_{\tau}y_2y_3y_5^2 - i\partial_{\tau}y_3x_1y_2^2 + 
\end{equation*}
\begin{equation*}
i\partial_{\tau}y_3x_1y_3^2 - i\partial_{\tau}y_3x_1y_5^2 - 2i\partial_{\tau}y_3x_2y_2y_3 - \partial_{\tau}y_3y_2^3 - \partial_{\tau}y_3y_2y_3^2 - \partial_{\tau}y_3y_2y_5^2 - 2i\partial_{\tau}y_3y_3y_5 + 
\end{equation*}
\begin{equation*}
2i\partial_{\tau}y_5x_1y_3y_5 - 2i\partial_{\tau}y_5x_2y_2y_5 + i\partial_{\tau}y_5y_2^2 + i\partial_{\tau}y_5y_3^2 - i\partial_{\tau}y_5y_5^2 - i\partial_{\sigma}x_1x_2y_2^2 - i\partial_{\sigma}x_1x_2y_3^2 - \end{equation*}\begin{equation*}  i\partial_{\sigma}x_1x_2y_5^2 - \partial_{\sigma}x_1y_2^2y_3 - \partial_{\sigma}x_1y_3^3 - \partial_{\sigma}x_1y_3y_5^2 + i\partial_{\sigma}x_2x_1y_2^2 + i\partial_{\sigma}x_2x_1y_3^2 + i\partial_{\sigma}x_2x_1y_5^2 + 
\end{equation*}
\begin{equation*}
\partial_{\sigma}x_2y_2^3 + \partial_{\sigma}x_2y_2y_3^2 + \partial_{\sigma}x_2y_2y_5^2 + i\partial_{\sigma}x_4y_2^2 + i\partial_{\sigma}x_4y_3^2 + i\partial_{\sigma}x_4y_5^2 + 2\partial_{\sigma}y_2x_1y_2y_3 - 
\end{equation*}
\begin{equation*}
\partial_{\sigma}y_2x_2y_2^2 + \partial_{\sigma}y_2x_2y_3^2 + \partial_{\sigma}y_2x_2y_5^2 - i\partial_{\sigma}y_2y_2^2y_3 - 2\partial_{\sigma}y_2y_2y_5 - i\partial_{\sigma}y_2y_3^3 - i\partial_{\sigma}y_2y_3y_5^2 - 
\end{equation*}
\begin{equation*}
\partial_{\sigma}y_3x_1y_2^2 + \partial_{\sigma}y_3x_1y_3^2 - \partial_{\sigma}y_3x_1y_5^2 - 2\partial_{\sigma}y_3x_2y_2y_3 + i\partial_{\sigma}y_3y_2^3 + i\partial_{\sigma}y_3y_2y_3^2 + i\partial_{\sigma}y_3y_2y_5^2 - 
\end{equation*}
\begin{equation*}
2\partial_{\sigma}y_3y_3y_5 + 2\partial_{\sigma}y_5x_1y_3y_5 - 2\partial_{\sigma}y_5x_2y_2y_5 + \partial_{\sigma}y_5y_2^2 + \partial_{\sigma}y_5y_3^2 - \partial_{\sigma}y_5y_5^2\,,
\end{equation*}
\begin{equation*}
\text{BPS}_{4} = \partial_{\tau}x_1x_1y_2^2 + \partial_{\tau}x_1x_1y_3^2 + 
\end{equation*}
\begin{equation*}
\partial_{\tau}x_1x_1y_5^2 - i\partial_{\tau}x_1y_2^3 - i\partial_{\tau}x_1y_2y_3^2 - i\partial_{\tau}x_1y_2y_5^2 + \partial_{\tau}x_2x_2y_2^2 + \partial_{\tau}x_2x_2y_3^2 + \partial_{\tau}x_2x_2y_5^2 - 
\end{equation*}
\begin{equation*}
i\partial_{\tau}x_2y_2^2y_3 - i\partial_{\tau}x_2y_3^3 - i\partial_{\tau}x_2y_3y_5^2 + \partial_{\tau}x_4x_4y_2^2 + \partial_{\tau}x_4x_4y_3^2 + \partial_{\tau}x_4x_4y_5^2 - i\partial_{\tau}x_4y_2^2y_5 - 
\end{equation*}
\begin{equation*}
i\partial_{\tau}x_4y_3^2y_5 - i\partial_{\tau}x_4y_5^3 + i\partial_{\tau}y_2x_1y_2^2 - i\partial_{\tau}y_2x_1y_3^2 - i\partial_{\tau}y_2x_1y_5^2 + 2i\partial_{\tau}y_2x_2y_2y_3 + 2i\partial_{\tau}y_2x_4y_2y_5 + 
\end{equation*}
\begin{equation*}
\partial_{\tau}y_2y_2^3 + \partial_{\tau}y_2y_2y_3^2 + \partial_{\tau}y_2y_2y_5^2 + 2i\partial_{\tau}y_3x_1y_2y_3 - i\partial_{\tau}y_3x_2y_2^2 + i\partial_{\tau}y_3x_2y_3^2 - i\partial_{\tau}y_3x_2y_5^2 + 
\end{equation*}
\begin{equation*}
2i\partial_{\tau}y_3x_4y_3y_5 + \partial_{\tau}y_3y_2^2y_3 + \partial_{\tau}y_3y_3^3 + \partial_{\tau}y_3y_3y_5^2 + 2i\partial_{\tau}y_5x_1y_2y_5 + 2i\partial_{\tau}y_5x_2y_3y_5 - i\partial_{\tau}y_5x_4y_2^2 - 
\end{equation*}
\begin{equation*}
i\partial_{\tau}y_5x_4y_3^2 + i\partial_{\tau}y_5x_4y_5^2 + \partial_{\tau}y_5y_2^2y_5 + \partial_{\tau}y_5y_3^2y_5 + \partial_{\tau}y_5y_5^3 - i\partial_{\sigma}x_1x_1y_2^2 - i\partial_{\sigma}x_1x_1y_3^2 - 
\end{equation*}
\begin{equation*}
i\partial_{\sigma}x_1x_1y_5^2 - \partial_{\sigma}x_1y_2^3 - \partial_{\sigma}x_1y_2y_3^2 - \partial_{\sigma}x_1y_2y_5^2 - i\partial_{\sigma}x_2x_2y_2^2 - i\partial_{\sigma}x_2x_2y_3^2 - i\partial_{\sigma}x_2x_2y_5^2 - 
\end{equation*}
\begin{equation*}
\partial_{\sigma}x_2y_2^2y_3 - \partial_{\sigma}x_2y_3^3 - \partial_{\sigma}x_2y_3y_5^2 - i\partial_{\sigma}x_4x_4y_2^2 - i\partial_{\sigma}x_4x_4y_3^2 - i\partial_{\sigma}x_4x_4y_5^2 - \partial_{\sigma}x_4y_2^2y_5 - 
\end{equation*}
\begin{equation*}
\partial_{\sigma}x_4y_3^2y_5 - \partial_{\sigma}x_4y_5^3 + \partial_{\sigma}y_2x_1y_2^2 - \partial_{\sigma}y_2x_1y_3^2 - \partial_{\sigma}y_2x_1y_5^2 + 2\partial_{\sigma}y_2x_2y_2y_3 + 2\partial_{\sigma}y_2x_4y_2y_5 - 
\end{equation*}
\begin{equation*}
i\partial_{\sigma}y_2y_2^3 - i\partial_{\sigma}y_2y_2y_3^2 - i\partial_{\sigma}y_2y_2y_5^2 + 2\partial_{\sigma}y_3x_1y_2y_3 - \partial_{\sigma}y_3x_2y_2^2 + \partial_{\sigma}y_3x_2y_3^2 - \partial_{\sigma}y_3x_2y_5^2 + 
\end{equation*}
\begin{equation*}
2\partial_{\sigma}y_3x_4y_3y_5 - i\partial_{\sigma}y_3y_2^2y_3 - i\partial_{\sigma}y_3y_3^3 - i\partial_{\sigma}y_3y_3y_5^2 + 2\partial_{\sigma}y_5x_1y_2y_5 + 2\partial_{\sigma}y_5x_2y_3y_5 - \partial_{\sigma}y_5x_4y_2^2 - 
\end{equation*}
\begin{equation*}
\partial_{\sigma}y_5x_4y_3^2 + \partial_{\sigma}y_5x_4y_5^2 - i\partial_{\sigma}y_5y_2^2y_5 - i\partial_{\sigma}y_5y_3^2y_5 - i\partial_{\sigma}y_5y_5^3\,,
\end{equation*}
and
\begin{equation*}
\widehat{\text{BPS}}_{1} = -\partial_{\tau}x_1x_1y_2^2 - \partial_{\tau}x_1x_1y_3^2 - \partial_{\tau}x_1x_1y_5^2 - i\partial_{\tau}x_1y_2^3 - i\partial_{\tau}x_1y_2y_3^2 - i\partial_{\tau}x_1y_2y_5^2 - 
\end{equation*}
\begin{equation*}
\partial_{\tau}x_2x_2y_2^2 - \partial_{\tau}x_2x_2y_3^2 - \partial_{\tau}x_2x_2y_5^2 - i\partial_{\tau}x_2y_2^2y_3 - i\partial_{\tau}x_2y_3^3 - i\partial_{\tau}x_2y_3y_5^2 - \partial_{\tau}x_4x_4y_2^2 - 
\end{equation*}
\begin{equation*}
\partial_{\tau}x_4x_4y_3^2 - \partial_{\tau}x_4x_4y_5^2 - i\partial_{\tau}x_4y_2^2y_5 - i\partial_{\tau}x_4y_3^2y_5 - i\partial_{\tau}x_4y_5^3 + i\partial_{\tau}y_2x_1y_2^2 - i\partial_{\tau}y_2x_1y_3^2 - 
\end{equation*}
\begin{equation*}
i\partial_{\tau}y_2x_1y_5^2 + 2i\partial_{\tau}y_2x_2y_2y_3 + 2i\partial_{\tau}y_2x_4y_2y_5 - \partial_{\tau}y_2y_2^3 - \partial_{\tau}y_2y_2y_3^2 - \partial_{\tau}y_2y_2y_5^2 + 2i\partial_{\tau}y_3x_1y_2y_3 - 
\end{equation*}
\begin{equation*}
i\partial_{\tau}y_3x_2y_2^2 + i\partial_{\tau}y_3x_2y_3^2 - i\partial_{\tau}y_3x_2y_5^2 + 2i\partial_{\tau}y_3x_4y_3y_5 - \partial_{\tau}y_3y_2^2y_3 - \partial_{\tau}y_3y_3^3 - \partial_{\tau}y_3y_3y_5^2 + 
\end{equation*}
\begin{equation*}
2i\partial_{\tau}y_5x_1y_2y_5 + 2i\partial_{\tau}y_5x_2y_3y_5 - i\partial_{\tau}y_5x_4y_2^2 - i\partial_{\tau}y_5x_4y_3^2 + i\partial_{\tau}y_5x_4y_5^2 - \partial_{\tau}y_5y_2^2y_5 - \partial_{\tau}y_5y_3^2y_5 - 
\end{equation*}
\begin{equation*}
\partial_{\tau}y_5y_5^3 - i\partial_{\sigma}x_1x_1y_2^2 - i\partial_{\sigma}x_1x_1y_3^2 - i\partial_{\sigma}x_1x_1y_5^2 + \partial_{\sigma}x_1y_2^3 + \partial_{\sigma}x_1y_2y_3^2 + \partial_{\sigma}x_1y_2y_5^2 - 
\end{equation*}
\begin{equation*}
i\partial_{\sigma}x_2x_2y_2^2 - i\partial_{\sigma}x_2x_2y_3^2 - i\partial_{\sigma}x_2x_2y_5^2 + \partial_{\sigma}x_2y_2^2y_3 + \partial_{\sigma}x_2y_3^3 + \partial_{\sigma}x_2y_3y_5^2 - i\partial_{\sigma}x_4x_4y_2^2 - 
\end{equation*}
\begin{equation*}
i\partial_{\sigma}x_4x_4y_3^2 - i\partial_{\sigma}x_4x_4y_5^2 + \partial_{\sigma}x_4y_2^2y_5 + \partial_{\sigma}x_4y_3^2y_5 + \partial_{\sigma}x_4y_5^3 - \partial_{\sigma}y_2x_1y_2^2 + \partial_{\sigma}y_2x_1y_3^2 + 
\end{equation*}
\begin{equation*}
\partial_{\sigma}y_2x_1y_5^2 - 2\partial_{\sigma}y_2x_2y_2y_3 - 2\partial_{\sigma}y_2x_4y_2y_5 - i\partial_{\sigma}y_2y_2^3 - i\partial_{\sigma}y_2y_2y_3^2 - i\partial_{\sigma}y_2y_2y_5^2 - 2\partial_{\sigma}y_3x_1y_2y_3 + 
\end{equation*}
\begin{equation*}
\partial_{\sigma}y_3x_2y_2^2 - \partial_{\sigma}y_3x_2y_3^2 + \partial_{\sigma}y_3x_2y_5^2 - 2\partial_{\sigma}y_3x_4y_3y_5 - i\partial_{\sigma}y_3y_2^2y_3 - i\partial_{\sigma}y_3y_3^3 - i\partial_{\sigma}y_3y_3y_5^2 - 
\end{equation*}
\begin{equation*}
2\partial_{\sigma}y_5x_1y_2y_5 - 2\partial_{\sigma}y_5x_2y_3y_5 + \partial_{\sigma}y_5x_4y_2^2 + \partial_{\sigma}y_5x_4y_3^2 - \partial_{\sigma}y_5x_4y_5^2 - i\partial_{\sigma}y_5y_2^2y_5 - i\partial_{\sigma}y_5y_3^2y_5 - 
\end{equation*}
\begin{equation*}
i\partial_{\sigma}y_5y_5^3\,,
\end{equation*}
\begin{equation*}
\widehat{\text{BPS}}_{2} = \partial_{\tau}x_1x_2y_2^2 + \partial_{\tau}x_1x_2y_3^2 + \partial_{\tau}x_1x_2y_5^2 + i\partial_{\tau}x_1y_2^2y_3 + i\partial_{\tau}x_1y_3^3 + i\partial_{\tau}x_1y_3y_5^2 - \partial_{\tau}x_2x_1y_2^2 - 
\end{equation*}
\begin{equation*}
\partial_{\tau}x_2x_1y_3^2 - \partial_{\tau}x_2x_1y_5^2 - i\partial_{\tau}x_2y_2^3 - i\partial_{\tau}x_2y_2y_3^2 - i\partial_{\tau}x_2y_2y_5^2 - \partial_{\tau}x_4y_2^2 - \partial_{\tau}x_4y_3^2 - 
\end{equation*}
\begin{equation*}
\partial_{\tau}x_4y_5^2 - 2i\partial_{\tau}y_2x_1y_2y_3 + i\partial_{\tau}y_2x_2y_2^2 - i\partial_{\tau}y_2x_2y_3^2 - i\partial_{\tau}y_2x_2y_5^2 + \partial_{\tau}y_2y_2^2y_3 + 2i\partial_{\tau}y_2y_2y_5 + 
\end{equation*}
\begin{equation*}
\partial_{\tau}y_2y_3^3 + \partial_{\tau}y_2y_3y_5^2 + i\partial_{\tau}y_3x_1y_2^2 - i\partial_{\tau}y_3x_1y_3^2 + i\partial_{\tau}y_3x_1y_5^2 + 2i\partial_{\tau}y_3x_2y_2y_3 - \partial_{\tau}y_3y_2^3 - 
\end{equation*}
\begin{equation*}
\partial_{\tau}y_3y_2y_3^2 - \partial_{\tau}y_3y_2y_5^2 + 2i\partial_{\tau}y_3y_3y_5 - 2i\partial_{\tau}y_5x_1y_3y_5 + 2i\partial_{\tau}y_5x_2y_2y_5 - i\partial_{\tau}y_5y_2^2 - i\partial_{\tau}y_5y_3^2 + 
\end{equation*}
\begin{equation*}
i\partial_{\tau}y_5y_5^2 + i\partial_{\sigma}x_1x_2y_2^2 + i\partial_{\sigma}x_1x_2y_3^2 + i\partial_{\sigma}x_1x_2y_5^2 - \partial_{\sigma}x_1y_2^2y_3 - \partial_{\sigma}x_1y_3^3 - \partial_{\sigma}x_1y_3y_5^2 - 
\end{equation*}
\begin{equation*}
i\partial_{\sigma}x_2x_1y_2^2 - i\partial_{\sigma}x_2x_1y_3^2 - i\partial_{\sigma}x_2x_1y_5^2 + \partial_{\sigma}x_2y_2^3 + \partial_{\sigma}x_2y_2y_3^2 + \partial_{\sigma}x_2y_2y_5^2 - i\partial_{\sigma}x_4y_2^2 - 
\end{equation*}
\begin{equation*}
i\partial_{\sigma}x_4y_3^2 - i\partial_{\sigma}x_4y_5^2 + 2\partial_{\sigma}y_2x_1y_2y_3 - \partial_{\sigma}y_2x_2y_2^2 + \partial_{\sigma}y_2x_2y_3^2 + \partial_{\sigma}y_2x_2y_5^2 + i\partial_{\sigma}y_2y_2^2y_3 - 
\end{equation*}
\begin{equation*}
2\partial_{\sigma}y_2y_2y_5 + i\partial_{\sigma}y_2y_3^3 + i\partial_{\sigma}y_2y_3y_5^2 - \partial_{\sigma}y_3x_1y_2^2 + \partial_{\sigma}y_3x_1y_3^2 - \partial_{\sigma}y_3x_1y_5^2 - 2\partial_{\sigma}y_3x_2y_2y_3 - 
\end{equation*}
\begin{equation*}
i\partial_{\sigma}y_3y_2^3 - i\partial_{\sigma}y_3y_2y_3^2 - i\partial_{\sigma}y_3y_2y_5^2 - 2\partial_{\sigma}y_3y_3y_5 + 2\partial_{\sigma}y_5x_1y_3y_5 - 2\partial_{\sigma}y_5x_2y_2y_5 + \partial_{\sigma}y_5y_2^2 + 
\end{equation*}
\begin{equation*}
\partial_{\sigma}y_5y_3^2 - \partial_{\sigma}y_5y_5^2\,,
\end{equation*}
\begin{equation*}
\widehat{\text{BPS}}_{3} = -\partial_{\tau}x_1x_4y_2^2 - \partial_{\tau}x_1x_4y_3^2 - \partial_{\tau}x_1x_4y_5^2 - i\partial_{\tau}x_1y_2^2y_5 - i\partial_{\tau}x_1y_3^2y_5 - 
\end{equation*}
\begin{equation*}
i\partial_{\tau}x_1y_5^3 - \partial_{\tau}x_2y_2^2 - \partial_{\tau}x_2y_3^2 - \partial_{\tau}x_2y_5^2 + \partial_{\tau}x_4x_1y_2^2 + \partial_{\tau}x_4x_1y_3^2 + \partial_{\tau}x_4x_1y_5^2 + 
\end{equation*}
\begin{equation*}
i\partial_{\tau}x_4y_2^3 + i\partial_{\tau}x_4y_2y_3^2 + i\partial_{\tau}x_4y_2y_5^2 + 2i\partial_{\tau}y_2x_1y_2y_5 - i\partial_{\tau}y_2x_4y_2^2 + i\partial_{\tau}y_2x_4y_3^2 + i\partial_{\tau}y_2x_4y_5^2 - 
\end{equation*}
\begin{equation*}
\partial_{\tau}y_2y_2^2y_5 + 2i\partial_{\tau}y_2y_2y_3 - \partial_{\tau}y_2y_3^2y_5 - \partial_{\tau}y_2y_5^3 + 2i\partial_{\tau}y_3x_1y_3y_5 - 2i\partial_{\tau}y_3x_4y_2y_3 - i\partial_{\tau}y_3y_2^2 + 
\end{equation*}
\begin{equation*}
i\partial_{\tau}y_3y_3^2 - i\partial_{\tau}y_3y_5^2 - i\partial_{\tau}y_5x_1y_2^2 - i\partial_{\tau}y_5x_1y_3^2 + i\partial_{\tau}y_5x_1y_5^2 - 2i\partial_{\tau}y_5x_4y_2y_5 + \partial_{\tau}y_5y_2^3 + 
\end{equation*}
\begin{equation*}
\partial_{\tau}y_5y_2y_3^2 + \partial_{\tau}y_5y_2y_5^2 + 2i\partial_{\tau}y_5y_3y_5 - i\partial_{\sigma}x_1x_4y_2^2 - i\partial_{\sigma}x_1x_4y_3^2 - i\partial_{\sigma}x_1x_4y_5^2 + \partial_{\sigma}x_1y_2^2y_5 + 
\end{equation*}
\begin{equation*}
\partial_{\sigma}x_1y_3^2y_5 + \partial_{\sigma}x_1y_5^3 - i\partial_{\sigma}x_2y_2^2 - i\partial_{\sigma}x_2y_3^2 - i\partial_{\sigma}x_2y_5^2 + i\partial_{\sigma}x_4x_1y_2^2 + i\partial_{\sigma}x_4x_1y_3^2 + 
\end{equation*}
\begin{equation*}
i\partial_{\sigma}x_4x_1y_5^2 - \partial_{\sigma}x_4y_2^3 - \partial_{\sigma}x_4y_2y_3^2 - \partial_{\sigma}x_4y_2y_5^2 - 2\partial_{\sigma}y_2x_1y_2y_5 + \partial_{\sigma}y_2x_4y_2^2 - \partial_{\sigma}y_2x_4y_3^2 - 
\end{equation*}
\begin{equation*}
\partial_{\sigma}y_2x_4y_5^2 - i\partial_{\sigma}y_2y_2^2y_5 - 2\partial_{\sigma}y_2y_2y_3 - i\partial_{\sigma}y_2y_3^2y_5 - i\partial_{\sigma}y_2y_5^3 - 2\partial_{\sigma}y_3x_1y_3y_5 + 2\partial_{\sigma}y_3x_4y_2y_3 + 
\end{equation*}
\begin{equation*}
\partial_{\sigma}y_3y_2^2 - \partial_{\sigma}y_3y_3^2 + \partial_{\sigma}y_3y_5^2 + \partial_{\sigma}y_5x_1y_2^2 + \partial_{\sigma}y_5x_1y_3^2 - \partial_{\sigma}y_5x_1y_5^2 + 2\partial_{\sigma}y_5x_4y_2y_5 + 
\end{equation*}
\begin{equation*}
i\partial_{\sigma}y_5y_2^3 + i\partial_{\sigma}y_5y_2y_3^2 + i\partial_{\sigma}y_5y_2y_5^2 - 2\partial_{\sigma}y_5y_3y_5\,,
\end{equation*}
\begin{equation*}
\widehat{\text{BPS}}_{4} = \partial_{\tau}x_1y_2^2 + \partial_{\tau}x_1y_3^2 + \partial_{\tau}x_1y_5^2 - \partial_{\tau}x_2x_4y_2^2 - 
\end{equation*}
\begin{equation*}
\partial_{\tau}x_2x_4y_3^2 - \partial_{\tau}x_2x_4y_5^2 - i\partial_{\tau}x_2y_2^2y_5 - i\partial_{\tau}x_2y_3^2y_5 - i\partial_{\tau}x_2y_5^3 + \partial_{\tau}x_4x_2y_2^2 + \partial_{\tau}x_4x_2y_3^2 + 
\end{equation*}
\begin{equation*}
\partial_{\tau}x_4x_2y_5^2 + i\partial_{\tau}x_4y_2^2y_3 + i\partial_{\tau}x_4y_3^3 + i\partial_{\tau}x_4y_3y_5^2 + 2i\partial_{\tau}y_2x_2y_2y_5 - 2i\partial_{\tau}y_2x_4y_2y_3 - i\partial_{\tau}y_2y_2^2 + 
\end{equation*}
\begin{equation*}
i\partial_{\tau}y_2y_3^2 + i\partial_{\tau}y_2y_5^2 + 2i\partial_{\tau}y_3x_2y_3y_5 + i\partial_{\tau}y_3x_4y_2^2 - i\partial_{\tau}y_3x_4y_3^2 + i\partial_{\tau}y_3x_4y_5^2 - \partial_{\tau}y_3y_2^2y_5 - 
\end{equation*}
\begin{equation*}
2i\partial_{\tau}y_3y_2y_3 - \partial_{\tau}y_3y_3^2y_5 - \partial_{\tau}y_3y_5^3 - i\partial_{\tau}y_5x_2y_2^2 - i\partial_{\tau}y_5x_2y_3^2 + i\partial_{\tau}y_5x_2y_5^2 - 2i\partial_{\tau}y_5x_4y_3y_5 + 
\end{equation*}
\begin{equation*}
\partial_{\tau}y_5y_2^2y_3 - 2i\partial_{\tau}y_5y_2y_5 + \partial_{\tau}y_5y_3^3 + \partial_{\tau}y_5y_3y_5^2 + i\partial_{\sigma}x_1y_2^2 + i\partial_{\sigma}x_1y_3^2 + i\partial_{\sigma}x_1y_5^2 - 
\end{equation*}
\begin{equation*}
i\partial_{\sigma}x_2x_4y_2^2 - i\partial_{\sigma}x_2x_4y_3^2 - i\partial_{\sigma}x_2x_4y_5^2 + \partial_{\sigma}x_2y_2^2y_5 + \partial_{\sigma}x_2y_3^2y_5 + \partial_{\sigma}x_2y_5^3 + i\partial_{\sigma}x_4x_2y_2^2 + 
\end{equation*}
\begin{equation*}
i\partial_{\sigma}x_4x_2y_3^2 + i\partial_{\sigma}x_4x_2y_5^2 - \partial_{\sigma}x_4y_2^2y_3 - \partial_{\sigma}x_4y_3^3 - \partial_{\sigma}x_4y_3y_5^2 - 2\partial_{\sigma}y_2x_2y_2y_5 + 2\partial_{\sigma}y_2x_4y_2y_3 + 
\end{equation*}
\begin{equation*}
\partial_{\sigma}y_2y_2^2 - \partial_{\sigma}y_2y_3^2 - \partial_{\sigma}y_2y_5^2 - 2\partial_{\sigma}y_3x_2y_3y_5 - \partial_{\sigma}y_3x_4y_2^2 + \partial_{\sigma}y_3x_4y_3^2 - \partial_{\sigma}y_3x_4y_5^2 - 
\end{equation*}
\begin{equation*}
i\partial_{\sigma}y_3y_2^2y_5 + 2\partial_{\sigma}y_3y_2y_3 - i\partial_{\sigma}y_3y_3^2y_5 - i\partial_{\sigma}y_3y_5^3 + \partial_{\sigma}y_5x_2y_2^2 + \partial_{\sigma}y_5x_2y_3^2 - \partial_{\sigma}y_5x_2y_5^2 + 
\end{equation*}
\begin{equation*}
2\partial_{\sigma}y_5x_4y_3y_5 + i\partial_{\sigma}y_5y_2^2y_3 + 2\partial_{\sigma}y_5y_2y_5 + i\partial_{\sigma}y_5y_3^3 + i\partial_{\sigma}y_5y_3y_5^2\,.
\end{equation*}
To check the above equations, we fix the boundary conditions to be circular and half-BPS, where the corresponding supersymmetric configuration is known and given by
\begin{equation*}
x_1(\tau,\sigma) = F(\tau)\cos(\sigma)\,,\qquad x_2(\tau,\sigma) = F(\tau)\sin(\sigma)\,, \qquad y_5(\tau,\sigma) = y(\tau)  
\end{equation*}
where
\begin{equation}
\label{AdS_2 minimal}
y(\tau) = \tanh(\tau)\,,\qquad F(\tau) = \frac{1}{\cosh(\tau)}\,.
\end{equation}
These functions, $y(\tau)$ and $F(\tau)$, satisfy the following relations that completely characterize them:
\begin{equation}
\dot y = 1-y^2 = F^2\,.
\end{equation}
This configuration is the familiar Ad$\mathbb{S}_2$ minimal surface \cite{Drukker:1999zq}. By substituting the minimal Ad$\mathbb{S}_2$ configuration given in Eq. \ref{AdS_2 minimal}, we verified, using Python and its open-source libraries such as NumPy and SymPy, that the equations \ref{BPS} are satisfied. 

\begin{figure}[h]
  \centering
  \includegraphics[width=1.0\textwidth]{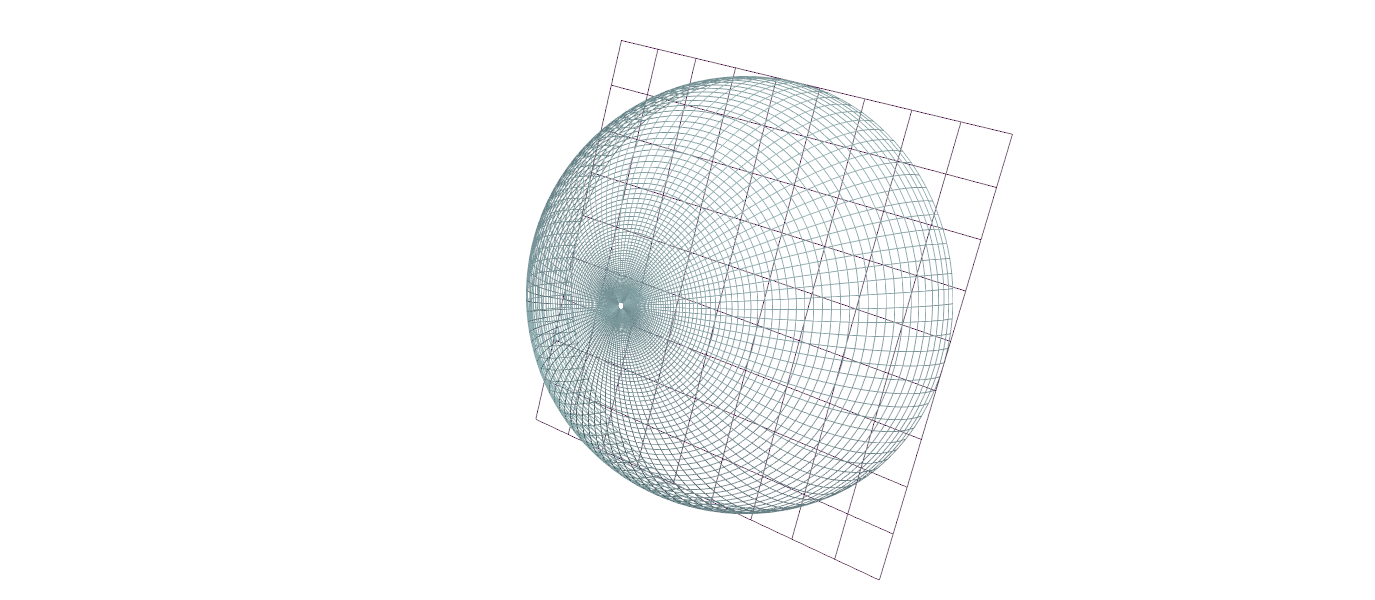}
  \caption{The Ad$\mathbb{S}_2$ minimal surface. The purple grid corresponds to the boundary of a Ad$\mathbb{S}_3$ slice of Ad$\mathbb{S}_5$ that contains the Ad$\mathbb{S}_2$ while the blue grid corredsponds to the Ad$\mathbb{S}_2$ minimal surface. The puncture arises from the fact that numerically one cannot reach $\tau =\infty$, but it is observed that the blue grid naturally converges to a point in the bulk, thus obeying the boundary conditions at $\tau\rightarrow\infty$.}
  \label{fig:minimal surface}
\end{figure}

\pagebreak

\section{Off-shell supersymmetric string configurations}

\label{off-shell}

\subsection{Constraints and gauge invariances for $\Omega_{a}$ and $\widehat\Omega_{a}$}
To obtain the off-shell supersymmetric configurations, we now consider the $\Omega_{a}$ and $\widehat\Omega_{a}$ in \ref{new susy eq}. They appear in the supersymmetric equations through the following combinations:
\begin{equation}
\label{important matrices eq}
\Omega_{a}(\gamma^{a}\lambda)_{\alpha}\,,\qquad \Omega_{a}(\gamma^{a}\widehat\lambda)_{\alpha} \,,\qquad \widehat\Omega_{a}(\gamma^{a}\lambda)_{\alpha} \,,\qquad \widehat\Omega_{a}(\gamma^{a}\widehat\lambda)_{\alpha}\,.    
\end{equation}
where $\Omega_{a}$ and $\widehat\Omega_{a}$ are constrained to be in the kernel of $(\gamma_{a}\widehat\lambda)_{\widehat\alpha}$ and $(\gamma_{a}\lambda)_{\alpha}$, respectively. Their images with respect to $(\gamma_{a}\lambda)_{\alpha}$ and $(\gamma_{a}\widehat\lambda)_{\widehat\alpha}$, respectively, deform the on-shell supersymmetric equations. For our choice of unbroken supercharge, in Ad$\mathbb{S}_4\times \mathbb{S}^2$, the matrices $(\gamma^{a}\lambda)_{\alpha}$ and $(\gamma^{a}\widehat\lambda)_{\widehat\alpha}$ have a kernel dimension of 5 in $a=1,\dots,10$. Also, in Ad$\mathbb{S}_4\times \mathbb{S}^2$, we observed that $j^{a}(\gamma_{a}\lambda)_{\alpha}$ have just 4 components. Thus, the 5 components in $\Omega_{a}(\gamma^{a}\lambda)_{\alpha}$ that are linearly independent from these four components serve as constraints for $\Omega_{a}$. Similarly, there will be constraints from the misalignment between the 4 components of $\overline j^{a}(\gamma_{a}\widehat\lambda)_{\widehat\alpha}$ and the 5 components in the image of $\Omega_{a}(\gamma^{a}\widehat\lambda)_{\widehat\alpha}$. Together with the constraints that $\Omega_{a}$ and $\widehat\Omega_{a}$ are in the kernel of $(\gamma_{a}\widehat\lambda)_{\widehat\alpha}$ and $(\gamma_{a}\lambda)_{\alpha}$, respectively, we have that
\begin{equation}
\label{obvious constraints for Omega}
\Omega_{3} = 0 \,\qquad \Omega_8 = 0 \,,\qquad \Omega_{5} = i\Omega_{10}\,,    
\end{equation}
\begin{equation*}
 \widehat\Omega_{3} = 0 \,\qquad \widehat\Omega_8 = 0 \,,\qquad \widehat\Omega_{5} = i\widehat\Omega_{10}\,,    
\end{equation*}
together with constraints imposing the two vectors:
\begin{equation}
\label{linear constraints}
  \left[\begin{matrix}\Omega_{1} + i \Omega_{6}\\\Omega_{1} - i \Omega_{6}\\\Omega_{2} + i \Omega_{7}\\\Omega_{2} - i \Omega_{7}\\\Omega_{4} + i \Omega_{9}\\\Omega_{4} - i \Omega_{9}\end{matrix}\right] \qquad \text{and}\qquad    \left[\begin{matrix}\widehat\Omega_{1} + i \widehat\Omega_{6}\\\widehat\Omega_{1} - i \widehat\Omega_{6}\\\widehat\Omega_{2} + i \widehat\Omega_{7}\\\widehat\Omega_{2} - i \widehat\Omega_{7}\\\widehat\Omega_{4} + i \widehat\Omega_{9}\\\widehat\Omega_{4} - i \widehat\Omega_{9}\end{matrix}\right]\,,
\end{equation}
to be in the kernel of the matrices:
\begin{equation*}
\left[\begin{matrix}\left(x_{1} + i y_{2}\right) & 0 & \left(x_{2} + i y_{3}\right) & 0 & \left(x_{4} + i y_{5}\right) & 0\\0 & - \left(x_{2} + i y_{3}\right) & 0 & \left(x_{1} + i y_{2}\right) & 1 & 0\\0 & \left(x_{4} + i y_{5}\right) & 1 & 0 & 0 & - \left(x_{1} + i y_{2}\right)\\-1 & 0 & 0 & \left(x_{4} + i y_{5}\right) & 0 & - \left(x_{2} + i y_{3}\right)\end{matrix}\right]\,,
\end{equation*}
and
\begin{equation*}
\left[\begin{matrix}0 & 1 & - \left(x_{4} + i y_{5}\right) & 0 & \left(x_{2} + i y_{3}\right) & 0\\\left(x_{4} + i y_{5}\right) & 0 & 0 & 1 & - \left(x_{1} + i y_{2}\right) & 0\\- \left(x_{2} + i y_{3}\right) & 0 & \left(x_{1} + i y_{2}\right) & 0 & 0 & 1\\0 & - \left(x_{1} + i y_{2}\right) & 0 & - \left(x_{2} + i y_{3}\right) & 0 & - \left(x_{4} + i y_{5}\right)\end{matrix}\right]  \,,
\end{equation*}
respectively. Before solving the constraints, it is illuminating to perform some simplifications in the sigma model action to check for gauge invariances of $\Omega_{a}$ and $\widehat\Omega_{a}$, specifically, components of these variables that decouple from both the supersymmetric equations and the sigma model action.Evaluating the sigma model action, as defined in \ref{sigma model action}, at the off-shell supersymmetric configurations and ignoring the fermions, yields:
\begin{equation}
\label{sigma model action restricted to bosons}
S=r^2\int \frac{d^2\sigma}{2 \pi\alpha'}\left[\frac{1}{2} j^{a}\widehat j^{b}\delta_{ab}+\overline\nabla\lambda^{\alpha}w_{\alpha}+\nabla\widehat\lambda^{\widehat\alpha}\widehat w_{\widehat\alpha}-\frac{1}{4}R_{abcd}N^{ab}\widehat N^{cd}\right]\,.   
\end{equation}
Due to the singular properties of pure Killing spinors, as discussed in section \ref{Derivation of the supersymmetric equations}, the term involving the curvature can be simplified:
\begin{equation}
-\frac{1}{4}R_{abcd}\widehat N^{dc}N^{ba}=\frac{1}{2}(\widehat w\eta)^{\alpha}\left[-\frac{1}{8}R_{abcd}(w\gamma^{dc}\lambda)(\gamma^{ba}\eta\widehat\lambda)_{\alpha}\right]=
\end{equation}
\begin{equation*}
=\frac{1}{2}(\widehat w \eta)^{\alpha}\Omega_{a}(\gamma^{a}\lambda)_{\alpha}=\Omega_{a}\widehat\Omega^{a} \,.  
\end{equation*}
The term involving the covariant derivatives acting on the pure Killing spinors can also be simplified by using equations \ref{pure killing} and \ref{new susy eq}, resulting in
\begin{equation}
\overline\nabla\lambda^{\alpha}w_{\alpha} = -\frac{1}{2}\widehat j^{a}(w\eta\gamma_{a}\widehat\lambda) = - \frac{1}{2} \widehat\Omega^{a}(w\eta\gamma_{a}\widehat\lambda) = -\widehat\Omega_{a}\Omega^{a}
\end{equation}
\begin{equation*}
\nabla\widehat \lambda^{\alpha}\widehat w_{\alpha} = -\frac{1}{2} j^{a}(\widehat w\eta\gamma_{a}\lambda) =-  \frac{1}{2} \widehat\Omega^{a}(\widehat w\eta\gamma_{a}\lambda) = -\widehat\Omega_{a}\Omega^{a}\,.
\end{equation*}
All this reduces the sigma model action to the following simple form:
\begin{equation}
S=r^2\int \frac{d^2\sigma}{2 \pi\alpha'}\left[\frac{1}{2}j^{a}\overline j^{b}\delta_{ab} -\widehat\Omega^{a}\Omega^{b}\delta_{ab}\right]\,.    
\end{equation}
It follows that, due to the constraints in \ref{obvious constraints for Omega}, the components $\widehat\Omega_{5}+i\widehat\Omega_{10}$ and $\widehat\Omega_{5}+i\widehat\Omega_{10}$ do not appear in the sigma model action. Additionally, these components also decouple from the supersymmetric equations; they are in the kernel of all matrices appearing in \ref{important matrices eq}, when restricted to the $Ad\mathbb{S}_4\times \mathbb{S}^2$ slice. Therefore, we choose to gauge-fix: 
\begin{equation}
\label{gauge choices of omega}
\Omega_5 = 0\,,\qquad \Omega_{10} = 0\,,\qquad \widehat\Omega_5 = 0\,,\qquad \widehat\Omega_{10} = 0\,.   
\end{equation}

\subsection{Solving the off-shell supersymmetric equations}

To solve for $\Omega_{a}$ and $\widehat\Omega_{a}$, it is essential to consider how the embedding coordinates are encoded in the pure Killing spinors $\lambda^{\alpha}$ and $\widehat\lambda^{\widehat\alpha}$, as explicitly discussed in section \ref{Killing spinors and their pure region}, particularly in equation \ref{complex structure}. The Killing spinors organize the spacetime coordinates into complex numbers. The associated complex structure labels five holographic directions as imaginary and one as real, while the boundary directions remain real (given our choice of supercharge). It was also noted that $\widehat\varepsilon$ includes the complex conjugates of $\varepsilon$, except for the coordinates $y_1$ and $y_6$. Given that the embedding surface is localized within the Ad$\mathbb{S}_4\times \mathbb{S}^2$ slice, where $y_1 = y_6 = 0$, we adopt $\lambda^{\alpha}$ and $\widehat\lambda^{\alpha}$ as complex conjugate spinors in this slice, thereby introducing a new conjugation matrix:
\begin{equation}
\label{new conjugation matrix}
\overline\lambda_{\alpha} = (\gamma_{3}\gamma_{5}\gamma_{8})_{\alpha\beta}\widehat\lambda^{\beta}  \,.
\end{equation}
For instance, the following conditions are maintained in the Ad$\mathbb{S}_4\times \mathbb{S}^2$ slice: 
\begin{equation}
\overline\lambda \lambda = \frac{x_1^2 + x_2^2 + x_4^2 + y_2^2 + y_3^2 + y_5^2 + 1}{y}>0\,.
\end{equation}
For the gauge-choice \ref{gauge choices of omega}, we can solve the supersymmetric equations \ref{new susy eq} for $\Omega_a$ and $\widehat\Omega_a$ by contracting it with $\frac{(\overline\lambda\gamma^{a})^{\alpha}}{(\lambda\overline\lambda)}$, and using the constraints presented in \ref{obvious constraints for Omega}:
\begin{equation}
\label{eliminate omega}
\Omega^{a} = \frac{(\overline\lambda\gamma^{a}\gamma^{b}\lambda)}{2\lambda\overline\lambda}j_{b}\,,\qquad   \widehat\Omega^{a} = \frac{(\lambda\gamma^{a}\gamma^{b}\overline\lambda)}{2(\overline\lambda\lambda)}\overline j_{b}\,.
\end{equation}
Here we used the fact that, due to the constraints and gauge choices, the matrices $\Omega^{a}\gamma_{a}$ and $\widehat\Omega_{a}\gamma_{a}$ anticommute with the conjugation matrix $\gamma_{3}\gamma_{5}\gamma_{8}$ proposed in \ref{new conjugation matrix}. Furthermore, requirement for the embedded surface to be within the Ad$\mathbb{S}_4\times\mathbb{S}^4$ implies that the matrices $j^{a}\gamma_{a}$ and $\widehat j^{a}\gamma_{a}$ anticommute with $\gamma_{3}\gamma_{5}\gamma_{8}$. After solving the supersymmetric equations for $\Omega_{a}$ and $\widehat\Omega_{a}$, we learn that there are no further constraints on the embedded surface. 

\subsection{Evaluating the action at off-shell supersymmetric configurations}
Having an infinite-dimensional space of supersymmetric configurations is disappointing, especially if one is aiming for exact results via supersymmetric localization.  However, we believe it is essential to listen to what the equations have to say, rather than overshadow their insights with our biases. We will now evaluate the worldsheet action at off-shell supersymmetric configurations respective to the supercharge chosen in subsection \ref{Killing spinors and their pure region}. The following manipulations are useful:
\begin{equation}
\Omega^{a}\delta_{ab}\widehat\Omega^{b}=\Omega^{a}\left[\frac{\lambda\gamma_{a}\gamma_{b}\overline\lambda}{2(\lambda\overline\lambda)}+\frac{\overline\lambda\gamma_{a}\gamma_{b}\lambda}{2(\lambda\overline\lambda)}\right]\widehat\Omega^{b} = \Omega_{a}\left[\frac{\lambda\gamma_{a}\gamma_{b}\overline\lambda}{2(\lambda\overline\lambda)}+0\right]\widehat\Omega_{b} = j^{a}\frac{\lambda\gamma_{a}\gamma_{b}\overline\lambda}{2(\lambda\overline\lambda)}\overline j^{b}\,,
\end{equation}
and
\begin{equation}
j^{a}\frac{\lambda\gamma_{a}\gamma_{b}\overline\lambda}{2(\lambda\overline\lambda)}\overline j^{b} = \frac{1}{2}j^{a}\overline j^{b}\delta_{ab}  + j^{a}\overline j^{b} \frac{(\lambda\gamma_{ab}\overline\lambda)}{2(\lambda\overline\lambda)}\,.
\end{equation}
We find that the sigma model action, when evaluated at off-shell supersymmetric configurations, also becomes an integral over a two-form, just as for the on-shell ones:
\begin{equation}
S = \int \frac{d^2\sigma}{2\pi\alpha'}\overline j^{a} j^{b} B_{ab}\,,\qquad B_{ab} =   \frac{(\overline\lambda\gamma_{ab}\lambda)}{2(\lambda\overline\lambda)}\,.
\end{equation}
This generalizes the result of \cite{Drukker:2007qr} to off-shell supersymmetric configuration. When restricted to Ad$\mathbb{S}_4\times\mathbb{S}^2$, that is, when $a=1,2,4,6,7,9$, the matrix $\frac{(\overline\lambda \gamma^{ab}\lambda)}{(\overline\lambda\lambda)}$ is:
\begin{equation*}
\frac{(\overline\lambda \gamma_{ab}\lambda)}{(\overline\lambda\lambda)}    = \left[\begin{matrix}0 & \frac{2 i \left(- x_{1} y_{3} + x_{2} y_{2} + y_{5}\right)}{x_{1}^{2} + x_{2}^{2} + x_{4}^{2} + y_{2}^{2} + y_{3}^{2} + y_{5}^{2} + 1} & \frac{2 i \left(- x_{1} y_{5} + x_{4} y_{2} - y_{3}\right)}{x_{1}^{2} + x_{2}^{2} + x_{4}^{2} + y_{2}^{2} + y_{3}^{2} + y_{5}^{2} + 1}\\\frac{2 i \left(x_{1} y_{3} - x_{2} y_{2} - y_{5}\right)}{x_{1}^{2} + x_{2}^{2} + x_{4}^{2} + y_{2}^{2} + y_{3}^{2} + y_{5}^{2} + 1} & 0 & \frac{2 i \left(- x_{2} y_{5} + x_{4} y_{3} + y_{2}\right)}{x_{1}^{2} + x_{2}^{2} + x_{4}^{2} + y_{2}^{2} + y_{3}^{2} + y_{5}^{2} + 1}\\\frac{2 i \left(x_{1} y_{5} - x_{4} y_{2} + y_{3}\right)}{x_{1}^{2} + x_{2}^{2} + x_{4}^{2} + y_{2}^{2} + y_{3}^{2} + y_{5}^{2} + 1} & \frac{2 i \left(x_{2} y_{5} - x_{4} y_{3} - y_{2}\right)}{x_{1}^{2} + x_{2}^{2} + x_{4}^{2} + y_{2}^{2} + y_{3}^{2} + y_{5}^{2} + 1} & 0\\\frac{i \left(x_{1}^{2} - x_{2}^{2} - x_{4}^{2} + y_{2}^{2} - y_{3}^{2} - y_{5}^{2} + 1\right)}{x_{1}^{2} + x_{2}^{2} + x_{4}^{2} + y_{2}^{2} + y_{3}^{2} + y_{5}^{2} + 1} & \frac{2 i \left(x_{1} x_{2} - x_{4} + y_{2} y_{3}\right)}{x_{1}^{2} + x_{2}^{2} + x_{4}^{2} + y_{2}^{2} + y_{3}^{2} + y_{5}^{2} + 1} & \frac{2 i \left(x_{1} x_{4} + x_{2} + y_{2} y_{5}\right)}{x_{1}^{2} + x_{2}^{2} + x_{4}^{2} + y_{2}^{2} + y_{3}^{2} + y_{5}^{2} + 1}\\\frac{2 i \left(x_{1} x_{2} + x_{4} + y_{2} y_{3}\right)}{x_{1}^{2} + x_{2}^{2} + x_{4}^{2} + y_{2}^{2} + y_{3}^{2} + y_{5}^{2} + 1} & \frac{i \left(- x_{1}^{2} + x_{2}^{2} - x_{4}^{2} - y_{2}^{2} + y_{3}^{2} - y_{5}^{2} + 1\right)}{x_{1}^{2} + x_{2}^{2} + x_{4}^{2} + y_{2}^{2} + y_{3}^{2} + y_{5}^{2} + 1} & \frac{2 i \left(- x_{1} + x_{2} x_{4} + y_{3} y_{5}\right)}{x_{1}^{2} + x_{2}^{2} + x_{4}^{2} + y_{2}^{2} + y_{3}^{2} + y_{5}^{2} + 1}\\\frac{2 i \left(x_{1} x_{4} - x_{2} + y_{2} y_{5}\right)}{x_{1}^{2} + x_{2}^{2} + x_{4}^{2} + y_{2}^{2} + y_{3}^{2} + y_{5}^{2} + 1} & \frac{2 i \left(x_{1} + x_{2} x_{4} + y_{3} y_{5}\right)}{x_{1}^{2} + x_{2}^{2} + x_{4}^{2} + y_{2}^{2} + y_{3}^{2} + y_{5}^{2} + 1} & \frac{i \left(- x_{1}^{2} - x_{2}^{2} + x_{4}^{2} - y_{2}^{2} - y_{3}^{2} + y_{5}^{2} + 1\right)}{x_{1}^{2} + x_{2}^{2} + x_{4}^{2} + y_{2}^{2} + y_{3}^{2} + y_{5}^{2} + 1}\end{matrix}\right.    
\end{equation*}
\begin{equation}
\left.\begin{matrix}\frac{i \left(- x_{1}^{2} + x_{2}^{2} + x_{4}^{2} - y_{2}^{2} + y_{3}^{2} + y_{5}^{2} - 1\right)}{x_{1}^{2} + x_{2}^{2} + x_{4}^{2} + y_{2}^{2} + y_{3}^{2} + y_{5}^{2} + 1} & \frac{2 i \left(- x_{1} x_{2} - x_{4} - y_{2} y_{3}\right)}{x_{1}^{2} + x_{2}^{2} + x_{4}^{2} + y_{2}^{2} + y_{3}^{2} + y_{5}^{2} + 1} & \frac{2 i \left(- x_{1} x_{4} + x_{2} - y_{2} y_{5}\right)}{x_{1}^{2} + x_{2}^{2} + x_{4}^{2} + y_{2}^{2} + y_{3}^{2} + y_{5}^{2} + 1}\\\frac{2 i \left(- x_{1} x_{2} + x_{4} - y_{2} y_{3}\right)}{x_{1}^{2} + x_{2}^{2} + x_{4}^{2} + y_{2}^{2} + y_{3}^{2} + y_{5}^{2} + 1} & \frac{i \left(x_{1}^{2} - x_{2}^{2} + x_{4}^{2} + y_{2}^{2} - y_{3}^{2} + y_{5}^{2} - 1\right)}{x_{1}^{2} + x_{2}^{2} + x_{4}^{2} + y_{2}^{2} + y_{3}^{2} + y_{5}^{2} + 1} & \frac{2 i \left(- x_{1} - x_{2} x_{4} - y_{3} y_{5}\right)}{x_{1}^{2} + x_{2}^{2} + x_{4}^{2} + y_{2}^{2} + y_{3}^{2} + y_{5}^{2} + 1}\\\frac{2 i \left(- x_{1} x_{4} - x_{2} - y_{2} y_{5}\right)}{x_{1}^{2} + x_{2}^{2} + x_{4}^{2} + y_{2}^{2} + y_{3}^{2} + y_{5}^{2} + 1} & \frac{2 i \left(x_{1} - x_{2} x_{4} - y_{3} y_{5}\right)}{x_{1}^{2} + x_{2}^{2} + x_{4}^{2} + y_{2}^{2} + y_{3}^{2} + y_{5}^{2} + 1} & \frac{i \left(x_{1}^{2} + x_{2}^{2} - x_{4}^{2} + y_{2}^{2} + y_{3}^{2} - y_{5}^{2} - 1\right)}{x_{1}^{2} + x_{2}^{2} + x_{4}^{2} + y_{2}^{2} + y_{3}^{2} + y_{5}^{2} + 1}\\0 & \frac{2 i \left(- x_{1} y_{3} + x_{2} y_{2} - y_{5}\right)}{x_{1}^{2} + x_{2}^{2} + x_{4}^{2} + y_{2}^{2} + y_{3}^{2} + y_{5}^{2} + 1} & \frac{2 i \left(- x_{1} y_{5} + x_{4} y_{2} + y_{3}\right)}{x_{1}^{2} + x_{2}^{2} + x_{4}^{2} + y_{2}^{2} + y_{3}^{2} + y_{5}^{2} + 1}\\\frac{2 i \left(x_{1} y_{3} - x_{2} y_{2} + y_{5}\right)}{x_{1}^{2} + x_{2}^{2} + x_{4}^{2} + y_{2}^{2} + y_{3}^{2} + y_{5}^{2} + 1} & 0 & \frac{2 i \left(- x_{2} y_{5} + x_{4} y_{3} - y_{2}\right)}{x_{1}^{2} + x_{2}^{2} + x_{4}^{2} + y_{2}^{2} + y_{3}^{2} + y_{5}^{2} + 1}\\\frac{2 i \left(x_{1} y_{5} - x_{4} y_{2} - y_{3}\right)}{x_{1}^{2} + x_{2}^{2} + x_{4}^{2} + y_{2}^{2} + y_{3}^{2} + y_{5}^{2} + 1} & \frac{2 i \left(x_{2} y_{5} - x_{4} y_{3} + y_{2}\right)}{x_{1}^{2} + x_{2}^{2} + x_{4}^{2} + y_{2}^{2} + y_{3}^{2} + y_{5}^{2} + 1} & 0\end{matrix}\right]\,, 
\end{equation}
and consequently, the components of the two-form $B_{ab}$, when restricted to the Ad$\mathbb{S}_4\times\mathbb{S}^2$, are given by:
\begin{equation*}
B_{\mu\nu} = i\frac{\varepsilon_{\mu\nu\rho}y_{(\rho+1)} - x_{\mu}y_{(\nu+1)} + x_{\nu}y_{(\mu+1)}}{|x|^2 + |y|^2 + 1}\,, \qquad  B_{(4+i)(4+j)} =i \frac{-\varepsilon_{ijk}y_{k} - x_{(i-1)}y_{j}+x_{(j-1)}y_{i}}{|x|^2 + |y|^2 + 1}\,,
\end{equation*}
\begin{equation*}
B_{\mu (4+i)} = \frac{i}{2}\delta_{(\mu+1)i}\left[\frac{|x|^2 + |y|^2 - 1 - (x_{\mu}^2 + y_{i}^2)}{|x|^2 + |y|^2 + 1}\right] +i\left[\frac{-\varepsilon_{\mu\nu\rho}x_{\rho} - x_{\mu}x_{(i-1)} - y_{i}y_{(\mu+1)}}{|x|^2 + |y|^2 + 1}\right]\,,  
\end{equation*}
where
\begin{equation}
\mu,\nu,\rho = 1,2,4\qquad i,j,k = 2,3,5 \,.    
\end{equation}
In curved indices, as estabilished in appendix \ref{coordinates}, the sigma model action evaluated at supersymmetric configurations is given by
\begin{equation}
S = \int \frac{d\sigma d\tau}{2\pi\alpha'} i\partial_{\tau}x^{m}\partial _{\sigma}x^{n}\widetilde B_{mn}\,,    
\end{equation}
where the two-form components are
\begin{equation*}
\widetilde B_{\mu\nu}  = \frac{B_{\mu\nu}}{|y|^2}\,,\qquad \widetilde B_{\mu(i+4)} = \left(\delta_{ij} - 2 \frac{y_{i}y_{j}}{y^2}\right)\frac{B_{\mu(j+4)}}{|y|^2}\,, 
\end{equation*}
\begin{equation}
 \widetilde B_{(i+4)(j+4)} = \left(\delta_{ik} - 2 \frac{y_{i}y_{k}}{y^2}\right)\left(\delta_{jl} - 2 \frac{y_{j}y_{l}}{y^2}\right)\frac{B_{(k+4)(l+4)}}{|y|^2} \,. 
\end{equation}
Given two surfaces within the same homotopy class, the difference in their actions can be expressed as:
\begin{equation}
S_1-S_0 = \int_0^{1}\frac{d\zeta d\sigma d\tau}{2\pi\alpha'}\,i\partial_{\zeta}x^{m} \partial_{\tau} x^{n}\partial_{\sigma} x^{p}\widetilde H_{mnp}\,,\qquad \widetilde H_{mnp} = \partial_{[m}\widetilde B_{np]}\,,
\end{equation}
where $\widetilde H_{mnp}$ is the field strength of the two-form $\widetilde B_{mn}$. The components of $\widetilde H_{mnp}$ are generally non-zero, indicating that the two-form $\widetilde B_{mn}$ is not closed in Ad$\mathbb{S}_4\times\mathbb{S}^2$. If $\widetilde B_{mn}$ were closed, the sigma model action would depend solely on boundary conditions, and consequently, the sigma model action evaluated on off-shell supersymmetric configurations would necessarily equate to the minimal area. Indeed, the following components are non-zero: 
\begin{equation}
\widetilde H_{\mu\nu\rho} = 4i\frac{(x_1y_2+x_2y_3+x_4y_5)}{(1+|x|^2+|y|^2)^2} \varepsilon_{\mu\nu\rho}\,,\quad \widetilde H_{(i+4)(j+4)(k+4)} = 2i\frac{(1+ |x|^2 - |y|^2)}{|y|^2(1+|x|^2+|y|^2)}\varepsilon_{ijk} \,,  
\end{equation}
where $\mu,\nu,\rho =1,2,4$ and $i,j,k=2,3,5$. The other non-zero components, $\widetilde H_{\mu\nu(i+4)}$ and $\widetilde H_{(i+4)(j+4)(k+4)}$, are more complex and not as illuminating.

\section{Conclusion}

\label{conclusion}

We saw that the pure spinor formalism allows for an off-shell generalization of the Green-Schwarz equations describing supersymmetric string configurations in Ad$\mathbb{S}_5\times \mathbb{S}^{5}$. This generalization is characterized by the inclusion of $\Omega_{a}$ and $\widehat\Omega_{a}$ variables, which deform the supersymmetric equations sufficiently to permit off-shell configurations. Additionally, it mandates that the corresponding Killing spinors be equated with the worldsheet variables obeying the pure spinor constraints, thereby constraining the embedded surfaces to regions where these conditions can be met. This scenario presents a puzzle, as there are choices of supercharges whose corresponding Killing spinors are impure everywhere in Ad$\mathbb{S}_5\times \mathbb{S}^5$. This puzzle persists in flat background.
\vskip 0.3cm
For our chosen supercharge, the supersymmetric string configurations are confined within an Ad$\mathbb{S}_4\times \mathbb{S}^2$ slice of Ad$\mathbb{S}_5\times \mathbb{S}^{5}$. Within this slice, the string configuration can vary freely, provided that the $\Omega_{a}$ and $\widehat\Omega_{a}$ variables are appropriately adjusted, leading to an infinite-dimensional space of off-shell supersymmetric configurations. Unfortunately, this is not optimal for establishing exact results via supersymmetric localization.
\vskip 0.3cm
Throughout this paper, we have maintained the signature of Ad$\mathbb{S}_5 \times \mathbb{S}^5$ spacetime as Euclidean. The natural next step might involve considering the complexification of spacetime to identify a finite-dimensional locus for supersymmetric configurations, which could enable exact results for the holographic computation. However, even after complexification, the $\Omega_{a}$ and $\widehat{\Omega}_{a}$ do not seem to decouple, and they are responsible for the infinite dimensionality of the loci.
\vskip 0.3cm
It should be noted that supersymmetric transformations of the worldsheet variables receive $\alpha'$ corrections, which are not accounted for in our paper \footnote{I thank Thiago Fleury for pointing this out.}. For example, the dilatation generator receives corrections, and the supercharges form an algebra that includes this generator. However, for certain choices of supercharges, particularly the nilpotent ones, these $\alpha'$ corrections are expected to manifest in terms that are higher order in fermions, thus not contributing to the determination of the supersymmetric locus. This issue might be related to the requirement for Killing spinors to be pure spinors.

\acknowledgments
\vskip 0.4cm
I acknowledge Nathan Berkovits and Jaume Gomis for giving me this project. I would like to thank Nathan Berkovits, Cassiano A. Daniel, and Thiago Fleury for their useful comments on the draft, the IIP - UFRN for its warm hospitality, and the Serrapilheira Institute Grant (Serra–R-2012-38185), FAPESP Grants 2021/09003-9 and 2018/07834-8, and the Simons Foundation Grant (1023171 - RC) for partial financial support.

\begin{appendices}

\section{Ad$\mathbb{S}_5\times \mathbb{S}^{5}$ coordinates and spin connection}
\label{coordinates}

We choose coordinates in which the Ad$\mathbb{S}_5\times \mathbb{S}^{5}$ metric is given by
\begin{equation}
ds^2 = \frac{dx^2+dy^2}{y^2}  \,,  
\end{equation}
where $x^{\mu}$, with $\mu=1,2,3,4$, correspond to the $\mathbb{R}^4$ coordinates parametrizing the $AdS_5$ boundary, and $y^{i}$, with $i=1,2,3,4,5,6$, correspond to the six holographic directions. The size of $SO(6)_{R}$ vector $y^{i}$ will be denoted by $y$ and is the fifth dimension of $AdS_5$. The boundary is located at $y=0$. The directions of $y^{i}$ parametrize the $S^5 = \frac{SO(6)}{SO(5)}$ compact space. The spacetime vielbeins forms $e^{a}$ and vectors $D_{a}$ are given by
\begin{equation}
e^{\mu}=\frac{dx^{\mu}}{y}\,,\qquad e^{4+i}=d\left(\frac{y^{i}}{y}\right)-\frac{y^{i}}{y}(y^{-1}d y) \,,
\end{equation}
\begin{equation}
D_{\mu}=y\frac{\partial}{\partial x^{\mu}}\,,\qquad D_{4+i}=y\left(\delta_{i}\,^{j}-2\frac{y_{i}y^{j}}{y^{2}}\right)\frac{\partial}{\partial y^{j}}\,.    
\end{equation}
The covariant derivatives are constructed by combining the vectors $D_{a}$ with an $SO(10)$ spin connecion $\Omega_{a}$, as follows:
\begin{equation}
\nabla_{a}=\left(D_{a}-\frac{1}{2}\Omega_{a}\,^{bc}t_{cb}\right)\,,    
\end{equation}
where $t_{cb}$ are the $SO(10)$ generators. For an arbitary choice of connection $\Omega_{a}$, one has
\begin{equation}
[\nabla_{a},\nabla_{b}] = T_{ab}\,^{c}\nabla_{c}+ \frac{1}{2} R_{ab}\,^{cd}t_{dc} \,,   
\end{equation}
where $T_{ab}\,^{c}$ are the torsion components and $R_{ab}\,^{cd}$ are the curvature components. We fix the connection to be torsion-less, i.e. we fix
\begin{equation}
T_{abc} = f_{abc} + \Omega_{[ab]c} =0 \iff \Omega_{abc}=\frac{1}{2}\left(f_{abc}-f_{bca}+f_{cab}\right)\,,
\end{equation}
where $f_{ab}\,^{c}$ are given by $[D_{a},D_{b}]=f_{ab}\,^{c}D_{c}$. So, we compute
\begin{equation}
[D_{\mu},D_{\nu}]=0\,,\qquad [D_{i},D_{\mu}]=-\frac{y_{i}}{y}D_{\mu}\,,\qquad [D_{i},D_{j}]=\frac{y_{[i}D_{j]}}{y}
\end{equation}
obtaining:
\begin{equation}
\Omega_{\mu\nu i}=\frac{y_{i}}{y}\eta_{\mu\nu}\,,\qquad\Omega_{ijk}=-\frac{\eta_{i[j}y_{k]}}{y}\,.
\end{equation}

\section{Ten-dimensional chiral gamma matrices}
\label{Conventions on spinors}

In our conventions, explicitly shown in appendix \ref{Gamma matrices}, the $32\times32$ Hermitian gamma matrices are structured as off-diagonal blocks. This arrangement results from our decision to employ a diagonal chirality matrix with the first $16$ eigenvalues being positive. By introducing the Greek indices $\alpha$, which correspond to the $16$ eigenvectors of the chirality matrix with positive eigenvalues, and the Greek indices with a dot, $\dot\alpha$, corresponding to the $16$ eigenvectors with negative eigenvalues, the index structure of our ten-dimensional $32\times32$ Hermitian gamma matrices is:
\begin{equation}
\label{gamma off}
\Gamma_a =  \left(\begin{array}{cc} 0&\gamma_{a}\,^{\alpha}\,_{\dot\alpha}\\ \gamma_{a}\,^{\dot\alpha}\,_{\alpha}&0\end{array}\right)\,.
\end{equation}
In our basis, the real symmetric gamma matrices are $\Gamma_1$,  $\Gamma_2$, $\Gamma_3$, $\Gamma_4$, and  $\Gamma_5$, while the imaginary antisymmetric ones are $\Gamma_6$,  $\Gamma_7$, $\Gamma_8$, $\Gamma_9$, and  $\Gamma_{10}$. Consequently, the $32\times32$ charge conjugation matrix is defined as $C = \Gamma_1\Gamma_2\Gamma_3\Gamma_4\Gamma_5$, such that it is also structured as off-diagonal blocks, but the index structure is different since they construct Lorentz scalars $(\zeta_1^{T}C\zeta_2)$ out of spinors with opposite chirality, $\zeta_1$ and $\zeta_2$. Its index structure is given by
\begin{equation}
\label{C off}
C =  \left(\begin{array}{cc} 0&\delta_{\alpha\dot\alpha}\\ \delta_{\dot\alpha\alpha}&0\end{array}\right) \,.  
\end{equation}
By construction, the product of the charge conjugation matrix with the gamma matrices results in symmetric matrices, i.e. $(C\Gamma_{a})=(C\Gamma_{a})^{T}$, which implies that
\begin{equation}
\delta_{\alpha\dot\alpha}\gamma_{a}\,^{\dot\alpha}\,_{\beta} = \gamma_{a}\,_{\alpha}\,^{\dot\alpha}\delta_{\dot\alpha\beta}  
\end{equation}
where $\gamma_{a}\,_{\alpha}\,^{\dot\alpha} = \gamma_{a}\,^{\dot\alpha}\,_{\alpha}$ is notation for taking the transpose of $\gamma_{a}\,^{\dot\alpha}\,_{\alpha}$ viewed as a matrix. Given that $\delta_{\alpha\dot\alpha}$ and $\delta_{\dot\alpha\alpha}$ are invertible matrices, we can eliminate the presence of anti-chiral indices in all formulas. Indeed, we have $\delta_{\alpha\dot\alpha} = \delta_{\dot\alpha\alpha}$ and $C^2=1$, making $\delta_{\alpha\dot\alpha}$ and its inverse, $\delta^{\alpha\dot\alpha}$, identical in our basis. By eliminating the anti-chiral indices using our Lorentz-invariant symbol $\delta_{\alpha\dot\alpha}$, we obtain symmetric $16\times16$ gamma matrices:
\begin{equation}
\gamma_{a}\,_{\alpha\beta} = \delta_{\alpha\dot\alpha}\gamma_{a}\,^{\dot\alpha}\,_{\beta} = \gamma_{a}\,_{\alpha}\,^{\dot\alpha}\delta_{\dot\alpha\beta} = \delta_{\beta\dot\alpha}\gamma_{a}\,^{\dot\alpha}\,_{\alpha} = \gamma_{a}\,_{\beta\alpha}\,,
\end{equation}
\begin{equation*}
\gamma_{a}^{\alpha\beta} =  \gamma_{a}\,^{\alpha}\,_{\dot\beta}\delta^{\dot\beta\beta} = \delta^{\alpha\dot\beta}\gamma^{a}\,_{\dot\beta}\,^{\beta} = \gamma^{a}\,_{\dot\beta}\,^{\beta}\delta^{\dot\beta\alpha} = \gamma_{a}^{\beta\alpha}   
\end{equation*}
which, according to the Clifford algebra, inherit the relation:
\begin{equation}
\gamma_{a}^{\alpha\beta}\gamma_{b \beta\gamma} + \gamma_{b}^{\alpha\beta}\gamma_{a\beta\gamma} = 2\delta_{\gamma}^{\alpha}\delta_{ab} \,,   
\end{equation}
from which all Fierz identities can be derived, assuming that these matrices are symmetric. Our gamma matrices are explicitly exhibited in appendix \ref{important matrices}.

\section{Killing spinors in Ramond-Ramond backgrounds}
\label{Killin spinors appendix}

In this appendix we will derive the Killing spinor equations for supergravity backgrounds from the type II superspace. Our starting point is the super-covariant derivatives
\begin{equation}
\nabla_{A}=E_{A}\,^{M}(Z)\frac{\partial}{\partial Z^{M}}+\frac{1}{2}\Omega_{A}\,^{ab}(Z)t_{ba}\,,    
\end{equation}
where $Z^{M}\in\mathbb{R}^{10|32}$ are super-coordinates that parametrize the superspace, $\frac{\partial}{\partial Z^{M}}$ span the tangent superspace, $t_{ab}$ are the $SO(10)$ generators, and $A$ range over the bosonic directions $a=1$ to $10$, and fermionic directions $\alpha=1$ to $16$ and $\widehat\alpha =1$ to $16$. The $E_{A}\,^{M}$ are the inverse of the super vielbeins $E_{M}\,^{A}$, and $\Omega_{A}\,^{ab}$ are the components of the superspace spin connectionthe. The superspace torsion $T_{AB}\,^{C}(Z)$ and curvature $R_{AB}\,^{ab}(Z)$ are obtained by computing the gradded commutator
\begin{equation}
\{\nabla_{A},\nabla_{B}]=T_{AB}\,^{C}(Z)\nabla_{C}+\frac{1}{2}R_{AB}\,^{ab}(Z)\Sigma_{ba}\,.    
\end{equation}
The Killing supervectors will be labelled by $\widetilde A$ and are defined as the gauge compensated supervector fields that (anti-)commutes with the covariant derivatives:
\begin{equation}
Q_{\widetilde A}=\varepsilon_{\widetilde A}\,^{A}(Z)\nabla_{A}+\frac{1}{2}\varepsilon_{\widetilde A}\,^{ab}(Z)\Sigma_{ba}\,,  \qquad  \{Q_{\widetilde A},\nabla_{A}]=0\,.  
\end{equation}
Here, $\varepsilon_{\widetilde A}\,^{A}$ are the components of the Killing supervectors and $\varepsilon_{\widetilde A}\,^{ab}$ are the components of the gauge compensators. In components, the above equation translates to
\begin{equation}
\nabla_{A}\varepsilon_{\widetilde A}\,^{B}(Z)=\varepsilon_{\widetilde  A}\,^{C}(Z)T_{CA}\,^{B}(Z)+\varepsilon_{\widetilde  AA}\,^{B}(Z)\,,\qquad \nabla_{ A}\varepsilon_{\widetilde  AB}\,^{C}=\varepsilon_{\widetilde  A}\,^{D}(Z)R_{DAB}\,^{C}(Z) \,,
\end{equation}
where
\begin{equation}
\frac{1}{2}\varepsilon_{\widetilde A}\,^{ab}(Z)[\Sigma_{ba},\nabla_{A}]=\varepsilon_{\widetilde A A}\,^{B}(Z)\nabla_{B}\,.    
\end{equation}
From now on, we will not write the arguments of the superfields for aesthetic reasons. To derive the Killing spinor equations, along with the Killing vector equations, we categorize the Killing supervector $Q_{\widetilde A}$ into three types: translation type $P_{\widetilde a}$, rotation type $M_{\widetilde a\widetilde b}$, and supersymmetry type $q_{\widetilde\alpha}$. We then impose that the fermions vanish, obtaining:
\begin{equation}
\label{killing vector - translation}
P_{\widetilde a}=\varepsilon_{\widetilde a}\,^{a}\nabla_{a}+\frac{1}{2}\varepsilon_{\widetilde a}\,^{ab}\Sigma_{ba}\,,\qquad \nabla_{a}\varepsilon_{\widetilde a}\,^{b}=\varepsilon_{\widetilde aa}\,^{b}\,,\qquad \nabla_{a}\varepsilon_{\widetilde a b}\,^{c}=\varepsilon_{\widetilde a}\,^{d}R_{dab}\,^{c} \,,   
\end{equation}
\begin{equation}
\label{killing vector - rotation}
M_{\widetilde{a}\widetilde{b}}=\varepsilon_{\widetilde{ a}\widetilde{b}}\,^{a}\nabla_{a}+\frac{1}{2}\varepsilon_{\widetilde{ a}\widetilde{b}}\,^{ab}\Sigma_{ba}\,,\qquad \nabla_{a}\varepsilon_{\widetilde a\widetilde b}\,^{b}=\varepsilon_{\widetilde a\widetilde ba}\,^{b}\,,\qquad \nabla_{a}\varepsilon_{\widetilde a\widetilde b b}\,^{c}=\varepsilon_{\widetilde a\widetilde b}\,^{d}R_{dab}\,^{c} \,,   
\end{equation}
\begin{equation}
\label{killing spinor - derivation}
Q_{\widetilde\alpha}=\varepsilon_{\widetilde\alpha}\,^{\alpha}\nabla_{\alpha}+\varepsilon_{\widetilde\alpha}\,^{\widehat\alpha}\nabla_{\widehat\alpha}\,,\qquad \nabla_{a}\varepsilon_{\widetilde\alpha}\,^{\alpha}=\varepsilon_{\widetilde\alpha}\,^{\widehat\alpha}T_{\widehat\alpha a}\,^{\beta}\,,\qquad \nabla_{a}\varepsilon_{\widetilde\alpha}\,^{\widehat\alpha}=\varepsilon_{\widetilde\alpha}\,^{\widehat\beta}T_{\widehat\beta a}\,^{\widehat\alpha}\,. 
\end{equation}
The equations in \ref{killing vector - translation} and \ref{killing vector - rotation} correspond to the Killing vector equations for translations and rotations, while the equations in \ref{killing spinor - derivation} correspond to the Killing spinor equations. Notice that in the presence of Ramond-Ramond fllux, the torsion components $T_{\widehat\beta a}\,^{\widehat\alpha}$ and $T_{\widehat\alpha a}\,^{\beta}$ are non-vanishing and given by
\begin{equation}
T_{\alpha b}\,^{\widehat\alpha}= \frac{1}{2}\gamma_{b\alpha\beta}\eta^{\beta\widehat\alpha}\,,\qquad T_{\widehat\alpha b}\,^{\alpha}=\frac{1}{2}\gamma_{b\widehat\alpha\widehat\beta}\eta^{\widehat\beta\beta}\,,
\end{equation}
with $\eta^{\widehat\alpha\alpha} = -\eta^{\alpha\widehat\alpha}$ being the Ramond-Ramond bi-spinor.

\section{Ad$\mathbb{S}_5\times \mathbb{S}^{5}$ Killing spinors}
\label{AdS Killing spinors}
As discussed in the previous appendix, the Killing spinor equation is modified in the presence of Ramond-Ramond flux. In the coordinates specified in \ref{coordinates}, the Ramond-Ramond self-dual five-form tensor is
\begin{equation}
F_{abcde} = \frac{1}{2}\left(\frac{1}{5!}\delta_{[a}^{1}\delta_{b}^{2}\delta_{c}^{3}\delta_{d}^{4}\delta_{e]}^{(i+4)} + \varepsilon_{abcde1234(4+i)}\right)\frac{y_{i}}{y}\,. 
\end{equation}
Consequently, the Ramond-Ramond bi-spinor $\eta^{\alpha\widehat\alpha}$ is given by
\begin{equation}
-\eta^{\widehat\beta\alpha}=\eta^{\alpha\widehat\beta}=i(I\gamma_{y})^{\alpha\widehat \beta}\,,\qquad I=\gamma^{1234}\,,\qquad \gamma_{y}=\frac{y^{i}}{y}\gamma_{(4+i)}=\frac{z^{i}}{z}\gamma_{(4+i)}\,.
\end{equation}
We can check that it is covariantly conserved $\nabla_{a}\eta^{\alpha\widehat\alpha} =0$. We can use the gamma matrices to simplify the vielbeins $e^{a}$ and vector $D_{a}$ of the appendix \ref{coordinates} as follows:
\begin{equation}
j^{a}\gamma_{a}^{\alpha\beta}=\partial x^{m}e_{m}\,^{a}\gamma_{a}^{\alpha\beta}=-\gamma_{y}^{\alpha\gamma}\frac{\partial x^{\mu}\gamma_{\mu\gamma\delta}+\partial y^{i}\gamma_{(i+4)\gamma\delta}}{y}\gamma_{y}^{\delta\beta}
\end{equation}
\begin{equation}
\gamma^{a}_{\alpha\beta}D_{a}=\gamma^{a}_{\alpha\beta}E_{a}\,^{m}\partial_{m}=-y(\gamma_{y}\gamma^{\mu}\gamma_{y})_{\alpha\beta}\frac{\partial}{\partial x^{\mu}}-y(\gamma_{y}\gamma^{i+4}\gamma_{y})_{\alpha\beta}\frac{\partial}{\partial y^{i}}\,.
\end{equation}
Let us introduce the notation $\widehat \varepsilon^{\alpha}=\varepsilon^{\widehat\alpha}$. The Killing spinor equations, acoording to \ref{killing spinor - derivation}, are given by
\begin{equation}
D_{a}\varepsilon^{\alpha}-\frac{1}{4}\Omega_{a}\,^{bc}(\gamma_{cb}\varepsilon)^{\alpha}=-\frac{i}{2}(I\gamma_{y}\gamma_{a}\widehat \varepsilon)^{\alpha} \,,
\end{equation}
\begin{equation}
D_{a}\widehat \varepsilon^{\alpha}-\frac{1}{4}\Omega_{a}\,^{bc}(\gamma_{cb}\widehat \varepsilon)^{\alpha}=+\frac{i}{2}(I\gamma_{y}\gamma_{a}\varepsilon)^{\alpha}\,.    
\end{equation}
These equations are solved by
\begin{equation}
\label{Killing spinor - the one}
\varepsilon=\frac{1}{2}y^{-\frac{1}{2}}\left[\varepsilon_{s}+x^{\mu}\gamma_{\mu}\varepsilon_{c}-y^{i}\gamma_{4+i}\varepsilon_{c}\right]
\qquad 
\widehat\varepsilon=\frac{i I}{2}y^{-\frac{1}{2}}\left[\varepsilon_{s}+x^{\mu}\gamma_{\mu}\varepsilon_{c}+y^{i}\gamma_{4+i}\varepsilon_{c}\right]\,.
\end{equation}
Notice that close to the boundary $y\rightarrow 0$, $\varepsilon$ and $- iI\widehat\varepsilon $ converge to the same spinor:
\begin{equation}
\varepsilon \rightarrow \frac{1}{2}y^{-\frac{1}{2}}\left[\varepsilon_{s}+x^{\mu}\gamma_{\mu}\varepsilon_{c}\right]\leftarrow - iI\widehat\varepsilon \,,
\end{equation}
which correspond to the ten-dimensional description of the super-Yang-Mills supersymmetries in $\mathbb{R}^{4}$, see \cite{Dymarsky:2009si}. This gives an holographic dictionary for the supersymmetries and justify our notation for the constant spinor $\varepsilon_{s}^{\alpha}$ and $\varepsilon_{c\alpha}$, where the chiral spinor $\varepsilon_{s}^{\alpha}$ parametrize the rigid supersymmetries of $\mathbb{R}^{4}$ (super-Poincaré) while the anti-chiral spinor $\varepsilon_{c\alpha}$ parametrize the non-rigid ones (superconformal), as in \cite{Dymarsky:2009si}.
\section{Ad$\mathbb{S}_5\times \mathbb{S}^{5}$ pure Killing spinors}
\label{pure killing spinors - appendix}
The fixed spinor $\varepsilon_{s}$ in \ref{Killing spinor - the one} can be expressed in terms of a fixed pure spinor $|0\rangle$ and a null 4-form $s_{IJKL}$ tensor as follows:
\begin{equation}
\label{ground}
\varepsilon_{s}=[1+s]|0\rangle,\,\qquad s = \frac{1}{4!}s_{IJKL}\gamma^{LKJI}\,.   
\end{equation}
Here, the $\gamma^{I}$, $I=1,2,3,4,5$, denote the null gamma matrices satisfying
\begin{equation}
\label{relations}
\gamma_{I}|0\rangle = 0\,,\qquad
\{\gamma^{I},\gamma^{J}\}=0\,,\qquad  \{\gamma_{I},\gamma_{J}\}=0  \,,\qquad \{\gamma^{I},\gamma_{J}\}=\delta_{J}^{I}\,,
\end{equation}
Since $s_{IJKL}$ is anti-symmetric, it possesses $5$ components. Thus, we have $16$ components as it should be, where $11$ comes from the choice of $|0\rangle$. Notice that $s|0\rangle$ is also a pure spinor, so one can always write a chiral spinor as a sum of two pure spinors. The fixed spinor $\varepsilon_{c}$ in \ref{Killing spinor - the one} can be expressed in terms of the same fixed pure spinor $|0\rangle$, a null 1-form $c_{I}$, a null 3-form $c_{IJK}$ and a null 5-form $c_{IJKLM}$, as follows:
\begin{equation}
\varepsilon_{c}=c|0\rangle=\left[c_{I}\gamma^{I}+\frac{1}{3!}c_{IJK}\gamma^{KJI}+\frac{1}{5!}c_{IJKLM}\gamma^{MLKJI}\right]|0\rangle\,, 
\end{equation}
where $c_{I}$ has $5$ components, $c_{IJK}$ has $10$ components and $c_{IJKLM}$ has $1$ component. The Killing spinor in \ref{Killing spinor - the one} is, projectively, given by
\begin{equation}
\varepsilon = \left[\left(1+c_{I}Q^{I}\right)+\frac{1}{2}(c_{[I}Q_{J]}+c_{IJK}Q^{K})\gamma^{JI}+\frac{1}{4!}\left(c_{[IJK}Q_{L]}+c_{IJKLM}Q^{M}\right)\gamma^{MNLKJI}\right]|0\rangle\,,  
\end{equation}
where $Q^{a}=x^{\mu}\delta_{\mu}\,^{a}-y^{i}\delta_{4+i}\,^{a}$. The pure spinor constraints can be expressed in terms of these null polyforms $Q_{I}$, $s$ and $c$, and null vectors $Q^{I}$, as follows:
\begin{equation}
\lambda \lambda_{IJKL}=-\frac{1}{8}\lambda_{[IJ}\lambda_{KL]}\,,\qquad \lambda_{I[J}\lambda_{KLMN]}=0  
\end{equation}
where
\begin{equation*}
\lambda = \left(1+c_{I}Q^{I}\right)\,,\qquad \lambda_{IJ}=(c_{[I}Q_{J]}+c_{IJK}Q^{K})\,,\qquad \lambda_{IJKL}=c_{[IJK}Q_{L]}+c_{IJKLM}Q^{M}\,.
\end{equation*}
In \cite{Dymarsky:2009si}, it was shown that the above constraint is satisfied for all real values of $Q^{a}$ if, and only if,
\begin{equation}
c_{IJK}=0\,,\qquad c_{IJKLM}=0\,.    
\end{equation}
If we impose a weaker constraint that the Killing spinor is pure only somewhere in Ad$\mathbb{S}_5\times \mathbb{S}^{5}$, then \cite{Dymarsky:2009si} asserts that the only possibility is to have a pure Killing spinor in a six-dimensional subspace, given by the equation
\begin{equation}
Q_{[I}c_{JKL]}+Q_{M}Q^{M}c_{[I}c_{JKL]}=0\,,    
\end{equation}
where $c_{IJK}$ should be rank one. This is the case in our paper. The six-dimensional region is given by
\begin{equation}
x_3 = 0\,,\qquad y_1=0\,,\qquad y_4=0\,,\qquad y_6 = 0\,,  
\end{equation}
which constitute a $Ad\mathbb{S}_4\times \mathbb{S}^2$ slice of Ad$\mathbb{S}_5\times \mathbb{S}^{5}$.

\section{PSU(2,2|4) and BRST invariance fixes the sigma model action}
\label{fixing sigma model action}

In this appendix we will fix the relative constants of the sigma model action using BRST symmetry. This is particularly important when we investigate the sigma model action evalutated at the off-shell supersymmetric configurations, where simplifications and cancellations may occur, and it is important to get each coefficient right, without factors of two and signs floating around.

\subsection{The normalization of $\mathfrak{psu}(2,2|4)$ algebra}

We start by fixing the normalization of the $\mathfrak{psu(2,2|4)}$ algebra. Before fixing normalization, the algebra of covariant derivatives that solves the supergravity constraints in the presence of a self-dual Ramond-Ramond flux is given by \cite{Berkovits:2001ue}
\begin{equation*}
[\nabla_{\alpha},\nabla_{\beta}]=\Phi\gamma^{c}_{\alpha\beta}\nabla_{c}\,,\qquad [\widehat \nabla_{\widehat \alpha},\widehat \nabla_{\widehat \beta}]=\Phi\gamma^{c}_{\widehat \alpha\widehat \beta}\nabla_{c}
\end{equation*}
\begin{equation*}
[\nabla_{\alpha},\nabla_{a}]=q\gamma_{a\alpha\beta}\eta^{\beta\widehat\beta}\nabla_{\widehat\beta}\,,\qquad [\overline\nabla_{\widehat\alpha},\nabla_{a}]=q\gamma_{a\widehat\alpha\widehat\beta}\eta^{\widehat\beta\beta}\nabla_{\beta}
\end{equation*}
\begin{equation*}
[\nabla_{c},t_{ab}]=\eta_{c[a}\nabla_{b]}\,,\qquad [\nabla_{\alpha},t_{ab}]=\frac{1}{2}\gamma_{ab\alpha}\,^{\beta}\nabla_{\beta}\,,\qquad [\overline\nabla_{\widehat\alpha},t_{ab}]=\frac{1}{2}\gamma_{ab\widehat\alpha}\,^{\widehat\beta}\nabla_{\widehat\beta}\,   
\end{equation*}
\begin{equation}
[\nabla_{\alpha},\overline\nabla_{\widehat\alpha}]=r(\gamma^{b}\eta\gamma^{c})_{\alpha\widehat\alpha}t_{cb}\,,\qquad [\nabla_{a},\nabla_{b}]=\frac{1}{2}R_{ab}\,^{cd}t_{dc}\,.
\end{equation}
The Bianchi identity
\begin{equation}
[\nabla_{(\alpha},\{\nabla_{\beta},\nabla_{\widehat\gamma)}\}]=0\iff
\frac{r}{4}(\gamma_{[a}\eta\gamma_{b]})_{\widehat\gamma(\alpha}\gamma^{ba}\,_{\beta)}\,^{\gamma}+\Phi q\gamma_{\alpha\beta}^{c}(\gamma_{c}\eta)_{\widehat\gamma}\,^{\gamma}=0
\end{equation}
fixes $r=\Phi q$ since
\begin{equation}
\frac{1}{16\times 5!\times 2}\gamma^{\alpha\beta}_{cdefg}(\gamma_{[a}\eta\gamma_{b]})_{\widehat\gamma (\alpha}\gamma^{ba}\,_{\beta)}\,^{\gamma}=0    
\end{equation}
and
\begin{equation*}
\frac{1}{16}\gamma^{\alpha\beta}_{c}(\gamma_{[a}\eta\gamma_{b]})_{\widehat\gamma (\alpha}\gamma^{ba}\,_{\beta)}\,^{\gamma}=\frac{2}{16}(\gamma_{[a}\eta\gamma_{b]}\gamma_{c}\gamma^{ba})_{\widehat\gamma}\,^{\gamma}=\frac{2\times 2}{16}(\gamma_{a}\eta\gamma_{b}\gamma_{c}\gamma^{ba})_{\widehat\gamma}\,^{\gamma}=
\end{equation*}
\begin{equation*}
=\frac{2\times 2}{16}(\gamma_{a}\eta\gamma_{b}\gamma_{c}\gamma^{b}\gamma^{a})_{\widehat\gamma}\,^{\gamma}=\frac{2\times 2\times (-8)}{16}(\gamma_{a}\eta\gamma_{c}\gamma^{a})_{\widehat\gamma}\,^{\gamma}=    
\end{equation*}
\begin{equation}
=\frac{2\times 2\times (-8)}{16}(\gamma_{a}\eta\{\gamma_{c},\gamma^{a}\})_{\widehat\gamma}\,^{\gamma}=\frac{2\times 2\times (-8)\times 2}{16}(\gamma_{c}\eta)_{\widehat\gamma}\,^{\gamma}=-4(\gamma_{c}\eta)_{\widehat\gamma}\,^{\gamma}\,,  
\end{equation}
such that
\begin{equation}
-4\frac{r}{4} +\Phi q = 0\iff r=\Phi q\,,
\end{equation}
The Bianchi identity
\begin{equation}
[\nabla_{[a},\{\nabla_{\alpha},\nabla_{\beta]}\}]=0 \iff \frac{\Phi}{2}R_{ab}\,^{cd}\Sigma_{dc}\gamma^{b}_{\alpha\beta}+\frac{q r}{2}(\gamma^{[c}\eta\gamma^{d]})_{\widehat\beta(\alpha}(\gamma_{a}\eta)_{\beta)}\,^{\widehat\beta}=0\,, \end{equation}
fixes
\begin{equation}
R_{ab}\,^{cd}=-\frac{qr}{8\Phi}(\gamma^{[c}\eta\gamma^{d]}\gamma_{b}\gamma_{a}\eta)_{\widehat\alpha}\,^{\widehat\alpha}\,.  
\end{equation}
Defining the signs $(\pm)^{[ab]}_{[cd]}$ as
\begin{equation}
(\gamma^{[a}\eta\gamma^{b]})_{\alpha\widehat\alpha}=(\pm)^{[ab]}_{[cd]}(\eta\gamma^{cd})_{\alpha\widehat\alpha}
\end{equation}
we obtain
\begin{equation}
R_{ab}\,^{cd}=-\frac{Qr}{8\Phi}(\pm)^{[cd]}_{[ef]}(\gamma^{ef}\gamma_{ba})_{\alpha}\,^{\alpha}=2\frac{qr}{\Phi}\delta^{[f}_{a}\delta^{e]}_{b}(\pm)_{[fe]}\,^{[cd]}\,.
\end{equation}
Now, plugging $r=\Phi q$ in the above equation returns:
\begin{equation}
R_{ab}\,^{cd}=\frac{(2q)^{2}}{2}\delta^{[f}_{a}\delta^{e]}_{b}(\pm)_{[fe]}\,^{[cd]} =(2q)^{2}(\pm)_{[ab]}\,^{[cd]} \,,
\end{equation}
so the $AdS_5\times S^{5}$ radius is
\begin{equation}
R=\frac{1}{2q}\,.
\end{equation}
In total we have the following torsion and curvature components:
\begin{equation}
T_{\alpha\beta}\,^{c}=\Phi\gamma^{c}_{\alpha\beta}\,,\quad T_{\widehat\alpha\widehat\beta}\,^{c}=\Phi\gamma^{c}_{\widehat\alpha\widehat\beta}\,,\quad T_{\alpha b}\,^{\widehat\alpha}=\frac{1}{R}\frac{1}{2}(\gamma_{b}\eta)_{\alpha}\,^{\widehat\alpha}\,,\quad T_{\widehat\alpha b}\,^{\alpha}=\frac{1}{R}\frac{1}{2}(\gamma_{b}\eta)_{\widehat\alpha}\,^{\alpha}    
\end{equation}
\begin{equation}
R_{\alpha\widehat\alpha}\,^{[bc]}=\frac{\Phi}{R} \frac{1}{2}(\gamma^{[b}\eta\gamma^{c]})_{\alpha\widehat\alpha}\,,\qquad R_{ab}\,^{cd}=\pm \frac{1}{R^{2}}\delta_{[a}^{c}\delta_{b]}^{d}\,.
\end{equation}
We choose the following normalization for the $\mathfrak{psu(2,2|4)}$ algebra:
\begin{equation}
\Phi=R=1\,,
\end{equation}
so that the physical Ad$\mathbb{S}_5\times \mathbb{S}^{5}$ curvature is mapped to an overal constant in the sigma model action. According to the Maldacena's conjecture, this overal constant is to be identified with the square root of the t' Hooft coupling.
\subsection{BRST transformations}
\label{BRST transformation - appedinx action}
The BRST generator $Q$ acts on $\frac{PSU(2,2|4)}{USp(2,2)\times USp(4)}$ superspace as follows: \begin{equation}
Q g =  (\lambda^{\alpha}\nabla_{\alpha}+\widehat\lambda^{\widehat\alpha}\nabla_{\widehat\alpha})g= g (\lambda^{\alpha}t_{\alpha}+\widehat\lambda^{\widehat\alpha}t_{\widehat\alpha})\,.
\end{equation}
where the superspace vielbein and spin connections can be extracted from the supercoset $g$ as follows:
\begin{equation}
g^{-1}\partial g=j-\Omega
\end{equation}
where
\begin{equation*}
j=j^{A}t_{A}=\partial Z^{M}E_{M}\,^{A}t_{A}\,,    
\qquad 
\Omega=\frac{1}{2}\Omega^{ab}t_{ba}=\frac{1}{2}\partial Z^{M}\Omega_{M}\,^{ab}t_{ba}\,.   
\end{equation*}
The BRST generator $Q$ will act on $j-\Omega$ as follows
\begin{equation}
Q(j-\Omega)=
\end{equation}
\begin{equation*}
=Q(g^{-1}\partial g)=(Qg^{-1})\partial g + g^{-1}\partial (Qg) = -g^{-1}(Qg)g^{-1}\partial g + g^{-1}\partial (Qg) = 
\end{equation*}
\begin{equation*}
=-g^{-1}g(\lambda^{A}t_{A})g^{-1}\partial g + g^{-1}\partial(g(\lambda^{A}t_{A}))=-(\lambda^{A}t_{A})g^{-1}\partial g + g^{-1}\partial g(\lambda^{A}t_{A})+\partial\lambda^{A}t_{A}=
\end{equation*}
\begin{equation*}
=[g^{-1}\partial g,\lambda^{A}t_{A}]+\partial\lambda^{A}t_{A}=[j-\Omega,\lambda^{A}t_{A}]+\partial\lambda^{A}t_{A}= [j,\lambda^{A}t_{A}]+\partial \lambda^{A}t_{A} +\lambda^{A}[ t_{A},\Omega]  
\end{equation*}
\begin{equation*}
=[j,\lambda^{A}t_{A}]+\partial \lambda^{A}t_{A}+[\Omega,\lambda^{A}]t_{A}=\lambda^{A}[j,t_{A}]+\nabla\lambda^{A}t_{A}=\lambda^{A}j^{B}[t_{B},t_{A}]+\nabla\lambda^{A}t_{A}=   
\end{equation*}
\begin{equation*}
=\lambda^{A}j^{B}T_{BA}\,^{C}t_{C}+\frac{1}{2}\lambda^{A}j^{B}R_{BA}\,^{ab}t_{ba}+\nabla\lambda^{A}t_{A}\,,
\end{equation*}
and from that we can extract the BRST transormations acting on the components of $j^{A}$ and $\Omega^{ab}$ by isolating the $\mathfrak{psu(2,2|4)}$ generators:
\begin{equation*}
Qj^{\alpha}=\nabla\lambda^{\alpha}+\widehat\lambda^{\widehat\alpha}j^{a}T_{a\widehat\alpha}\,^{\alpha}\,,\qquad Qj^{\widehat\alpha}=\nabla\widehat\lambda^{\widehat\alpha}+\lambda^{\alpha}j^{a}T_{a\alpha}\,^{\widehat\alpha}\,,
\end{equation*}
\begin{equation*}
Qj^{a}=\lambda^{\alpha}j^{\beta}T_{\beta\alpha}\,^{a}+\widehat\lambda^{\widehat\alpha}j^{\widehat\beta}T_{\widehat\beta\widehat\alpha}\,^{a}\,,    
\end{equation*}
and
\begin{equation}
Q\Omega^{ab}=-\lambda^{\alpha}j^{\widehat\alpha}R_{\widehat\alpha\alpha}\,^{ab}-\widehat\lambda^{\widehat\alpha}j^{\alpha}R_{\alpha\widehat\alpha}\,^{ab}\,.
\end{equation}
In plugging the values for the torsion components and curvature, we obtain
\begin{equation*}
Q j^{\alpha}=\nabla\lambda^{\alpha}+\frac{1}{2}j^{a}(\eta\gamma_{a} \widehat\lambda)^{\alpha}\,,\qquad Q j^{\widehat\alpha}=\nabla\widehat\lambda^{\widehat\alpha}+\frac{1}{2}j^{a}(\eta\gamma_{a}\lambda)^{\widehat\alpha}\,,
\end{equation*}
\begin{equation*}
Q j^{a}=(\lambda\gamma^{a})_{\alpha}j^{\alpha}+(\widehat\lambda\gamma^{a})_{\widehat\alpha}j^{\widehat\alpha}\,,    
\end{equation*}
and
\begin{equation}
Q\Omega^{ab}=-\frac{1}{2}(\lambda \gamma^{[a}\eta\gamma^{b]})_{\widehat\alpha}j^{\widehat\alpha}-\frac{1}{2}(\widehat\lambda \gamma^{[a}\eta\gamma^{b]})_{\alpha}j^{\alpha}\,.
\end{equation}
The BRST transformation of the spin connection is useful in that it appears in the sigma model action \ref{sigma model action} through $\overline\nabla\lambda^{\alpha}$ and $\nabla\widehat\lambda^{\widehat\alpha}$. We have 
\begin{equation*}
Q(\nabla\lambda^{\alpha})=-\frac{1}{4}(\nabla_{\lambda}\Omega^{ab})\gamma_{ba}\,^{\alpha}\,_{\beta}\lambda^{\beta}=\frac{1}{4}\left((\lambda R^{ab})_{\widehat\gamma}j^{\widehat\gamma}+(\widehat\lambda R^{ab})_{\gamma}j^{\gamma}\right)\gamma_{ba}\,^{\alpha}\,_{\beta}\lambda^{\beta}=
\end{equation*}
\begin{equation}
= \frac{1}{4}(\widehat\lambda R^{ab})_{\gamma}j^{\gamma}\gamma_{ba}\,^{\alpha}\,_{\beta}\lambda^{\beta}.
\end{equation}
and similar for $\nabla\widehat\lambda^{\alpha}$, $\overline\nabla \lambda^{\alpha}$ and $\overline\nabla\widehat\lambda^{\widehat\alpha}$. In the above manipulation, we have the following identity
\begin{equation}
R_{\widehat\gamma\gamma}\,^{ab}\gamma_{ba}\,^{\alpha}\,_{\beta}\lambda^{\beta}\lambda^{\gamma}=R_{\widehat\gamma\gamma}\,^{ab}\gamma_{ba}\,^{\alpha}\,_{\beta}\lambda^{((\beta}\lambda^{\gamma))}= R_{\widehat\gamma((\gamma}\,^{ab}\gamma_{ba}\,^{\alpha}\,_{\beta))}\lambda^{\beta}\lambda^{\gamma}=0\,,
\end{equation}
which is true in any Type II supergravity background. In our case this can be seen explicitly:
\begin{equation}
\frac{1}{2}(\gamma^{[a}\eta\gamma^{b]}\lambda)_{\widehat\alpha}(\gamma_{ab}\lambda)^{\alpha}= (\gamma^{a}\eta\gamma^{b}\lambda)_{\widehat\alpha}(\gamma_{a}\gamma_{b}\lambda)^{\alpha} =0  
\end{equation}
since $(\gamma^{a}\eta\gamma_{a})_{\alpha\widehat\alpha}=0$ and $(\gamma_{b}\lambda)_{\alpha}(\gamma^{b}\lambda)_{\beta}=0$.

\subsection{Maurer-Cartan identity}

In considering two worldsheet vectors, $\partial_{i}$ and their worldsheet currents $j_{i}=\partial_{i}Z^{M}E_{M}^{A}t_{A}$, where $i=\sigma,\tau$, because $\partial_{[i}\partial_{j]}=0$, we have that
\begin{equation}
\partial_{[i}j_{j]}-\partial_{[i}\Omega_{j]}=    
\end{equation}
\begin{equation*}
=\partial_{[i}(g^{-1}\partial_{j]}g)=\partial_{[i}g^{-1}\partial_{j]}g=-(g^{-1}\partial_{[i}g)( g^{-1}\partial_{j]}g)=-[g^{-1}\partial_{i}g,g^{-1}\partial_{j}g]=  
\end{equation*}
\begin{equation*}
=-[j_{i}-\Omega_{i},E_{j}-\Omega_{j}]=-[j_{i},j_{j}]+[j_{[i},\Omega_{j]}]-[\Omega_{i},\Omega_{j}]\,.
\end{equation*}
Isolating the generators we obtain
\begin{equation*}
\nabla_{[i}j_{j]}\,^{A}=\partial_{[i}j_{j]}\,^{A}+[\Omega_{[i},j_{j]}\,^{A}]=j_{i}\,^{B}j_{j}\,^{C}T_{CB}\,^{A}\,,
\end{equation*}
\begin{equation}
\partial_{[i}\Omega_{j]}\,^{ab}+[\Omega_{i},\Omega_{j}\,^{ab}] =-j_{i}\,^{A}j_{j}\,^{B}R_{BA}\,^{ab}\,.
\end{equation}
If we take the derivatives to be $\partial$ and $\overline\partial$ we obtain
\begin{equation}
\nabla \overline j^{A}-\overline\nabla j^{A} = j^{B} \overline j^{C}T_{CB}\,^{A}\,.
\end{equation}
Specializing to the $AdS_5\times S^{5}$ superspace the above equation becomes
\begin{equation*}
\nabla \overline j^{\alpha}-\overline\nabla j^{\alpha}=(j^{\widehat\beta}\overline j^{c}-\overline j^{\widehat\beta}j^{c})T_{c\widehat\beta}\,^{\alpha}=-\frac{1}{2}(j^{\widehat\beta}\overline j^{c}-\overline j^{\widehat\beta}j^{c})\gamma_{c\widehat\beta\widehat\alpha}\eta^{\widehat\alpha\alpha}\,,   
\end{equation*}
\begin{equation*}
\nabla \overline j^{\widehat\alpha}-\overline\nabla j^{\widehat\alpha}=(j^{\beta}\overline j^{c}-\overline j^{\beta}j^{c})T_{c\beta}\,^{\widehat\alpha}=-\frac{1}{2}(j^{\beta}\overline j^{c}-\overline j^{\beta}j^{c})\gamma_{c\beta\alpha}\eta^{\alpha\widehat\alpha}\,,
\end{equation*}
\begin{equation}
\label{MC identity}
\nabla \overline j^{a} -\overline\nabla j^{a}=j^{\alpha}\overline j\,^{\beta}T_{\beta\alpha}\,^{a}+j^{\widehat\alpha}\overline j^{\widehat\beta}T_{\widehat\beta\widehat\alpha}\,^{c}=\left(j^{\alpha}\overline j\,^{\beta}\gamma_{\beta\alpha}^{a}+j^{\widehat\alpha}\overline j^{\widehat\beta}\gamma_{\widehat\beta\widehat\alpha}^{c}\right)\,.
\end{equation}
\subsection{The Ad$\mathbb{S}_5\times \mathbb{S}^{5}$ sigma model in pure spinor formalism}

We are now armed with sufficient ammunition to fix the relative coefficients in the sigma model Lagrangian:
\begin{equation}
\frac{2\pi\alpha'}{r^{2}}L=\frac{1}{2}j^{a}\overline j_{a}+g_1(j^{\alpha} \overline j^{\widehat\alpha}+\overline j^{\alpha}j^{\widehat\alpha})\eta_{\widehat\alpha\alpha}+g_2(j^{\alpha} \overline j\,^{\widehat\alpha}-\overline j\,^{\alpha} j^{\widehat\alpha})\eta_{\widehat\alpha\alpha}+   
\end{equation}
\begin{equation}
+\overline\nabla\lambda^{\alpha}w_{\alpha}+\nabla\widehat\lambda^{\widehat\alpha}\widehat w_{ \widehat\alpha}+g_3\frac{1}{4}R_{abcd}\widehat N^{dc}N^{ba} +g_4w^{*}_{\alpha}\widehat w^{*}_{\widehat\alpha}\eta^{\widehat\alpha \alpha}
\end{equation}
where
\begin{equation}
\lambda^{\alpha}\gamma^{a}_{\alpha\beta}\lambda^{\beta}=0\,,\qquad  \widehat\lambda^{\widehat\alpha}\gamma^{a}_{\widehat\alpha\widehat\beta}\widehat\lambda^{\widehat\beta}=0\,. 
\end{equation}
To fix the coupling constants $g_{1}$, $g_2$, $g_{3}$ and $g_4$ we impose BRST invariance. The BRST transformation of the above Lagrangian is given by
\begin{equation}
Q\frac{2\pi\alpha'}{r^{2}}L =\frac{1}{2}\left[\lambda^{\alpha}(j^{\beta}\overline j^{c}+\overline j^{\beta}j^{c})\gamma_{c\beta\alpha}+\widehat\lambda^{\widehat\alpha}(j^{\widehat\beta}\overline j^{c}+\overline j^{\widehat\beta}j^{c})\gamma_{c\widehat\beta\widehat\alpha}\right]+   
\end{equation}
\begin{equation}
+\frac{g_1}{2}\left[\lambda^{\alpha}(j^{\beta}\overline j^{c}+\overline j^{\beta}j^{c})\gamma_{c\beta\alpha}+\widehat\lambda^{\widehat\alpha}(E^{\widehat\beta}\overline j^{c}+\overline j^{\widehat\beta}j^{c})\gamma_{c\widehat\beta\widehat\alpha}\right] +   
\end{equation}
\begin{equation}
+\frac{g_2}{2}\left[\lambda^{\alpha}(j^{\beta}\overline j^{c}-\overline j^{\beta}j^{c})\gamma_{c\beta\alpha}-\widehat\lambda^{\widehat\alpha}(j^{\widehat\beta}\overline j^{c}-\overline j^{\widehat\beta}j^{c})\gamma_{c\widehat\beta\widehat\alpha}\right] +   
\end{equation}
\begin{equation}
+(g_{1}+g_2)\left[\nabla\lambda^{\alpha}\overline j^{\widehat\alpha}\eta_{\widehat\alpha\alpha}+\overline\nabla\lambda^{\widehat\alpha}j^{\alpha}\eta_{\alpha\widehat\alpha}\right]+(g_{1}-g_2)\left[\overline\nabla\lambda^{\alpha}j^{\widehat\alpha}\eta_{\widehat\alpha\alpha}+\nabla\lambda^{\widehat\alpha}\overline j^{\alpha}\eta_{\alpha\widehat\alpha}\right]+     
\end{equation}
\begin{equation}
+Q\left[w_{\alpha}\overline\nabla\lambda^{\alpha}+\widehat w_{ \widehat\alpha}\nabla\widehat\lambda^{\widehat\alpha}+g_3\frac{1}{4}R_{abcd}\widehat N^{dc}N^{ba} +g_4w^{*}_{\alpha}\widehat w^{*}_{\widehat\alpha}\eta^{\widehat\alpha \alpha}\right]\,.   
\end{equation}
Imposing tBRST invariance immediately fixes
\begin{equation}
g_1=-1\,.    
\end{equation}
Using the Maurer-Cartan identities exhibited in \ref{MC identity} we have
\begin{equation}
Q\frac{2\pi\alpha'}{r^{2}}L=-g_2\left[\lambda^{\alpha}\left(\nabla \overline j^{\widehat\alpha}\eta_{\widehat\alpha\alpha}-\overline\nabla j^{\widehat\alpha}\eta_{\widehat\alpha\alpha}\right)-\widehat\lambda^{\widehat\alpha}\left(\nabla \overline j^{\alpha}\eta_{\alpha\widehat\alpha}-\overline\nabla j^{\alpha}\eta_{\alpha\widehat\alpha}\right)\right]  +  
\end{equation}
\begin{equation}
+(g_{1}+g_2)\left[\nabla\lambda^{\alpha}\overline j^{\widehat\alpha}\eta_{\widehat\alpha\alpha}+\overline\nabla\lambda^{\widehat\alpha}j^{\alpha}\eta_{\alpha\widehat\alpha}\right]+(g_{1}-g_2)\left[\overline\nabla\lambda^{\alpha}E^{\widehat\alpha}\eta_{\widehat\alpha\alpha}+\nabla\lambda^{\widehat\alpha}\overline j^{\alpha}\eta_{\alpha\widehat\alpha}\right]+     
\end{equation}
\begin{equation}
+Q\left[w_{\alpha}\overline\nabla\lambda^{\alpha}+\widehat w_{ \widehat\alpha}\nabla\widehat\lambda^{\widehat\alpha}+g_3\frac{1}{4}R_{abcd}\widehat N^{dc}N^{ba} +g_4w^{*}_{\alpha}\widehat w^{*}_{\widehat\alpha}\eta^{\widehat\alpha \alpha}\right]\,,   
\end{equation}
such that, up to total derivaitves, we end up with
\begin{equation}
Q\frac{2\pi\alpha'}{r^{2}}L\cong (g_{1}+2g_2)\left[\nabla\lambda^{\alpha}\overline j^{\widehat\alpha}\eta_{\widehat\alpha\alpha}+\overline\nabla\lambda^{\widehat\alpha}j^{\alpha}\eta_{\alpha\widehat\alpha}\right]+
\end{equation}
\begin{equation}
+(g_{1}-2g_2)\left[\overline\nabla\lambda^{\alpha}j^{\widehat\alpha}\eta_{\widehat\alpha\alpha}+\nabla\lambda^{\widehat\alpha}\overline j^{\alpha}\eta_{\alpha\widehat\alpha}\right]+     
\end{equation}
\begin{equation}
+Q\left[w_{\alpha}\overline\nabla\lambda^{\alpha}+\widehat w_{ \widehat\alpha}\nabla\widehat\lambda^{\widehat\alpha}+g_3\frac{1}{4}R_{abcd}\widehat N^{dc}N^{ba} +g_4w^{*}_{\alpha}\widehat w^{*}_{\widehat\alpha}\eta^{\widehat\alpha \alpha}\right]\,.   
\end{equation}
Imposing BRST invariance in what is left fixes one more coupling constant
\begin{equation}
g_2=-\frac{1}{2}g_1=\frac{1}{2}    
\end{equation}
and fixes the BRST transformations for the canonical conjugates for the pure spinor variables:
\begin{equation}
\nabla_{\lambda}w_{\alpha}=2 j^{\widehat\alpha}\eta_{\widehat\alpha\alpha}+w_{\alpha}^{*}\,,\qquad \nabla_{\lambda}\widehat w_{\widehat\alpha}=2\overline j^{\alpha}\eta_{\alpha\widehat\alpha}+\widehat w_{\widehat \alpha}^{*}\,.   
\end{equation}
We are now left with the ghost sector:
\begin{equation}
Q\frac{2\pi\alpha'}{r^{2}}L=\left[w_{\alpha}Q\overline\nabla\lambda^{\alpha}+\widehat w_{\widehat\alpha}Q\nabla\widehat\lambda^{\widehat\alpha}\right]+
\end{equation}
\begin{equation}
+\left[\frac{g_3}{4}R_{abcd}Q(\widehat N^{dc}N^{ba})\right]+
\end{equation}
\begin{equation}
+\left[w_{\alpha}^{*}\overline\nabla\lambda^{\alpha}+\widehat w_{ \widehat\alpha}^{*}\nabla\widehat\lambda^{\widehat\alpha}+Q\left(g_4w^{*}_{\alpha}\widehat w^{*}_{\widehat\alpha}\eta^{\widehat\alpha \alpha}\right)\right]\,.
\end{equation}
Where the BRST transformation acting in $\overline\nabla\lambda^{\alpha}$ and $\nabla\widehat\lambda^{\widehat\alpha}$ sees the spin connection in:
\begin{equation}
\overline\nabla\lambda^{\alpha}=\overline\partial\lambda^{\alpha}-\frac{1}{4}\overline\Omega^{ab}\gamma_{ba}\,^{\alpha}\,_{\beta}\lambda^{\beta}\,,\qquad  \nabla\widehat\lambda^{\widehat\alpha}=\overline\partial\widehat\lambda^{\widehat\alpha}-\frac{1}{4}\Omega^{ab}\gamma_{ba}\,^{\widehat\alpha}\,_{\widehat\beta}\widehat\lambda^{\widehat\beta}\,,   
\end{equation}
as discussed in \ref{BRST transformation - appedinx action}. Now, using the identities
\begin{equation}
(\gamma_{ba}\lambda)^{\alpha}(\lambda\gamma^{[a}\eta\gamma^{b]})_{\widehat\alpha}=0\,,\qquad (\gamma_{ba}\widehat\lambda)^{\widehat\alpha}(\widehat\lambda\gamma^{[a}\eta\gamma^{b]})_{\alpha}=0     
\end{equation}
we are left with
\begin{equation}
Q\frac{2\pi\alpha'}{r^{2}}L=\frac{1}{4}\left[N_{ab}(\widehat\lambda\gamma^{[b}\eta\gamma^{a]})_{\alpha}\overline j^{\alpha}+\widehat N_{ab}(\lambda\gamma^{[b}\eta\gamma^{a]})_{\widehat\alpha}j^{\widehat \alpha}\right]+    
\end{equation}
\begin{equation}
+\left[\frac{g_3}{4}R_{abcd}Q(\widehat N^{dc}N^{ba})\right]+
\end{equation}
\begin{equation}
+\left[w_{\alpha}^{*}\overline\nabla\lambda^{\alpha}+\widehat w_{ \widehat\alpha}^{*}\nabla\widehat\lambda^{\widehat\alpha}+Q\left(g_4w^{*}_{\alpha}\widehat w^{*}_{\widehat\alpha}\eta^{\widehat\alpha \alpha}\right)\right]\,.
\end{equation}
Now, plugging the expressions for $N_{ab}$ and $\widehat N_{ab}$:
\begin{equation}
N_{ab}=\frac{1}{2}(w\gamma_{ab}\lambda)\,,\qquad \widehat N_{ab}=\frac{1}{2}(\widehat w\gamma_{ab}\widehat\lambda)    
\end{equation}
and using the identity
\begin{equation}
\frac{1}{2}R_{ab}\,^{cd}(\gamma_{dc}\eta)_{\alpha\widehat\alpha}=\frac{\kappa}{2R^{2}}\delta_{a}^{[c}\delta_{b}^{d]}(\gamma_{dc}\eta)_{\alpha\widehat\alpha}=-\frac{\kappa}{R^{2}}(\gamma_{ab}\eta)_{\alpha\widehat\alpha}=+\frac{1}{R^{2}}(\gamma_{[a}\eta\gamma_{b]})_{\alpha\widehat\alpha} 
\end{equation}
we obtain
\begin{equation}
Q\frac{2\pi\alpha'}{r^{2}}L=\frac{1}{4}\left[\lambda^{\alpha}j^{\widehat \alpha}\widehat N^{ab}(\gamma_{[b}\eta\gamma_{a]})_{\alpha\widehat\alpha}+\widehat\lambda^{\widehat\alpha} \overline j^{\alpha} N^{ab}(\gamma_{[b}\eta\gamma_{a]})_{\alpha\widehat\alpha}\right]+
\end{equation}
\begin{equation}
+\frac{g_3}{4}\left[\lambda^{\alpha}j^{\widehat\alpha}\widehat N^{ab}(\gamma_{[b}\eta\gamma_{a]})_{\alpha\widehat\alpha}+\widehat\lambda^{\widehat\alpha}\overline j^{\alpha}N^{ab}(\gamma_{[b}\eta\gamma_{a]})_{\alpha\widehat\alpha}\right]+
\end{equation}
\begin{equation}
+\frac{g_3}{8}R_{abcd}\left[(w^{*}\gamma^{dc}\lambda)\widehat N^{ba}+(\widehat w^{*}\gamma^{dc}\widehat\lambda) N^{ba}\right]+    
\end{equation}
\begin{equation}
+\left[w_{\alpha}^{*}\overline\nabla\lambda^{\alpha}+\widehat w_{ \widehat\alpha}^{*}\nabla\widehat\lambda^{\widehat\alpha}+Q\left(g_4w^{*}_{\alpha}\widehat w^{*}_{\widehat\alpha}\eta^{\widehat\alpha \alpha}\right)\right]\,,
\end{equation}
and BRST invariance fixes one more coupling constant:
\begin{equation}
g_3=-1\,. 
\end{equation}
We are now left with
\begin{equation}
Q\frac{2\pi\alpha'}{R^{2}}L=w^{*}_{\alpha}\left[\overline\nabla\lambda^{\alpha}-\frac{1}{8}R_{abcd}\widehat N^{dc}(\gamma^{ba}\lambda)^{\alpha}\right]+\widehat w^{*}_{\widehat\alpha}\left[\nabla\widehat\lambda^{\widehat\alpha}-\frac{1}{8}R_{abcd}N^{dc}(\gamma^{ba}\widehat\lambda)^{\widehat\alpha}\right] +   
\end{equation}
\begin{equation}
+g_4\left[\widehat w^{*}_{\widehat\alpha}\eta^{\widehat\alpha \alpha}Qw^{*}_{\alpha}+w^{*}_{\alpha}\eta^{\alpha \widehat\alpha}Q\widehat w^{*}_{\widehat\alpha}\right]\,,
\end{equation}
where BRST invariance fixes the anti-ghost BRST transformations to be:
\begin{equation}
Q\widehat w_{\widehat \alpha}^{*}= \eta_{\widehat\alpha\alpha}\overline\nabla\lambda^{\alpha}-\frac{1}{8}R_{abcd}\widehat N^{dc}(\eta\gamma^{ba}\lambda)_{\widehat\alpha}=\eta_{\widehat\alpha\alpha}\overline\nabla\lambda^{\alpha}-\frac{1}{4}\widehat N^{ab}(\gamma_{b}\eta\gamma_{a}\lambda)_{\widehat\alpha} \,,  
\end{equation}
\begin{equation}
Q w_{ \alpha}^{*}= \eta_{\alpha\widehat\alpha}\nabla\widehat\lambda^{\widehat\alpha}-\frac{1}{8}R_{abcd} N^{dc}(\eta\gamma^{ba}\widehat\lambda)_{\alpha}=\eta_{\alpha\widehat\alpha}\nabla\widehat\lambda^{\widehat\alpha}-\frac{1}{4} N^{ab}(\gamma_{b}\eta\gamma_{a}\widehat\lambda)_{\alpha}\,,
\end{equation}
and the last coupling constant to be $g_4 = -1$\,.

\section{Gamma matrices}

\label{important matrices}
\label{Gamma matrices}

In this appendix we will exhibit the $16\times16$ off-diagonal blocks of our hermitian gamma matrices presented in \ref{gamma off}, without using charge conjugation matrix $\ref{C off}$ to eliminate the anti-chiral spinorial dotted indices $\dot\alpha$, together with the off-diagonal blocks of the charge conjugation matrix presented in \ref{C off}. They are given as follows:
\begin{equation*}
\gamma_{1}\,^{\dot\alpha}\,_{\alpha} = \left[\begin{array}{cccccccccccccccc}+1 & 0 & 0 & 0 & 0 & 0 & 0 & 0 & 0 & 0 & 0 & 0 & 0 & 0 & 0 & 0\\0 & -1 & 0 & 0 & 0 & 0 & 0 & 0 & 0 & 0 & 0 & 0 & 0 & 0 & 0 & 0\\0 & 0 & -1 & 0 & 0 & 0 & 0 & 0 & 0 & 0 & 0 & 0 & 0 & 0 & 0 & 0\\0 & 0 & 0 & +1 & 0 & 0 & 0 & 0 & 0 & 0 & 0 & 0 & 0 & 0 & 0 & 0\\0 & 0 & 0 & 0 & -1 & 0 & 0 & 0 & 0 & 0 & 0 & 0 & 0 & 0 & 0 & 0\\0 & 0 & 0 & 0 & 0 & +1 & 0 & 0 & 0 & 0 & 0 & 0 & 0 & 0 & 0 & 0\\0 & 0 & 0 & 0 & 0 & 0 & +1 & 0 & 0 & 0 & 0 & 0 & 0 & 0 & 0 & 0\\0 & 0 & 0 & 0 & 0 & 0 & 0 & -1 & 0 & 0 & 0 & 0 & 0 & 0 & 0 & 0\\0 & 0 & 0 & 0 & 0 & 0 & 0 & 0 & -1 & 0 & 0 & 0 & 0 & 0 & 0 & 0\\0 & 0 & 0 & 0 & 0 & 0 & 0 & 0 & 0 & +1 & 0 & 0 & 0 & 0 & 0 & 0\\0 & 0 & 0 & 0 & 0 & 0 & 0 & 0 & 0 & 0 & +1 & 0 & 0 & 0 & 0 & 0\\0 & 0 & 0 & 0 & 0 & 0 & 0 & 0 & 0 & 0 & 0 & -1 & 0 & 0 & 0 & 0\\0 & 0 & 0 & 0 & 0 & 0 & 0 & 0 & 0 & 0 & 0 & 0 & +1 & 0 & 0 & 0\\0 & 0 & 0 & 0 & 0 & 0 & 0 & 0 & 0 & 0 & 0 & 0 & 0 & -1 & 0 & 0\\0 & 0 & 0 & 0 & 0 & 0 & 0 & 0 & 0 & 0 & 0 & 0 & 0 & 0 & -1 & 0\\0 & 0 & 0 & 0 & 0 & 0 & 0 & 0 & 0 & 0 & 0 & 0 & 0 & 0 & 0 & +1\end{array}\right]
\end{equation*}

\begin{equation*}
\gamma_{2}\,^{\dot\alpha}\,_{\alpha}=   \left[\begin{array}{cccccccccccccccc}0 & +1 & 0 & 0 & 0 & 0 & 0 & 0 & 0 & 0 & 0 & 0 & 0 & 0 & 0 & 0\\+1 & 0 & 0 & 0 & 0 & 0 & 0 & 0 & 0 & 0 & 0 & 0 & 0 & 0 & 0 & 0\\0 & 0 & 0 & -1 & 0 & 0 & 0 & 0 & 0 & 0 & 0 & 0 & 0 & 0 & 0 & 0\\0 & 0 & -1 & 0 & 0 & 0 & 0 & 0 & 0 & 0 & 0 & 0 & 0 & 0 & 0 & 0\\0 & 0 & 0 & 0 & 0 & -1 & 0 & 0 & 0 & 0 & 0 & 0 & 0 & 0 & 0 & 0\\0 & 0 & 0 & 0 & -1 & 0 & 0 & 0 & 0 & 0 & 0 & 0 & 0 & 0 & 0 & 0\\0 & 0 & 0 & 0 & 0 & 0 & 0 & +1 & 0 & 0 & 0 & 0 & 0 & 0 & 0 & 0\\0 & 0 & 0 & 0 & 0 & 0 & +1 & 0 & 0 & 0 & 0 & 0 & 0 & 0 & 0 & 0\\0 & 0 & 0 & 0 & 0 & 0 & 0 & 0 & 0 & -1 & 0 & 0 & 0 & 0 & 0 & 0\\0 & 0 & 0 & 0 & 0 & 0 & 0 & 0 & -1 & 0 & 0 & 0 & 0 & 0 & 0 & 0\\0 & 0 & 0 & 0 & 0 & 0 & 0 & 0 & 0 & 0 & 0 & +1 & 0 & 0 & 0 & 0\\0 & 0 & 0 & 0 & 0 & 0 & 0 & 0 & 0 & 0 & +1 & 0 & 0 & 0 & 0 & 0\\0 & 0 & 0 & 0 & 0 & 0 & 0 & 0 & 0 & 0 & 0 & 0 & 0 & +1 & 0 & 0\\0 & 0 & 0 & 0 & 0 & 0 & 0 & 0 & 0 & 0 & 0 & 0 & +1 & 0 & 0 & 0\\0 & 0 & 0 & 0 & 0 & 0 & 0 & 0 & 0 & 0 & 0 & 0 & 0 & 0 & 0 & -1\\0 & 0 & 0 & 0 & 0 & 0 & 0 & 0 & 0 & 0 & 0 & 0 & 0 & 0 & -1 & 0\end{array}\right]
\end{equation*}

\begin{equation*}
\gamma_{3}\,^{\dot\alpha}\,_{\alpha}=    \left[\begin{array}{cccccccccccccccc}0 & 0 & +1 & 0 & 0 & 0 & 0 & 0 & 0 & 0 & 0 & 0 & 0 & 0 & 0 & 0\\0 & 0 & 0 & +1 & 0 & 0 & 0 & 0 & 0 & 0 & 0 & 0 & 0 & 0 & 0 & 0\\+1 & 0 & 0 & 0 & 0 & 0 & 0 & 0 & 0 & 0 & 0 & 0 & 0 & 0 & 0 & 0\\0 & +1 & 0 & 0 & 0 & 0 & 0 & 0 & 0 & 0 & 0 & 0 & 0 & 0 & 0 & 0\\0 & 0 & 0 & 0 & 0 & 0 & -1 & 0 & 0 & 0 & 0 & 0 & 0 & 0 & 0 & 0\\0 & 0 & 0 & 0 & 0 & 0 & 0 & -1 & 0 & 0 & 0 & 0 & 0 & 0 & 0 & 0\\0 & 0 & 0 & 0 & -1 & 0 & 0 & 0 & 0 & 0 & 0 & 0 & 0 & 0 & 0 & 0\\0 & 0 & 0 & 0 & 0 & -1 & 0 & 0 & 0 & 0 & 0 & 0 & 0 & 0 & 0 & 0\\0 & 0 & 0 & 0 & 0 & 0 & 0 & 0 & 0 & 0 & -1 & 0 & 0 & 0 & 0 & 0\\0 & 0 & 0 & 0 & 0 & 0 & 0 & 0 & 0 & 0 & 0 & -1 & 0 & 0 & 0 & 0\\0 & 0 & 0 & 0 & 0 & 0 & 0 & 0 & -1 & 0 & 0 & 0 & 0 & 0 & 0 & 0\\0 & 0 & 0 & 0 & 0 & 0 & 0 & 0 & 0 & -1 & 0 & 0 & 0 & 0 & 0 & 0\\0 & 0 & 0 & 0 & 0 & 0 & 0 & 0 & 0 & 0 & 0 & 0 & 0 & 0 & +1 & 0\\0 & 0 & 0 & 0 & 0 & 0 & 0 & 0 & 0 & 0 & 0 & 0 & 0 & 0 & 0 & +1\\0 & 0 & 0 & 0 & 0 & 0 & 0 & 0 & 0 & 0 & 0 & 0 & +1 & 0 & 0 & 0\\0 & 0 & 0 & 0 & 0 & 0 & 0 & 0 & 0 & 0 & 0 & 0 & 0 & +1 & 0 & 0\end{array}\right]
\end{equation*}

\begin{equation*}
\gamma_{4}\,^{\dot\alpha}\,_{\alpha}=    \left[\begin{array}{cccccccccccccccc}0 & 0 & 0 & 0 & +1 & 0 & 0 & 0 & 0 & 0 & 0 & 0 & 0 & 0 & 0 & 0\\0 & 0 & 0 & 0 & 0 & +1 & 0 & 0 & 0 & 0 & 0 & 0 & 0 & 0 & 0 & 0\\0 & 0 & 0 & 0 & 0 & 0 & +1 & 0 & 0 & 0 & 0 & 0 & 0 & 0 & 0 & 0\\0 & 0 & 0 & 0 & 0 & 0 & 0 & +1 & 0 & 0 & 0 & 0 & 0 & 0 & 0 & 0\\+1 & 0 & 0 & 0 & 0 & 0 & 0 & 0 & 0 & 0 & 0 & 0 & 0 & 0 & 0 & 0\\0 & +1 & 0 & 0 & 0 & 0 & 0 & 0 & 0 & 0 & 0 & 0 & 0 & 0 & 0 & 0\\0 & 0 & +1 & 0 & 0 & 0 & 0 & 0 & 0 & 0 & 0 & 0 & 0 & 0 & 0 & 0\\0 & 0 & 0 & +1 & 0 & 0 & 0 & 0 & 0 & 0 & 0 & 0 & 0 & 0 & 0 & 0\\0 & 0 & 0 & 0 & 0 & 0 & 0 & 0 & 0 & 0 & 0 & 0 & -1 & 0 & 0 & 0\\0 & 0 & 0 & 0 & 0 & 0 & 0 & 0 & 0 & 0 & 0 & 0 & 0 & -1 & 0 & 0\\0 & 0 & 0 & 0 & 0 & 0 & 0 & 0 & 0 & 0 & 0 & 0 & 0 & 0 & -1 & 0\\0 & 0 & 0 & 0 & 0 & 0 & 0 & 0 & 0 & 0 & 0 & 0 & 0 & 0 & 0 & -1\\0 & 0 & 0 & 0 & 0 & 0 & 0 & 0 & -1 & 0 & 0 & 0 & 0 & 0 & 0 & 0\\0 & 0 & 0 & 0 & 0 & 0 & 0 & 0 & 0 & -1 & 0 & 0 & 0 & 0 & 0 & 0\\0 & 0 & 0 & 0 & 0 & 0 & 0 & 0 & 0 & 0 & -1 & 0 & 0 & 0 & 0 & 0\\0 & 0 & 0 & 0 & 0 & 0 & 0 & 0 & 0 & 0 & 0 & -1 & 0 & 0 & 0 & 0\end{array}\right]
\end{equation*}

\begin{equation*}
\gamma_{5}\,^{\dot\alpha}\,_{\alpha}=    \left[\begin{array}{cccccccccccccccc}0 & 0 & 0 & 0 & 0 & 0 & 0 & 0 & +1 & 0 & 0 & 0 & 0 & 0 & 0 & 0\\0 & 0 & 0 & 0 & 0 & 0 & 0 & 0 & 0 & +1 & 0 & 0 & 0 & 0 & 0 & 0\\0 & 0 & 0 & 0 & 0 & 0 & 0 & 0 & 0 & 0 & +1 & 0 & 0 & 0 & 0 & 0\\0 & 0 & 0 & 0 & 0 & 0 & 0 & 0 & 0 & 0 & 0 & +1 & 0 & 0 & 0 & 0\\0 & 0 & 0 & 0 & 0 & 0 & 0 & 0 & 0 & 0 & 0 & 0 & +1 & 0 & 0 & 0\\0 & 0 & 0 & 0 & 0 & 0 & 0 & 0 & 0 & 0 & 0 & 0 & 0 & +1 & 0 & 0\\0 & 0 & 0 & 0 & 0 & 0 & 0 & 0 & 0 & 0 & 0 & 0 & 0 & 0 & +1 & 0\\0 & 0 & 0 & 0 & 0 & 0 & 0 & 0 & 0 & 0 & 0 & 0 & 0 & 0 & 0 & +1\\+1 & 0 & 0 & 0 & 0 & 0 & 0 & 0 & 0 & 0 & 0 & 0 & 0 & 0 & 0 & 0\\0 & +1 & 0 & 0 & 0 & 0 & 0 & 0 & 0 & 0 & 0 & 0 & 0 & 0 & 0 & 0\\0 & 0 & +1 & 0 & 0 & 0 & 0 & 0 & 0 & 0 & 0 & 0 & 0 & 0 & 0 & 0\\0 & 0 & 0 & +1 & 0 & 0 & 0 & 0 & 0 & 0 & 0 & 0 & 0 & 0 & 0 & 0\\0 & 0 & 0 & 0 & +1 & 0 & 0 & 0 & 0 & 0 & 0 & 0 & 0 & 0 & 0 & 0\\0 & 0 & 0 & 0 & 0 & +1 & 0 & 0 & 0 & 0 & 0 & 0 & 0 & 0 & 0 & 0\\0 & 0 & 0 & 0 & 0 & 0 & +1 & 0 & 0 & 0 & 0 & 0 & 0 & 0 & 0 & 0\\0 & 0 & 0 & 0 & 0 & 0 & 0 & +1 & 0 & 0 & 0 & 0 & 0 & 0 & 0 & 0\end{array}\right]
\end{equation*}
\begin{equation*}
\gamma_{6}\,^{\dot\alpha}\,_{\alpha}=    \left[\begin{array}{cccccccccccccccc}- i & 0 & 0 & 0 & 0 & 0 & 0 & 0 & 0 & 0 & 0 & 0 & 0 & 0 & 0 & 0\\0 & - i & 0 & 0 & 0 & 0 & 0 & 0 & 0 & 0 & 0 & 0 & 0 & 0 & 0 & 0\\0 & 0 & - i & 0 & 0 & 0 & 0 & 0 & 0 & 0 & 0 & 0 & 0 & 0 & 0 & 0\\0 & 0 & 0 & - i & 0 & 0 & 0 & 0 & 0 & 0 & 0 & 0 & 0 & 0 & 0 & 0\\0 & 0 & 0 & 0 & - i & 0 & 0 & 0 & 0 & 0 & 0 & 0 & 0 & 0 & 0 & 0\\0 & 0 & 0 & 0 & 0 & - i & 0 & 0 & 0 & 0 & 0 & 0 & 0 & 0 & 0 & 0\\0 & 0 & 0 & 0 & 0 & 0 & - i & 0 & 0 & 0 & 0 & 0 & 0 & 0 & 0 & 0\\0 & 0 & 0 & 0 & 0 & 0 & 0 & - i & 0 & 0 & 0 & 0 & 0 & 0 & 0 & 0\\0 & 0 & 0 & 0 & 0 & 0 & 0 & 0 & - i & 0 & 0 & 0 & 0 & 0 & 0 & 0\\0 & 0 & 0 & 0 & 0 & 0 & 0 & 0 & 0 & - i & 0 & 0 & 0 & 0 & 0 & 0\\0 & 0 & 0 & 0 & 0 & 0 & 0 & 0 & 0 & 0 & - i & 0 & 0 & 0 & 0 & 0\\0 & 0 & 0 & 0 & 0 & 0 & 0 & 0 & 0 & 0 & 0 & - i & 0 & 0 & 0 & 0\\0 & 0 & 0 & 0 & 0 & 0 & 0 & 0 & 0 & 0 & 0 & 0 & - i & 0 & 0 & 0\\0 & 0 & 0 & 0 & 0 & 0 & 0 & 0 & 0 & 0 & 0 & 0 & 0 & - i & 0 & 0\\0 & 0 & 0 & 0 & 0 & 0 & 0 & 0 & 0 & 0 & 0 & 0 & 0 & 0 & - i & 0\\0 & 0 & 0 & 0 & 0 & 0 & 0 & 0 & 0 & 0 & 0 & 0 & 0 & 0 & 0 & - i\end{array}\right]
\end{equation*}

\begin{equation*}
\gamma_{7}\,^{\dot\alpha}\,_{\alpha}=    \left[\begin{array}{cccccccccccccccc}0 & +i & 0 & 0 & 0 & 0 & 0 & 0 & 0 & 0 & 0 & 0 & 0 & 0 & 0 & 0\\- i & 0 & 0 & 0 & 0 & 0 & 0 & 0 & 0 & 0 & 0 & 0 & 0 & 0 & 0 & 0\\0 & 0 & 0 & - i & 0 & 0 & 0 & 0 & 0 & 0 & 0 & 0 & 0 & 0 & 0 & 0\\0 & 0 & +i & 0 & 0 & 0 & 0 & 0 & 0 & 0 & 0 & 0 & 0 & 0 & 0 & 0\\0 & 0 & 0 & 0 & 0 & - i & 0 & 0 & 0 & 0 & 0 & 0 & 0 & 0 & 0 & 0\\0 & 0 & 0 & 0 & +i & 0 & 0 & 0 & 0 & 0 & 0 & 0 & 0 & 0 & 0 & 0\\0 & 0 & 0 & 0 & 0 & 0 & 0 & +i & 0 & 0 & 0 & 0 & 0 & 0 & 0 & 0\\0 & 0 & 0 & 0 & 0 & 0 & - i & 0 & 0 & 0 & 0 & 0 & 0 & 0 & 0 & 0\\0 & 0 & 0 & 0 & 0 & 0 & 0 & 0 & 0 & - i & 0 & 0 & 0 & 0 & 0 & 0\\0 & 0 & 0 & 0 & 0 & 0 & 0 & 0 & +i & 0 & 0 & 0 & 0 & 0 & 0 & 0\\0 & 0 & 0 & 0 & 0 & 0 & 0 & 0 & 0 & 0 & 0 & +i & 0 & 0 & 0 & 0\\0 & 0 & 0 & 0 & 0 & 0 & 0 & 0 & 0 & 0 & - i & 0 & 0 & 0 & 0 & 0\\0 & 0 & 0 & 0 & 0 & 0 & 0 & 0 & 0 & 0 & 0 & 0 & 0 & +i & 0 & 0\\0 & 0 & 0 & 0 & 0 & 0 & 0 & 0 & 0 & 0 & 0 & 0 & - i & 0 & 0 & 0\\0 & 0 & 0 & 0 & 0 & 0 & 0 & 0 & 0 & 0 & 0 & 0 & 0 & 0 & 0 & - i\\0 & 0 & 0 & 0 & 0 & 0 & 0 & 0 & 0 & 0 & 0 & 0 & 0 & 0 & +i & 0\end{array}\right]
\end{equation*}
\begin{equation*}
\gamma_{8}\,^{\dot\alpha}\,_{\alpha}=    \left[\begin{array}{cccccccccccccccc}0 & 0 & +i & 0 & 0 & 0 & 0 & 0 & 0 & 0 & 0 & 0 & 0 & 0 & 0 & 0\\0 & 0 & 0 & +i & 0 & 0 & 0 & 0 & 0 & 0 & 0 & 0 & 0 & 0 & 0 & 0\\- i & 0 & 0 & 0 & 0 & 0 & 0 & 0 & 0 & 0 & 0 & 0 & 0 & 0 & 0 & 0\\0 & - i & 0 & 0 & 0 & 0 & 0 & 0 & 0 & 0 & 0 & 0 & 0 & 0 & 0 & 0\\0 & 0 & 0 & 0 & 0 & 0 & - i & 0 & 0 & 0 & 0 & 0 & 0 & 0 & 0 & 0\\0 & 0 & 0 & 0 & 0 & 0 & 0 & - i & 0 & 0 & 0 & 0 & 0 & 0 & 0 & 0\\0 & 0 & 0 & 0 & +i & 0 & 0 & 0 & 0 & 0 & 0 & 0 & 0 & 0 & 0 & 0\\0 & 0 & 0 & 0 & 0 & +i & 0 & 0 & 0 & 0 & 0 & 0 & 0 & 0 & 0 & 0\\0 & 0 & 0 & 0 & 0 & 0 & 0 & 0 & 0 & 0 & - i & 0 & 0 & 0 & 0 & 0\\0 & 0 & 0 & 0 & 0 & 0 & 0 & 0 & 0 & 0 & 0 & - i & 0 & 0 & 0 & 0\\0 & 0 & 0 & 0 & 0 & 0 & 0 & 0 & +i & 0 & 0 & 0 & 0 & 0 & 0 & 0\\0 & 0 & 0 & 0 & 0 & 0 & 0 & 0 & 0 & +i & 0 & 0 & 0 & 0 & 0 & 0\\0 & 0 & 0 & 0 & 0 & 0 & 0 & 0 & 0 & 0 & 0 & 0 & 0 & 0 & +i & 0\\0 & 0 & 0 & 0 & 0 & 0 & 0 & 0 & 0 & 0 & 0 & 0 & 0 & 0 & 0 & +i\\0 & 0 & 0 & 0 & 0 & 0 & 0 & 0 & 0 & 0 & 0 & 0 & - i & 0 & 0 & 0\\0 & 0 & 0 & 0 & 0 & 0 & 0 & 0 & 0 & 0 & 0 & 0 & 0 & - i & 0 & 0\end{array}\right]
\end{equation*}

\begin{equation*}
\gamma_{9}\,^{\dot\alpha}\,_{\alpha}=    \left[\begin{array}{cccccccccccccccc}0 & 0 & 0 & 0 & +i & 0 & 0 & 0 & 0 & 0 & 0 & 0 & 0 & 0 & 0 & 0\\0 & 0 & 0 & 0 & 0 & +i & 0 & 0 & 0 & 0 & 0 & 0 & 0 & 0 & 0 & 0\\0 & 0 & 0 & 0 & 0 & 0 & +i & 0 & 0 & 0 & 0 & 0 & 0 & 0 & 0 & 0\\0 & 0 & 0 & 0 & 0 & 0 & 0 & +i & 0 & 0 & 0 & 0 & 0 & 0 & 0 & 0\\- i & 0 & 0 & 0 & 0 & 0 & 0 & 0 & 0 & 0 & 0 & 0 & 0 & 0 & 0 & 0\\0 & - i & 0 & 0 & 0 & 0 & 0 & 0 & 0 & 0 & 0 & 0 & 0 & 0 & 0 & 0\\0 & 0 & - i & 0 & 0 & 0 & 0 & 0 & 0 & 0 & 0 & 0 & 0 & 0 & 0 & 0\\0 & 0 & 0 & - i & 0 & 0 & 0 & 0 & 0 & 0 & 0 & 0 & 0 & 0 & 0 & 0\\0 & 0 & 0 & 0 & 0 & 0 & 0 & 0 & 0 & 0 & 0 & 0 & - i & 0 & 0 & 0\\0 & 0 & 0 & 0 & 0 & 0 & 0 & 0 & 0 & 0 & 0 & 0 & 0 & - i & 0 & 0\\0 & 0 & 0 & 0 & 0 & 0 & 0 & 0 & 0 & 0 & 0 & 0 & 0 & 0 & - i & 0\\0 & 0 & 0 & 0 & 0 & 0 & 0 & 0 & 0 & 0 & 0 & 0 & 0 & 0 & 0 & - i\\0 & 0 & 0 & 0 & 0 & 0 & 0 & 0 & +i & 0 & 0 & 0 & 0 & 0 & 0 & 0\\0 & 0 & 0 & 0 & 0 & 0 & 0 & 0 & 0 & +i & 0 & 0 & 0 & 0 & 0 & 0\\0 & 0 & 0 & 0 & 0 & 0 & 0 & 0 & 0 & 0 & +i & 0 & 0 & 0 & 0 & 0\\0 & 0 & 0 & 0 & 0 & 0 & 0 & 0 & 0 & 0 & 0 & +i & 0 & 0 & 0 & 0\end{array}\right]
\end{equation*}

\begin{equation*}
\gamma_{10}\,^{\dot\alpha}\,_{\alpha}=    \left[\begin{array}{cccccccccccccccc}0 & 0 & 0 & 0 & 0 & 0 & 0 & 0 & +i & 0 & 0 & 0 & 0 & 0 & 0 & 0\\0 & 0 & 0 & 0 & 0 & 0 & 0 & 0 & 0 & +i & 0 & 0 & 0 & 0 & 0 & 0\\0 & 0 & 0 & 0 & 0 & 0 & 0 & 0 & 0 & 0 & +i & 0 & 0 & 0 & 0 & 0\\0 & 0 & 0 & 0 & 0 & 0 & 0 & 0 & 0 & 0 & 0 & +i & 0 & 0 & 0 & 0\\0 & 0 & 0 & 0 & 0 & 0 & 0 & 0 & 0 & 0 & 0 & 0 & +i & 0 & 0 & 0\\0 & 0 & 0 & 0 & 0 & 0 & 0 & 0 & 0 & 0 & 0 & 0 & 0 & +i & 0 & 0\\0 & 0 & 0 & 0 & 0 & 0 & 0 & 0 & 0 & 0 & 0 & 0 & 0 & 0 & +i & 0\\0 & 0 & 0 & 0 & 0 & 0 & 0 & 0 & 0 & 0 & 0 & 0 & 0 & 0 & 0 & +i\\- i & 0 & 0 & 0 & 0 & 0 & 0 & 0 & 0 & 0 & 0 & 0 & 0 & 0 & 0 & 0\\0 & - i & 0 & 0 & 0 & 0 & 0 & 0 & 0 & 0 & 0 & 0 & 0 & 0 & 0 & 0\\0 & 0 & - i & 0 & 0 & 0 & 0 & 0 & 0 & 0 & 0 & 0 & 0 & 0 & 0 & 0\\0 & 0 & 0 & - i & 0 & 0 & 0 & 0 & 0 & 0 & 0 & 0 & 0 & 0 & 0 & 0\\0 & 0 & 0 & 0 & - i & 0 & 0 & 0 & 0 & 0 & 0 & 0 & 0 & 0 & 0 & 0\\0 & 0 & 0 & 0 & 0 & - i & 0 & 0 & 0 & 0 & 0 & 0 & 0 & 0 & 0 & 0\\0 & 0 & 0 & 0 & 0 & 0 & - i & 0 & 0 & 0 & 0 & 0 & 0 & 0 & 0 & 0\\0 & 0 & 0 & 0 & 0 & 0 & 0 & - i & 0 & 0 & 0 & 0 & 0 & 0 & 0 & 0\end{array}\right]
\end{equation*}
\begin{equation*}
\delta_{\alpha\dot\alpha} = \left[\begin{array}{cccccccccccccccc}0 & 0 & 0 & 0 & 0 & 0 & 0 & 0 & 0 & 0 & 0 & 0 & 0 & 0 & 0 & +1\\0 & 0 & 0 & 0 & 0 & 0 & 0 & 0 & 0 & 0 & 0 & 0 & 0 & 0 & -1 & 0\\0 & 0 & 0 & 0 & 0 & 0 & 0 & 0 & 0 & 0 & 0 & 0 & 0 & +1 & 0 & 0\\0 & 0 & 0 & 0 & 0 & 0 & 0 & 0 & 0 & 0 & 0 & 0 & -1 & 0 & 0 & 0\\0 & 0 & 0 & 0 & 0 & 0 & 0 & 0 & 0 & 0 & 0 & -1 & 0 & 0 & 0 & 0\\0 & 0 & 0 & 0 & 0 & 0 & 0 & 0 & 0 & 0 & +1 & 0 & 0 & 0 & 0 & 0\\0 & 0 & 0 & 0 & 0 & 0 & 0 & 0 & 0 & -1 & 0 & 0 & 0 & 0 & 0 & 0\\0 & 0 & 0 & 0 & 0 & 0 & 0 & 0 & +1 & 0 & 0 & 0 & 0 & 0 & 0 & 0\\0 & 0 & 0 & 0 & 0 & 0 & 0 & +1 & 0 & 0 & 0 & 0 & 0 & 0 & 0 & 0\\0 & 0 & 0 & 0 & 0 & 0 & -1 & 0 & 0 & 0 & 0 & 0 & 0 & 0 & 0 & 0\\0 & 0 & 0 & 0 & 0 & +1 & 0 & 0 & 0 & 0 & 0 & 0 & 0 & 0 & 0 & 0\\0 & 0 & 0 & 0 & -1 & 0 & 0 & 0 & 0 & 0 & 0 & 0 & 0 & 0 & 0 & 0\\0 & 0 & 0 & -1 & 0 & 0 & 0 & 0 & 0 & 0 & 0 & 0 & 0 & 0 & 0 & 0\\0 & 0 & +1 & 0 & 0 & 0 & 0 & 0 & 0 & 0 & 0 & 0 & 0 & 0 & 0 & 0\\0 & -1 & 0 & 0 & 0 & 0 & 0 & 0 & 0 & 0 & 0 & 0 & 0 & 0 & 0 & 0\\+1 & 0 & 0 & 0 & 0 & 0 & 0 & 0 & 0 & 0 & 0 & 0 & 0 & 0 & 0 & 0\end{array}\right] \,.   
\end{equation*}

\end{appendices}

\bibliography{ref.bib}

\end{document}